\documentclass[aps,prx, 
twocolumn,
nofootinbib,
amsmath,amssymb,longbibliography,superscriptaddress, notitlepage]{revtex4-2}
\usepackage[titles]{tocloft}

\usepackage{currfile} 
\usepackage{braket}
\usepackage{amsmath}
\usepackage{amssymb}
\usepackage{enumitem} 
\usepackage{soul}
\usepackage{graphicx}
\usepackage{tikz}
\usetikzlibrary{calc,fadings,decorations.pathreplacing,shapes,shapes.multipart,arrows,shapes.misc,intersections,positioning}
\usepackage{wrapfig}
\usepackage{ytableau}
\usepackage{multirow, tabularx, longtable, makecell, diagbox}
\usepackage{comment}

\usepackage[normalem]{ulem}

\usepackage{hyperref}
\usepackage{xcolor}
\hypersetup{
    colorlinks,
    linkcolor={blue!70!black},
    citecolor={blue!75!},
    urlcolor={blue!80!black}
}

\newcommand{\I}{\mathbb{I}}

\renewcommand{\vec}[1]{\boldsymbol{\mathbf{#1}}}

\newcommand{\bit}{\begin{itemize}}
\newcommand{\eit}{\end{itemize}}

\newcommand{\f}{\frac}

\newcommand{\ba}{\begin{align}}
\newcommand{\ea}{\end{align}}
\newcommand{\be}{\begin{equation}}
\newcommand{\ee}{\end{equation}}
\newcommand{\bi}{\begin{itemize}}
\newcommand{\ei}{\end{itemize}}
\newcommand{\lf}{\left(}
\newcommand{\ri}{\right)}
\newcommand{\dd}{\mathrm{d}}

\newcommand{\id}{\mathbb{I}}
\newcommand{\Tr}{\operatorname{tr}}
\newcommand{\tr}{\operatorname{tr}}

\newcommand{\conj}[1]{{[{#1}]}}

\newcommand{\rrangle}{\rangle\hspace{-0.8mm}\rangle}
\newcommand{\llangle}{\langle\hspace{-0.8mm}\langle}

\newcommand{\kket}[1]
{|{#1}\rrangle }
\newcommand{\bbra}[1]
{\llangle {#1}| }

\newcommand{\bbraket}[1]{\llangle {#1} \rrangle}

\usepackage{xargs}
\usepackage{tikz}
\usetikzlibrary{calc,fadings,decorations.pathreplacing,shapes,shapes.multipart,arrows,shapes.misc,intersections,positioning} 
\graphicspath{{./}{./images/}}
\newcommandx{\fineq}[4][1=-.8ex,2=1,3=1]{
\begin{tikzpicture}[baseline={([yshift=#1]current  bounding  box.center)}, scale = #2, every node/.style={scale = #3}]
#4
\end{tikzpicture}
}

\newcommandx{\swap}[3][1=0,2=0,3=-0.2]{
 \begin{scope}[shift={(#1,#2)}]
    \draw (0,0)--(0,#3);
    \draw[fill=white] (0,0) circle (0.05);
    \draw[fill=white] (0,#3) circle (0.05);
\end{scope}
}	

\begin{document}
\title{Universality classes for purification in nonunitary quantum processes}
\author{Andrea De Luca$^*$}
\affiliation{Laboratoire de Physique Th\'eorique et Mod\'elisation, CY Cergy Paris Universit\'e, \\
\hphantom{$^\dag$}~CNRS, F-95302 Cergy-Pontoise, France}

\author{Chunxiao Liu$^*$}
\affiliation {Department of Physics, University of California, Berkeley, California 94720, USA}

\author{Adam Nahum$^*$}
\affiliation{Laboratoire de Physique de l’\'Ecole Normale Sup\'erieure, CNRS, ENS \& Universit\'e PSL, Sorbonne Universit\'e, Universit\'e Paris Cit\'e, 75005 Paris, France}

\author{Tianci Zhou$^*$}
\affiliation{Department of Physics, Virginia Tech, Blacksburg, Virginia 24061, USA}
\date{\today}

\begin{abstract}
\noindent
We consider universal aspects of two problems:
(i) the slow purification of a large number of qubits by repeated quantum measurements, and
(ii) the singular value structure of a product ${m_t m_{t-1}\ldots m_1}$ of many large random matrices.
Each kind of process is associated with the decay of natural measures of entropy as a function of time or of the number of matrices in the product.
We argue that, for a broad class of models, each process is described by universal scaling forms for purification, and that (i) and (ii) represent distinct ``universality classes'' with distinct scaling functions. 
Using the replica trick, these universality classes correspond to one-dimensional effective statistical mechanics models for a gas of ``kinks'', representing domain walls between elements of the permutation group. (This is an instructive low-dimensional limit of the effective statistical mechanics models for random circuits and tensor networks.) 

These results apply to long-time purification in spatially local monitored circuit models on the entangled side of the measurement phase transition.
\end{abstract}

\maketitle
\def\thefootnote{*}\footnotetext{Alphabetical order of authors' names}\def\thefootnote{\arabic{footnote}}

\section{Introduction}
This paper will mainly be concerned with effective models involving products of random matrices \cite{furstenberg_noncommuting_1963,furstenberg1960products,bellman1954limit,oseledec_multiplicative_1968, bouchard_rigorous_1986}, either drawn independently or with correlations associated with Born's rule for quantum measurements. However, we start with the more structured quantum dynamical processes which are one of the motivations.

Consider a system of ${\mathcal{V}\gg 1}$ qubits undergoing a combination of unitary dynamics and repeated quantum measurement. Such systems can be in either a  ``strongly monitored'' or a ``weakly monitored'' phase \cite{skinner2019measurement,li2018quantum} which can be distinguished via the time-dependence of the total entropy of the qubits, starting from a maximally mixed state with entropy ${\mathcal{V}\log 2}$ \cite{gullans2020dynamical,gullans2020scalable,choi2020quantum}. The repeated measurements typically extract information about the state, leading to a decrease in entropy (i.e. to the purification of the quantum state). In the strongly monitored phase, the entropy decays exponentially with a characteristic decay time of order 1. However, in the weakly monitored phase the characteristic timescale $t_*$ for purification of the state grows exponentially with the system volume $\mathcal{V}$, as has been explored in models in various limits \cite{gullans2020dynamical,li2021statistical,fidkowski2021dynamical,nahum2021measurement,bentsen_measurement-induced_2021, giachetti2023elusive}.

Purification is simply related to the properties of the nonunitary time-evolution operator, $M_t$, that relates the quantum state after evolution for time $t$ to the initial state.
$M_t$ is a random operator since it depends on the measurement outcomes obtained during the dynamics (and on any randomness in the unitary parts of the dynamics).
In a given basis, it is a ${2^\mathcal{V} \times 2^\mathcal{V}}$ matrix, which can be expressed as a product of matrices representing the time sequence of unitary operations and measurements.
If the initial state is maximally mixed then the von Neumann or R\'enyi entropies of the qubits, at time $t$, are simply the natural entropies that can be expressed in terms of the singular values of the matrix $M_t$.

Heuristically, these entropies quantify how much the map ${\psi \rightarrow \f{M_t \psi}{|M_t \psi|}}$ concentrates the uniform distribution on normalized state vectors $\psi$. 
One limit is a unitary map, which leaves the uniform distribution unchanged. The other limit is where one singular value of $M$ is much larger than all the others so that all inputs $\psi$ are mapped to the same output (up to the unphysical overall phases).

We will view $M_t$ as a product of $2^\mathcal{V}\times 2^\mathcal{V}$ matrices, 
\begin{equation}
M = m_t m_{t-1}\ldots m_1.
\end{equation}
Here each $m_i$ is associated with a ``timestep'' in the dynamics.
This is effectively a coarse-graining of the original model (which might for example have a spatial structure in $d$ spatial dimensions plus time, i.e. in $d+1$ dimensions) to an effective $(0+1)$-dimensional model.

In general, the matrices $m_i$ have a highly nontrivial probability distribution that depends on the structure of the qubit problem. We will argue that despite this there is typically a scaling regime at large times where the entropies take universal scaling forms. For example, the average von Neumann entropy $S_1$ may be expressed as a universal function of the natural scaling variable~${x=t/t^*}$.
In turn, these universal forms can be understood from much simpler ``0+1 dimensional''  models which are closely related to the problem of multiplying random matrices drawn from some simple unitarily invariant distribution.
However, if we are modeling true quantum measurements, then the $m_i$ are not independent even in the effective 0+1D  model. Born's rule implies that there are correlations between the $m_i$. Fortunately, these correlations take a simple form which can be handled using the replica trick.

\begin{figure}
    \centering
\includegraphics[width=0.92\columnwidth]{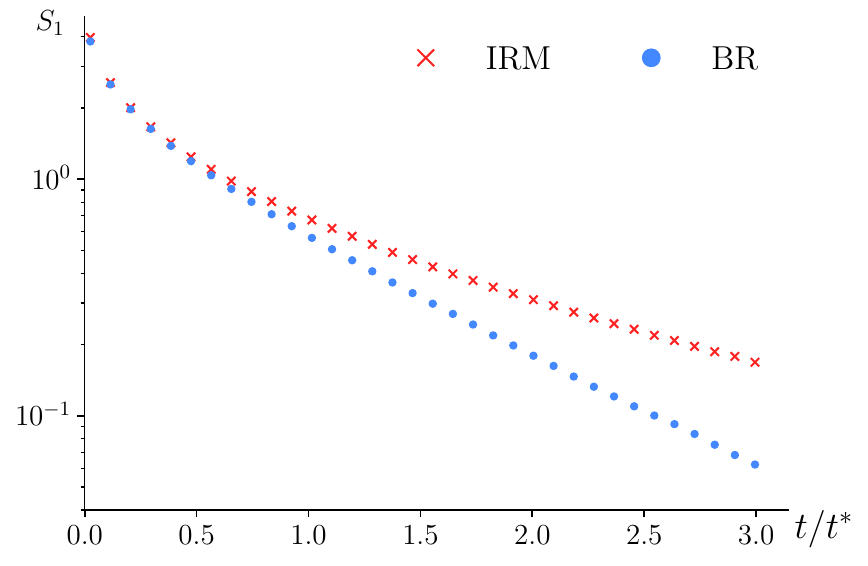}
    \caption{As examples of universal scaling functions we show the decay of the average von Neumann entropy. The $x$ axis is the scaling variable $x=t/t_*$. The red points (crosses) correspond to the multiplication of independent random matrices (IRM), and the blue points (dots) to purification with Born rule measurements (BR). The protocols and error bars of the numerical data are explained in Sec.~\ref{sec:numerics}.}
    \label{fig:intro}
\end{figure}

In fact in the replica formalism it is possible to handle two problems in parallel: 
(i) effective models for monitored qubits, which involve products of random matrices ${m_1, \ldots, m_t}$ with the characteristic correlations induced by Born's rule,\footnote{In the effective model, the probability of a sequence ${(m_1, \ldots, m_t)}$ is proportional to $\Tr [(m_t m_{t-1}\ldots m_1)^\dag(m_t m_{t-1}\ldots m_1)]$.} and 
(ii) the more standard problem of a product $M_t=m_tm_{t-1}\ldots m_1$ of \textit{independent} random matrices $m_i$.
The scaling of the entropies is a natural question in the second setting as well as in the first.
We argue that the two problems are in different universality classes, and that this difference is manifested in the scaling functions describing the mean values and statistical fluctuations of the entropies on timescales of order $t_*$.
As an example, in Fig.~\ref{fig:intro} we show the average von Neumann entropy (as a function of the scaling variable $t/t_*$) for the two types of problem, using numerical data that is explained in Sec.~\ref{sec:numerics}.

The replica formalism involves an effective one-dimensional statistical mechanics problem for a chain of interacting local ``spins'' ${\sigma_1, \sigma_2, \ldots, \sigma_t}$,  which take values in a permutation group $S_N$. A replica limit must be taken, which is $N\to 0$ for a product of independent random matrices and $N\to 1$ for the measurement process. 
This one-dimensional model is the simplest example of a larger family of replica models for interacting permutations in a general number of dimensions, various versions of which describe random tensor networks \cite{vasseur2019entanglement}, random unitary circuits \cite{zhou2019emergent}, and monitored random circuits \cite{jian2020measurement,bao2020theory} (see also related models away from the replica limits \cite{hayden2016holographic,nahum2018operator,von2018operator,chan2018solution,zhou2020entanglement,nahum2017quantum}, as well as discussions of purification in terms of $d+1$ dimensional effective models \cite{li2021statistical,nahum2021measurement}).
In general, these models are not directly solvable, because of the extreme difficulty of taking the replica limit, and this is another reason for trying to understand the one-dimensional case.

\begin{figure}
    \centering
    \includegraphics{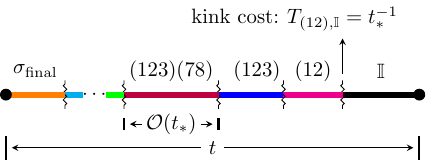}
    \caption{
    The replica calculation leads to a partition function for a one-dimensional gas of kinks --- the figure illustrates a possible configuration. Each kink separates two permutations in $S_N$ that differ by a transposition. The transfer matrix $T$ assigns each kink a cost $1/t_*$.
    As a result, the typical separation between kinks is of order $t_*$, the purification time. }
    \label{fig:kink_gas}
\end{figure}

In one dimension we have a picture in terms of ``kinks'' (point-like domain walls) that separate domains with a given value of the local spin $\sigma_i$: see Fig.~\ref{fig:kink_gas}.
A priori, there is an infinite and complex spectrum of kinks.\footnote{Each kink can itself be labeled by a permutation group element, and for the replica trick it is necessary to consider the permutation group $S_N$ with arbitrarily large $N$.}
However, universality arises from two features: first, the diluteness of the kinks (their typical separation is of order $t_*$), and second that, in the scaling regime, it is sufficient to consider only the simplest ``elementary'' kink.\footnote{Elementary kinks are associated with transpositions, the simplest nontrivial permutations in $S_N$.}
As a result, only a single model-dependent constant is needed, which is the cost of the elementary kink. The inverse of this cost sets the purification timescale $t_*$.

We show that this universality holds even for spatially local qubit models in the appropriate scaling regime. Interestingly, in generic local models, the purification timescale is sensitive to rare events, i.e. to atypical, ``strongly purifying'' terms in the product ${m_t m_{t-1}\cdots m_1}$ of coarse-grained timesteps.

The replica formalism allows some explicit results for the scaling functions. Expansions of the scaling functions at small ${x=t/t_*}$ reduce to combinatorial problems involving walks on the permutation group. At low orders, the counting can be done directly (Sec.~\ref{subsec:S2_order_x2}), but higher orders may also be obtained using results from the mathematical literature, as described below. For certain observables, we are also able to obtain all-orders resummations of the perturbative series. 

The small-$x$ asymptotics of the scaling functions are particularly simple, and imply that for ${t\ll t_*}$ the entropies evolve approximately deterministically. When ${t\ll t_*}$, or equivalently when ${x\ll 1}$, we may write the $n$th R\'enyi entropy as:
\begin{equation}\label{eq:introsmallxasymptotics}
S_n(t) =  - \ln \frac{t}{t_*} - \frac{1}{n-1}\ln \left[ \frac{n^{(n-2)}}{(n-1)!} \right]
+ \Delta_n(t).
\end{equation}
This formula includes the von Neumann entropy as the ${n\to 1}$ limit:
${S_1(t) =  - \ln \frac{t}{t_*}  + 1 - \gamma + \Delta_1(t)}$.
Here $\Delta_n(t)$ is random, i.e. realization-dependent, and its distribution differs between the two universality classes described above. However, $\Delta_n(t)$ is small when $x$ is small (its mean and variance are of order $x^2$).
 Therefore in this limit the scaling forms for the two universality classes coincide.
When the scaling variable ${x=t/t_*}$ is not small, differences between different realizations and the two different universality classes become significant.

The study of products of large numbers of random matrices has a long history  
\cite{bellman1954limit,furstenberg1960products, furstenberg_noncommuting_1963,oseledec_multiplicative_1968,
derrida_singular_1983, bouchard_rigorous_1986, gudowska2003infinite}.
Recently, the eigenvalue statistics of finite product matrix ensembles have been extensively studied in physics~\cite{crisanti2012products} and mathematics~\cite{orlov2017hurwitz}, including for Ginibre, i.e. complex Gaussian, random matrices~\cite{PhysRevE.81.041132}
(including infinite~\cite{PhysRevE.83.061118}, finite size square~\cite{Akemann_2012,Akemann_2013} and  rectangular~\cite{PhysRevE.88.052118, neuschel2014plancherel} Ginibre matrices), as well as products combining Ginibre matrices with inverse Ginibre random matrices~\cite{Forrester_2014, Kartick_2016_determinantal} or with truncated Haar random matrices~\cite{kuijlaars2014singular, gawronski2016jacobi}. For more recent developments, see Refs.~\cite{forrester2016singular,e22090972,halmagyi2020mixed,zavatone2022replica,2021arXiv210907375A}.
Most relevant to the present work are the recent Refs.~\cite{liu_lyapunov_2022,akemann_universality_2020} obtaining universal behaviors of the bulk and edge singular spectra for products of independent Ginibre matrices, classifying them in different scaling regimes using the same scaling variable $x$ as in this paper.
A scaling regime for products of large random matrices (spatial transfer matrices) also arises in the study of spectral form factors~\cite{PhysRevLett.121.060601, Chan2022,PhysRevLett.130.140403}.

The evaluation of partition functions for kinks is closely related to counting paths on the permutation group. This path counting and related combinatorial problems are extensively studied \cite{goulden_labelled_1993,lewis_reflection_2014,mackiw_permutations_1995,moszkowski_solution_1989,poulalhon_factorizations_2002,strehl_minimal_1996,goulden_labelled_1993,lewis_reflection_2014,mackiw_permutations_1995,moszkowski_solution_1989,poulalhon_factorizations_2002,strehl_minimal_1996}.
For the power-series expansion of the scaling functions, we make use of mathematical results on the character approach of Refs.~\cite{STANLEY1981255,babai1979spectra,jackson1988some,murty2020ramanujan,stone_mathematics_2009},
which diagonalizes the transfer matrix in terms of irreducible representations of the permutation group. 
A complete enumeration of the character table allows us to explicitly compute the expansion around small $x$ to $8$th order and analytically continue to the required $N = 0$ or $N = 1$ limit for various R\'enyi entropies. Numerical ratio tests of the expansion coefficients suggest that the small $x$ expansions may be only asymptotic series, so that (in the absence of resummation) they do not give accurate results for arbitrary $x$.
However, for the average of the von Neumann entropy in a quantum measurement process, an analytical technique to sum characters over Plancherel measure~\cite{borodin2000asymptotics} gives rapidly convergent results for any $x$, so that the full scaling function may be obtained semi-analytically. We compare results for the scaling functions with numerical simulations (Sec.~\ref{sec:numerics}), confirming the existence of two universality classes.

\begin{figure}
    \centering
\includegraphics[width=0.92\columnwidth]{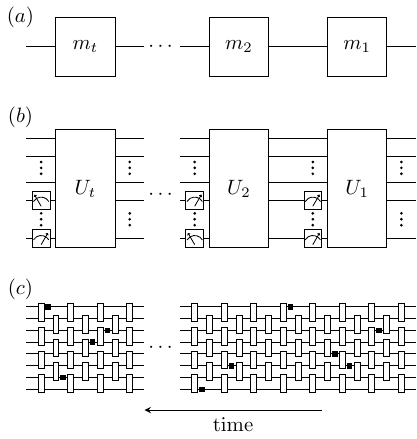}
    \caption{
Models considered in this paper. 
(a)~A product of independent large random matrices $m_i$ drawn from a simple unitary invariant distribution (e.g. Ginibre).
(b)~A simple measurement process for $\mathcal{V}$ qubits, where each timestep involves a Haar-random unitary $U_i$ on the many-body Hilbert space followed by projective measurement of a fraction of the qubits~\cite{fidkowski2021dynamical,nahum2021measurement}.
(c)~Spatially local (or $k$-local) monitored quantum circuits in the entangling (weak-monitoring) phase: rectangles and squares represent unitaries and measurements respectively. In the appropriate scaling regime, (b) and (c) are governed by the same scaling functions.}
    \label{fig:protocols}
\end{figure}

\addtocontents{toc}{\protect\setcounter{tocdepth}{2}}
\tableofcontents

\section{Replica descriptions and universality}\label{sec:replicaanduniversality}

In this section, we describe the reduction of models for random matrix multiplication and monitored dynamics of qubits to effective one-dimensional replica models. 
The one-dimension here is referred to as ``time''. In the matrix-product setting, this is the integer coordinate labeling the successive matrices in the product.

We proceed in three steps, illustrated in Fig.~\ref{fig:protocols}.
We start with the most straightforward setting, which is a product of independent complex random matrices drawn from a simple distribution (Sec.~\ref{sec:ginibre}--\ref{sec:transfermatrixbasics}), see Fig.~\ref{fig:protocols}~(a). As an illustrative example, we discuss Ginibre random matrices (but the results apply to more general cases, as we will discuss).

Next in Sec.~\ref{sec:mea} we consider a very simple model~\cite{fidkowski2021dynamical,nahum2021measurement} of a monitored quantum system, 
in which there is no notion of spatial locality, and no preferred tensor product structure for the Hilbert space [Fig.~\ref{fig:protocols}~(c)]. This brings in the additional ingredient of Born's rule.

For each of these two cases, we discuss the formulation of the replica problem in terms of a transfer matrix $\mathcal{T}$ whose rows and columns are labeled by permutations ${\sigma, \sigma'\in S_N}$. 
The element $\mathcal{T}_{\sigma', \sigma}$ gives the Boltzmann weight (fugacity) for a kink with $\sigma'$ on the left and $\sigma$ on the right. 
We discuss general properties of the transfer matrix (Sec.~\ref{sec:transfermatrixbasics}) and note that the structure of the transfer matrix can give rise to universality, so long as the weights $\mathcal{T}_{\sigma', \sigma}$ for more complex kinks are small compared to the weight of the minimal kink. 

Following the definition of the effective 1D replica statistical mechanics models in Subsections \ref{sec:ginibre}--\ref{sec:mea}, some readers may wish to skip ahead to Sec.~\ref{sec:computations} where we use these effective descriptions to compute scaling functions.

In the final part of the present Section (Sec.~\ref{sec:universality}) we consider more general models, 
including spatially local models [e.g. a 1D chain of qubits subjected to brickwork quantum circuit dynamics --- Fig.~\ref{fig:protocols}~(c)]. 
We argue that a coarse-graining argument reduces them to effective one-dimensional models, which can again be characterized by a replica transfer matrix.  
We argue that for typical models, the resulting transfer matrix satisfies the universality property, 
giving a scaling regime where the universal forms hold. 

Sec.~\ref{sec:universality} also discusses ``short-time'' crossovers that occur in spatially local models,
prior to the onset of the universal regime. Caution is required as for spatially local models, the timescale $t_1$ for the onset of the universal regime (for a given observable) may be very large, 
though still much smaller than the characteristic purification time $t_*$
\begin{equation}
1 \ll t_1 \ll t_*.
\end{equation}

\subsection{Products of independent random matrices}
\label{sec:ginibre}

Consider a $q$-dimensional Hilbert space $\mathcal{H} = \mathbb{C}^q$ and a product of $t$ independent random matrices acting on $\mathcal{H}$
\begin{equation}
\label{eq:M_m}
M = m_t m_{t-1}\ldots m_1.
\end{equation}
We suppress the time argument on ${M=M_t}$.
Each matrix $m=m_i$ is complex, and each of its elements is an independent complex Gaussian variable with mean zero: for all $\alpha,\beta,\gamma,\delta = 1,\ldots, q$, one has
\begin{align}
\label{eq:mcov}
\overline{
m_{\alpha \beta} m^*_{\gamma\delta}} & = \delta_{\alpha \gamma} \delta_{\beta \delta }, 
&
\overline{
m_{\alpha \beta} m_{\gamma\delta}} & = 0,
\end{align}
where we use overlines for disorder averages. 

Let us define a normalized density matrix $\rho$ by
\begin{align}
\rho & = \f{\check \rho}{\tr \check \rho}, 
& 
\check \rho = M M^\dag.
\end{align}
We have also defined the un-normalized density matrix $\check \rho$ which will play an important role for the measurement model of qubits later.

We will wish to average various functionals $F[\rho]$ of the density matrix. In particular, we consider the R\'enyi entropies
\begin{align}
\label{eq:renyi_entropy}
S_n & = \f{1}{1-n} \ln \tr \rho^n,
& 
S_1 & = - \tr \rho \ln \rho,
\end{align}
which probe the (normalized) singular values $p_a$ of the matrix $M$.
To express the disorder average of these quantities, it is useful to introduce the replica formalism (see e.g.~\cite{vasseur2019entanglement,zhou2019emergent,jian2020measurement,bao2020theory,nahum2021measurement,potter2022entanglement,PhysRevX.13.041045} for the higher-dimensional setting). 
Regarding the density matrix as a state in $\mathcal{H} \otimes \mathcal{H}^{\ast}$ (where $\mathcal{H}^\ast$ denotes the dual Hilbert space), we will consider $N$ replicas of the system 
to give a Hilbert space $\mathcal{H}^{\otimes N} \otimes \mathcal{H}^{\ast \otimes N}$.
We then define states $\kket{\sigma}$ in this space, for ${\sigma\in S_N}$, by 
\begin{equation} \label{eq:permutationstatesdefn}
\llangle \alpha_1,\bar\alpha_1,\ldots, \alpha_N, \bar\alpha_N | \sigma \rrangle = \prod_{j=1}^N \delta_{\alpha_j, \bar\alpha_{\sigma_j}}
\end{equation}
where $\alpha,\bar\alpha$ label basis states for $\mathcal{H}$ and $\mathcal{H}^\ast$ respectively. 
The overlaps of these (nonorthogonal) states are 
\begin{equation}\label{eq:overlaps}
\llangle \sigma | \mu \rrangle = q^{N-|\sigma^{-1}\mu|}.
\end{equation}
The exponent is the number of cycles in $\sigma^{-1}\mu$, written using the ``length'' function
$|\bullet|$, which is the minimal number of transpositions needed to express a given permutation. 
A basic point is that $|\tau|$ depends only on the conjugacy class of $\tau$, i.e. on its cycle structure.
We denote this by ${\conj{\tau} = (1^{r_1} 2^{r_2} \ldots )}$, indicating $r_j$ cycles of length $j$.\footnote{We have  $\sum_j j r_j = N$ and ${\sum_j (j-1) r_j = |\tau|}$.}
Some examples are 
\begin{equation}\notag
\begin{array}{|l|c|l|}
\hline
   \,\,\,\, \tau & \,\,\,\,|\tau|\,\,\,\, & \,\,\,\, \,\,\,\,\conj{\tau}\,\,\, \,\,\,\,\,\, \\ \hline
      \,\I  &  0 & 1^N \\ \hline
      (12)   & 1 & 1^{N-2}\,2^1 \\ \hline
\end{array}\,\, , \,\,\,\,\,\,\,\,
\begin{array}{|l|c|l|}
\hline
   \,\,\,\, \tau & \,\,\,\,|\tau|\,\,\,\, & \,\,\,\, \,\,\,\,\conj{\tau}\,\,\, \,\,\,\,\,\, \\ \hline
      (12)(34)   & 2 & 1^{N-4}\,2^2 \\ \hline
      (123)  & 2 & 1^{N-3}\,3^1 \\  \hline
\end{array}\,.
\end{equation}
We will often encounter expressions such as $\bbraket{\I|\sigma}$ whose values depend only on the cycle structure of $\sigma$, so we allow ourselves to write e.g. ${\bbraket{\I|(12)}=\bbraket{1^N|1^{N-2} \, 2}}$.

As a final piece of notation, the density matrix $\check\rho$ corresponds to the state $\kket{\check\rho}$ in ${\mathcal{H}\otimes \mathcal{H}^*}$ such that ${\bbraket{\alpha \bar \alpha | \check \rho} = {\check\rho}_{\alpha\bar\alpha}}$. At ${t=0}$, ${\kket{\check\rho}= \kket{\I}}$.\footnote{In Sec.~\ref{sec:mea}, where we discuss a physical measurement process, it will be more convenient to use the normalization   ${\kket{\check\rho_0}= \kket{\I}}/q_{\rm tot}$  (see text above Eq.~\ref{eq:evolutionofcheckrho}).}
Note that we suppress the time argument on $\check \rho=\check\rho_t$.

It is easy to express the disorder averages (and higher moments) of R\'enyi entropies within this formalism. They are encoded in generating functions (with $k$ as the parameter) of the form
\begin{equation}\label{eq:genfnfirst}
\overline{e^{-k (n-1)S_n}}
= \overline{\Tr[\check \rho^{n}]^k
\Tr[\check \rho]^{ - n k}}.
\end{equation}
As usual, expanding the log of the generating function around ${k=0}$ (or taking derivatives) gives cumulants of the entropies,
\begin{equation}
\overline{e^{-k (n-1)S_n}}
= e^{- k (n-1) \overline{S_n} + \f{k^2(n-1)^2}{2} \overline{(S_n)^2}^c + \ldots}. 
\end{equation}
The replica trick is to express the right hand side of Eq.~\eqref{eq:genfnfirst} as the continuation ${N\to 0}$ of the expression
\begin{equation}\label{eq:unaveragedtraces}
\Tr[\check \rho^{n}]^k
\Tr[\check \rho]^{N - n k}
=
\bbraket{1^{N - nk} n^k  | \check\rho^{\otimes N}}.
\end{equation}
Having obtained the generating function or the moments for a general R\'enyi entropy index $n$, the result for the von Neumann entropy may be obtained by a further limit ${n\to 1}$.

We now consider the time evolution. In a given timestep, ${\kket{\check\rho^{\otimes N}}}$ is evolved by an operator of the form ${{m\otimes m^* \otimes \cdots \otimes m \otimes m^*}}$. It is easy to verify using \eqref{eq:mcov} that the average of this operator is
\begin{equation}\label{eq:msigma}
\overline{m\otimes m^* \otimes \cdots \otimes m \otimes m^* } = \sum_{ \sigma \in S_N } \kket{ \sigma} \bbra{ \sigma } \;.
\end{equation}
Therefore, after evolving by $t$ steps, $\overline{\kket{\check \rho^{\otimes N}}}$ is given by
\begin{align}
 \overline{\kket{ \check\rho^{\otimes N}}} = \sum_{\{\sigma_i\}} 
 \kket{ \sigma_t}  \bbraket{ \sigma_t |  \sigma_{t-1} } 
\cdots 
 \bbraket{ \sigma_2 |  \sigma_{1} }  \bbraket{ \sigma_{1} | 1^N }  \;.
\end{align}
Introducing the transfer matrix as 
\begin{equation}\label{eq:transferginibrefirst}
T_{\sigma', \sigma}  = \llangle \sigma' | \sigma \rrangle, 
\end{equation}
with elements given by (\ref{eq:overlaps}), we can express the generating function for a given R\'enyi entropy (cf. Eqs.~\ref{eq:genfnfirst},~\ref{eq:unaveragedtraces}) as:
\begin{equation}\label{eq:genfnTonlyN0}
    \overline{e^{-k (1-n) S_n}} = \lim_{N\to 0} \lf T^{t+1}\ri_{\sigma_{\text{final}},\,\I},
\end{equation}
if we choose the ``left boundary condition'' $\sigma_\text{final}$ to have the cycle structure ${\conj{\sigma_{\text{final}}} = (1^{N - nk} n^k}$). This is the key object we will analyze.

We will also write the matrix element using the alternative notations
\begin{align}\label{eq:matrixelementnotations}
\lf T^{t+1}\ri_{\sigma_{\text{final}},\,\I}
&=\braket{ \sigma_{\text{final}} | T^{t+1} \, | \I} 
= \braket{ 1^{N - nk} n^k | T^{t+1} \, | 1^N } \; .
\end{align}
Note that the orthonormal basis states $\ket{\sigma}$ in Eq.~\eqref{eq:matrixelementnotations} 
should not be confused with the states $\kket{\sigma}$ defined in Eq.~\eqref{eq:permutationstatesdefn}, which live in a different Hilbert space, and are not orthonormal.

In the replica limit $N\to 0$ the factor $q^N$ in Eq.~\eqref{eq:overlaps} becomes unity and does not contribute to averages, so it is convenient to redefine the transfer matrix as  
 \be\label{eq:transferginibrenormalized}
T_{\sigma', \sigma} := q^{-|\sigma'\sigma^{-1}|} =  q^{-N} \llangle \sigma' | \sigma \rrangle.
\end{equation}
Note that with this convention $T_{\sigma,\sigma}=1$.
For our explicit calculations with transfer matrices we will impose this normalization.

At this point, Eq.~\eqref{eq:genfnTonlyN0} expresses the generating function in terms of a 1D ``spin model'' partition function defined by the transfer matrix $T$. In other words, the matrix element in Eq.~\eqref{eq:genfnTonlyN0} may be written as 
\begin{align}\label{eq:partitionfunctionfirst}
\mathcal{Z}(N,n,k) 
& \equiv 
\langle  \sigma_{\text{final}} | T^{t+1} \, | \I \rangle
\phantom{\bigg(\bigg)}
\notag \\
& = \sum_{\sigma_1,\ldots,\sigma_t}
T_{\sigma_\text{final},\sigma_t}\cdots 
T_{\sigma_2,\sigma_{1}}T_{\sigma_1,\I},
\end{align}
corresponding to a 1D classical spin model, for spins ${\sigma_i\in S_N}$, with fixed boundary conditions:
\begin{equation}\notag
\begin{tikzpicture}[scale = 0.85*0.8,every node/.style={scale = 0.85}]
\draw (0,0)--(5,0);
\node () at (5.5,0) {$\cdots$};
\draw[fill=white] (0,0) circle (0.5) node () {$\sigma_{\rm final}$};
\draw[fill=white] (2,0) circle (0.5) node () {$\sigma_{t}$};
\draw[fill=white] (4,0) circle (0.5) node () {$\sigma_{t-1}$};
\begin{scope}[shift={(7,0)}]
\draw (-1,0)--(4,0);
\draw[fill=white] (0,0) circle (0.5) node () {$\sigma_{2}$};
\draw[fill=white] (2,0) circle (0.5) node () {$\sigma_{1}$};
\draw[fill=white] (4,0) circle (0.5) node () {$\mathbb{I}$};
\end{scope}
\end{tikzpicture}
\end{equation}
However, this transfer matrix can be simplified in the scaling limit, as we discuss below.

\subsection{Scaling limit \& symmetries of effective model}\label{sec:transfermatrixbasics}

Configurations for the partition function in Eq.~\eqref{eq:partitionfunctionfirst},
\begin{align}\label{eq:partitionfunctionsecond}
\mathcal{Z}(N,n,k) 
& = \sum_{\sigma_1,\ldots,\sigma_t}
T_{\sigma_\text{final},\sigma_t}\cdots 
T_{\sigma_2,\sigma_{1}}T_{\sigma_1,\I},
\end{align}
may be visualized in terms of domain walls (kinks) between different spin values.
$T_{\sigma',\sigma}$ is the fugacity for a kink with $\sigma'$ on the left and $\sigma$ on the right.
We describe some basic features of this partition function.

\subsubsection{Simplification in the scaling limit}
\label{sec:simplificationscalinglimit}

In the scaling limit, it is enough to consider only ``elementary'' kinks in which $\sigma'$ and $\sigma$ differ by a single transposition. In the present model, it is straightforward to see this algebraically. 
By the definition of $T$ in Eq.~\eqref{eq:transferginibrenormalized}, 
\begin{equation}\label{eq:TidentityplusA}
T_{\sigma',\sigma}=\delta_{\sigma',\sigma} + A_{\sigma',\sigma}/q + o(1/q),
\end{equation}
where $A_{\sigma', \sigma}$ is the adjacency matrix of the ``transposition network''~\cite{Friedman2000,liu_eigenvalues_2022}: it is equal to 1 if $\sigma'\sigma^{-1}$ is a single transposition, and zero otherwise.
When we consider $T^t$ for $t$ of order $q$, we can drop the $o(q^{-1})$ terms, corresponding to allowing only elementary kinks. 
To be more precise, we define the scaling limit as ${t, q\to\infty}$ with a fixed ratio $t/q = x$:
\begin{equation}
\label{eq:transfmatlimit}
T^t = \Big( \id + A/q + o(q^{-1}) \Big)^t\longrightarrow e^{x A}  .
\end{equation}
In the present model, the fact that the higher terms in Eq.~\eqref{eq:TidentityplusA} are $o(1/q)$ is a trivial algebraic fact, but it holds in a much larger class of models, including spatially local models (with an appropriate identification of $q$) as we will discuss in Sec.~\ref{sec:universality}. This simplification is the algebraic reason for universality. A heuristic way of understanding the ``irrelevance'' of higher kinks, when Eq.~\eqref{eq:TidentityplusA} is satisfied, is by an energy/entropy argument: splitting a higher kink into multiple elementary kinks increases the positional entropy of the kink gas --- App.~\ref{app:energyentropy}.

As a result of Eq.~\eqref{eq:transfmatlimit}, the generating functions and therefore the entropy averages depend only on $x = t/q$ in the scaling limit, instead of depending on the full transfer matrix.
For example, the averages of entropies will take scaling forms,
\begin{align}\label{eq:scalingfnpreview}
\overline{S_1} &  = F_1^{\text{IRM}}(x),
&
\overline{S_2} & = F_2^{\text{IRM}}(x),
\end{align}
where the superscript indicates that we are now considering products of independent random matrices.
Similar forms will hold for the measurement processes discussed in the next section, but with different scaling functions $F^{\text{BR}}$ (BR indicates that measurements are sampled using Born's rule).

We will give explicit results for scaling functions in Sec.~\ref{sec:computations}.
Anticipating slightly, we note that the small $x$ expansions of the scaling functions in Eq.~\eqref{eq:scalingfnpreview} start  with ${\ln (1/x) + \ldots}$ (we will refer to this result in Sec.~\ref{sec:universality}). The small $x$ expansions are related to expansions in the number of kinks.

\subsubsection{Global symmetry and kink labeling}
\label{subsub:global_symm}

Returning to Eq.~\eqref{eq:partitionfunctionsecond}, the effective statistical mechanics model has a global symmetry: the local Boltzmann weights $T_{\sigma',\sigma}$ are unchanged under global symmetry operations\footnote{The symmetry is explicitly broken by the fixed boundary conditions.} of the form ${\sigma  \rightarrow \tau_L\,\sigma \,\tau_R}$, for all spins, where $\tau_L$ and $\tau_R$ are arbitrary permutations, and also under global inversion, ${\sigma  \rightarrow \sigma^{-1}}$. These operations generate the symmetry group ${\mathcal{G} = (S_N\times S_N) \rtimes Z_2}$. This symmetry is shared by all the models we will consider and is a general consequence of the replicated structure
~\cite{zhou2020entanglement}.

\begin{figure}
    \centering
\includegraphics[width=0.8\columnwidth]{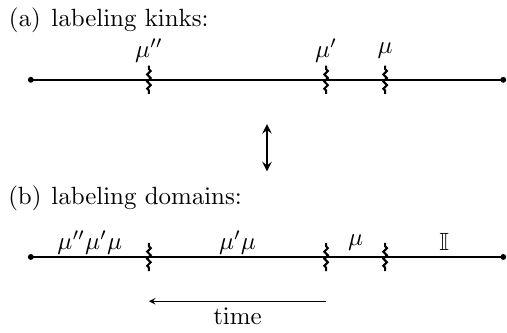}
    \caption{
    We may label configurations by assigning permutation labels to the kinks, 
    as in (a), rather than to the domains as in (b). This figure shows the relation between the two conventions.}
    \label{fig:kinklabelling}
\end{figure}

A kink between domains $\sigma'$ and $\sigma$ can be labeled with a ``kink type'' ${\mu = \sigma'\sigma^{-1}}$. It is sometimes convenient to specify a configuration by labeling the kink types, rather than the domain types: the relation between the two labeling conventions is shown in Fig.~\ref{fig:kinklabelling}. Note that the product of kink types is fixed by the boundary conditions: schematically, ${\overleftarrow{\prod}\mu = \sigma_{\rm final}}$.

A kink of type $\mu$ has a fugacity (Boltzmann weight cost) $T_{\mu, \I}$. As a result of ${S_N\times S_N}$ symmetry, this fugacity depends only on the conjugacy class $[{\mu}]$.
In the present model, the fugacity of an elementary kink such as $\mu = (12)$ is $1/q$.
The next-simplest kinks have the structure ${(12)(34)}$ or ${(123)}$, and in the present model both of these have fugacity~$q^{-2}$. 
In more general models, we will \textit{define} the parameter $q$ via the weight of the \textit{elementary} kink,\footnote{In the present model, a weight like $T_{(12),\I}$ (which exists in the effective model for all  ${N>1}$) is manifestly independent of $N$. More generally, this need not be the case. The quantity ${q^{-1}=T_{(12),\I}}$ that is relevant for scaling forms in the ${N\to 0}$ limit will then be the ${N\to 0}$ continuation of the weights $T_{(12),\I}$, which are defined in the effective model for positive integer $N$. A similar statement applies for the ${N\to 1}$ limit considered in the next subsection.} 
\begin{equation}\label{eq:qdefn}
q^{-1} \equiv T_{(12), \I}
\end{equation} 
(we always normalize so $T_{\I,\I}=1$).
Loosely speaking, $q$ is also an ``effective'' Hilbert space dimension which may be smaller than the actual size of the original matrices.

\subsection{Measurement processes}
\label{sec:mea}

We now describe how transfer matrix expressions similar to those in Sec.~\ref{sec:ginibre} describe quantum trajectories of monitored systems. 
The full transfer matrix for such systems is (necessarily) different from that for the Ginibre ensemble, 
but in the scaling limit it may again reduce to the simplified ``universal'' transfer matrix for elementary kinks Eq.~\eqref{eq:TidentityplusA}. The needed limit is however $N\to 1$.

In this section, we treat the simplest model (discussed in Refs.~\cite{nahum2021measurement,
fidkowski2021dynamical}) in which the dynamics has a statistical invariance under unitary rotations of the entire Hilbert space --- see Fig.~\ref{fig:protocols}~(b). Sec.~\ref{sec:universality} considers models with additional locality structure.

We take a collection of $\mathcal{V}$ qubits, with total Hilbert space dimension
\begin{equation}
q_{\rm tot} = 2^{\mathcal{V}},
\end{equation}
that are initially in a maximally mixed state ${\rho_0 =q_{\rm tot}^{-1} \I}$.
In each timestep $i$, we first apply a Haar-random $U(q_{\rm tot})$ unitary to the qubits, then measure a fraction $(1-s)$ of them in the computational basis. Let us define
\begin{equation}
q_s = 2^{s \mathcal{V}},
\end{equation}
which is the Hilbert space dimension associated with the unmeasured qubits in a given timestep. We will study the ``purification'' of the qubits by measurements, i.e. the decrease with time of their mixed-state entropy~\cite{gullans2020dynamical,choi2020quantum}.

In this model, the key parameter $q$, which plays the role of an effective Hilbert space dimension (and sets the fugacity for an elementary kink) will be 
\begin{equation}
q = \lf \f{1}{q_s} - \f{1}{q_{\rm tot}} \ri^{-1}
\end{equation}
as will be clear below Eq.~\eqref{eq:T12_eff_q}. If the number of qubits measured in each timestep is large, then $q$ is close to $q_s$, the number of unmeasured spins in each timestep.

Let us now derive the effective model describing this dynamics.

The measurements at timestep $i$ implement a projection ${\rho \rightarrow \f{P_i \rho P_i}{\tr P_i \rho P_i}}$. Here $P_i$ is an orthogonal projector of rank $q_s$. This projector is random, since it depends on the measurement outcomes of the $(1-s)\mathcal{V}$  qubits, and there are ${2^{(1-s)\mathcal{V}}=q_{\rm tot}/q_s}$ possible outcomes.

Given the unitaries, the measurement outcomes have a nontrivial probability distribution. Born's rule states that the probability of a given set of outcomes in timestep $i$ (which lead to the projector $P_i$) is ${\tr(P_i \rho_{i-1}' P_i)}$, where $\rho_{i-1}'$ is the normalized density matrix immediately prior to these measurements.
This is the probability for the outcomes in a single timestep, conditioned on the earlier part of the trajectory. As usual, iterating Born's rule shows that the joint probability for the complete sequence of measurement outcomes up to time $t$ (the quantum trajectory) is given by
\begin{align}\label{eq:Poutcomes}
P(\text{outcomes})  = \tr \check\rho_t.
\end{align}
Here $\check\rho_t$ is the \textit{un}normalized density matrix
\begin{equation}\label{eq:evolutionofcheckrho}
\check \rho_t = (P_tU_t\cdots P_1U_1) \rho_0 (P_tU_t\cdots P_1U_1)^\dag .
\end{equation}
The physical density matrix at time $t$ is $\rho_t = \check\rho_t/(\tr \check \rho_t)$.
(We normalize the initial density matrix, $\rho_0=\check\rho_0$, in the standard fashion, $\tr \rho_0=1$.)
By Eq.~\eqref{eq:Poutcomes}, the Born-rule average of a function of $\rho_t$ is given by
\begin{equation}\label{eq:sumtrajectories}
\overline{F[\rho_t]}_{\mathrm{\,BR}}
= \sum_{\text{outcomes}} \overline{(\tr \check \rho_t) F[\rho_t]}
\end{equation}
(where the outcomes label the complete quantum trajectory from time $0$ to time $t$).
Note that the overlines on the right and left hand sides have different meanings (indicated by the subscript on the LHS).
On the right-hand side, the overline represents averaging over the independent Haar-random unitaries $U_i$.

Now we notice that the Haar-invariance of $U_i$ means that the summand in (\ref{eq:sumtrajectories}) 
is independent of the measurement outcomes,
\begin{equation}\label{eq:summedtrajectories}
\overline{F[\rho_t]}_{\mathrm{\,BR}}
=
\lf{q_{\rm tot}}/{q_s}\ri^t \times 
\overline{(\tr \check \rho_t) F[\rho_t]}.
\end{equation}

We may now repeat the steps from Sec.~\ref{sec:ginibre} to write the generating function for entropies in terms of a transfer matrix. Readers not interested in these details may wish to skip ahead to (\ref{eq:BRavgenfn}). (Below we suppress the subscript on $\rho_t$, as in Sec.~\ref{sec:ginibre}.)
\begin{align}\label{eq:genfnagain}
\overline{e^{-k (n-1)S_n}}_{\, \mathrm{BR}}
& = 
\lf\f{q_{\rm tot}}{q_s}\ri^t \times
\overline{\Tr[\check \rho^{n}]^k
\Tr[\check \rho]^{1 - n k}}
\\ \label{eq:genfnagainline2}
& = 
\lim_{N\to 1}\,
\lf\f{q_{\rm tot}}{q_s}\ri^{Nt} 
\bbraket{1^{N - nk} n^k \, | \,\overline{\check\rho^{\otimes N}}}.
\end{align}
The evolution of $\kket{\overline{\check\rho^{\otimes N}}}$ in a timestep is  
\begin{equation}
\kket{\overline{\check\rho^{\otimes N}}}
\rightarrow
(P\otimes\cdots \otimes P)
\overline{(U\otimes\cdots \otimes U^*)}\,
\kket{\overline{\check\rho^{\otimes N}}}
\end{equation} 
where $P$ is a fixed rank-$q_s$ projector.
Using standard results, the Haar average may be written
\begin{equation}
\overline{(U\otimes\cdots \otimes U^*)}
=
\sum_\sigma \kket{\sigma}\bbra{\sigma^*},
\end{equation}
where the dual states $\kket{\sigma^*}$ are written in term of the Weingarten functions~\cite{Collins_2022}, 
${\kket{\sigma^*}\equiv \sum_\tau \text{Wg}(\tau^{-1}\sigma) \kket{\tau}}$.
These states satisfy the biorthogonality property $\bbraket{\tau|\sigma^*}=\bbraket{\sigma|\tau^*}=\delta_{\sigma\tau}$ and $\bbra{\nu} = \sum_{\tau}\bbraket{\nu|\tau}\bbra{\tau^*}$. Thus, $\sum_\tau \kket{\tau}\bbra{\tau^*}$ is simply the projector onto the space spanned by permutation states. 

Therefore we define a transfer matrix by
\begin{equation}\label{eq:measurementtransfermatrix}
\mathcal{T}_{\sigma,\tau} =\, 
 {  \bbra{\sigma^*}} P\otimes\cdots \otimes P \kket{\tau} \lf \f{q_{\rm tot}}{q_s}\ri^N.
\end{equation}

The appearance of the dual state in this expression is crucial: for example, it guarantees that in the unitary limit $s \to 1$, where $P$ becomes simply an identity matrix, we have $\mathcal{T}_{\sigma,\tau} = \delta_{\sigma,\tau}$. That is, in the unitary limit, the fugacity for kinks vanishes, and the purification timescale diverges.

With the above definition, we may write a matrix element like that in Eq.~\eqref{eq:genfnagainline2} as (here $\nu$ is an arbitrary permutation in $S_N$)
\begin{align}
 (q_{\rm tot}/q_s)^{Nt}
\bbraket{\nu|
\overline{\check\rho^{\otimes N}}
}
& = (q_{\rm tot}/q_s)^{Nt}
\sum_{\tau}\bbraket{\nu|\tau}\bbraket{\tau^*|
\overline{\check\rho^{\otimes N}}
}
\\
& = \sum_{\tau,\sigma_1} \bbraket{\nu|\tau}\lf \mathcal{T}^t\ri_{\tau,\sigma_1}
\bbraket{\sigma_1^*|\overline{\check\rho_0^{\otimes N}}} \\
& = 
q_{\rm tot}^N \sum_{\sigma_1} \lf T \, \mathcal{T}^t\ri_{\nu,\sigma_1}
\bbraket{\sigma_1^*|\overline{\check\rho_0^{\otimes N}}}.
\end{align}
In the first equality, we have inserted the resolution of identity $\sum_{\tau} \kket{\tau} \bbra{\tau^*}$ for the subspace of permutation states.
We have used the fact that $\bbraket{\nu|\tau}=q_{\rm tot}^{N}T_{\nu,\tau}$, where $T$ is the transition matrix in Eq.~\eqref{eq:transferginibrenormalized} for the Ginibre model. 
Finally the initial-time replicated   density matrix is
$\kket{\overline{\check\rho_0^{\otimes N}}}=q_{\rm tot}^{-N}\kket{\id}$, so by biorthogonality  we have
\begin{equation}
\label{eq:sigma_star_rho_0}
\bbraket{\sigma_1^*|\overline{\check\rho_0^{\otimes N}}} = 
q_{\rm tot}^{-N} \delta_{\sigma_1, \I}.
\end{equation}
Therefore 
\begin{align}
 (q_{\rm tot}/q_s)^t
\bbraket{\nu|
\overline{\check\rho^{\otimes N}}
} 
= \lf T\, \mathcal{T}^t\ri_{\nu,\I}.
\end{align}

Applied to the generating function (\ref{eq:genfnagain}),
\begin{align}\label{eq:BRavgenfn}
\overline{e^{-k (n-1)S_n}}_{\, \mathrm{BR}}
& = \lim_{N\to 1}\,
\lf T \, \mathcal{T}^t \ri_{\sigma_{\text{final}},\I},
\end{align}
where as for Eq.~\eqref{eq:genfnTonlyN0} in the Ginibre model, ${\conj{\sigma_{\text{final}}} = 1^{N - nk} n^k}$.
Again we have a 1D lattice model (compare Eq.~\eqref{eq:partitionfunctionfirst}), with a modified transfer matrix on all but one of the links:
\begin{equation}\notag
\begin{tikzpicture}[scale = 0.85*0.8,every node/.style={scale = 0.85}]
\draw (0,0)--(5,0);
\node () at (5.5,0) {$\cdots$};
\node () at (1,0.5) {$T$};
\node () at (3,0.5) {$\mathcal{T}$};
\node () at (8,0.5) {$\mathcal{T}$};
\node () at (10,0.5) {$\mathcal{T}$};
\draw[fill=white] (0,0) circle (0.5) node () {$\sigma_{\rm final}$};
\draw[fill=white] (2,0) circle (0.5) node () {$\sigma_{t}$};
\draw[fill=white] (4,0) circle (0.5) node () {$\sigma_{t-1}$};
\begin{scope}[shift={(7,0)}]
\draw (-1,0)--(4,0);
\draw[fill=white] (0,0) circle (0.5) node () {$\sigma_{2}$};
\draw[fill=white] (2,0) circle (0.5) node () {$\sigma_{1}$};
\draw[fill=white] (4,0) circle (0.5) node () {$\mathbb{I}$};
\end{scope}
\end{tikzpicture}
\end{equation}

Next we consider the transfer matrix $\mathcal{T}$.
Its structure is slightly different from that of the Ginibre model in the previous section. 
However we will see that in the scaling limit $\mathcal{T}$ again has the universal structure discussed around Eq.~\eqref{eq:TidentityplusA}.
Additionally, the fact that one of the transfer matrix factors in Eq.~\eqref{eq:BRavgenfn} is different from the rest is unimportant at large $t$.
However, there is one important difference from Sec.~\ref{sec:ginibre}, which is that Born-rule averages are given by ${N\to 1}$ and not ${N\to 0}$.

If we  use the same transfer matrix, but take the other limit ${N\to 0}$, then we are describing a dynamics with ``forced'' measurements that are \textit{not} sampled with Born's rule.\footnote{Physically  this forced measurement dynamics would mean for example retaining only ``runs'' in which all measurements are ``up'' and discarding all others.
This must be done in such a way that the probability distribution of the unitaries is not biased (i.e., if a run is discarded part way through, we try again with the same sequence of unitaries). This discarding process is exponentially costly in runs.}
The forced measurement problem is an example of the multiplication of independent random matrices, 
similar to the Ginibre problem in Sec.~\ref{sec:ginibre}: these will share the same scaling functions in the scaling regime, which differ from those of the true measurement problem as we show below.

The elements of the transfer matrix $\mathcal{T}$ introduced in Eq.~\eqref{eq:measurementtransfermatrix} can also be written (by symmetry it is sufficient to take the row element to be $\I$)
\begin{equation}
\mathcal{T}_{\I , \sigma} =
\lf
\f{q_{\rm tot}}{q_s}
\ri^{N}
\sum_{\tau \in S_N} 
\text{Wg} ( \tau^{-1} \sigma )
\langle \tau  | P^{ \otimes 2N}  |\I \rangle  
\end{equation}
We have used the property ${\mathcal{T}_{ \I,\sigma,}=\mathcal{T}_{\sigma, \I}}$ which is a general consequence of the replica symmetry $G$ discussed in Sec.~\ref{sec:transfermatrixbasics}.

Using the standard large $q_{\rm tot}$ expansion of the Weingarten functions (e.g. see Ref.~\cite{Collins_2022}, or App.~D of Ref.~\cite{zhou2019emergent}) and the basic property
\begin{equation}
\langle \I | P^{\otimes 2N}| \tau  \rangle  = q_s^{N - |\tau |},
\end{equation}
we obtain for example (when $q_s\gg 1$)
\begin{equation}
\begin{aligned}\label{eq:measurementweights}
T_{\I , \I} &\simeq 1 , 
& 
T_{\I ,(12)(34)} &\simeq  \left(\frac{1}{q_s} - \frac{1}{q_{\rm tot}} \right)^2  ,
\\
T_{\I , (12)} &\simeq \left( \frac{1}{q_s} - \frac{1}{q_{\rm tot}}  \right) ,
& 
T_{\I , (123)} &\simeq \frac{1}{q_s^2}  - \frac{3}{q_{\rm tot} q_s} + \frac{2}{q_{\rm tot}^2}.
\end{aligned}
\end{equation}
It is then appropriate to define the effective dimension 
\begin{equation}
\label{eq:T12_eff_q}
\frac{1}{q} \equiv T_{\I, (12)} \simeq \frac{1}{q_s} - \frac{1}{q_{\rm tot}}. 
\end{equation}
Note that the scalings of the higher domain wall weights in Eq.~\eqref{eq:measurementweights} are consistent with universality Eq.~\eqref{eq:transfmatlimit}.

\subsection{Universality and models with spatial structure} 
\label{sec:universality}

Now we argue that the transfer matrix analysis gives results that are universal, i.e. valid for a much larger class of models than the two above. (To go directly to results for scaling functions of the universal regime, readers may skip ahead to Sec.~\ref{sec:computations}.)

Above, unitary invariance allowed a reduction to a transfer matrix with explicit kink weights $\mathcal{T}_{\I, \mu}$. 
We saw algebraically (see  Eq.~\eqref{eq:transfmatlimit}) that a universal transfer matrix emerged when $t$ was of order $q$, as a result of the fugacities of higher kinks $\mu$ being much smaller than that of the elementary kink:
\begin{equation}\label{eq:higherkinksuppressioncondition}
T_{\I, \mu} \ll q^{-1} \text{ at large $q$ \quad\quad (for $|\mu|>1$)}.
\end{equation}
We argue below that this condition is satisfied in most models of  physical interest.

Although higher kinks are typically irrelevant when $t$ is of order $q$, 
in many models they may be  relevant up to timescales that are  much larger than 1, though still much smaller than $q$. 
This leads to a crossover timescale which is much larger than 1 but much smaller than the typical timescale for purification.

As we discuss next, the typical timescale ${t_*=q}$ for purification scales as
\begin{equation}\label{eq:qexprpurity}
q \asymp 1/ \overline{\mathcal{P}} ,
\end{equation}
where ${\mathcal{P}=\lf e^{-S_2}\ri_\text{1 step}}$ is the purity after \textit{one timestep}, and the average is over all forms of randomness (including measurements if they are present).
In many interesting models, $\mathcal{P}$ has a very broad distribution.
This includes standard spatially local (or $k$-local) hybrid quantum circuit models, which we can relate to the one-dimensional models here by a simple coarse-graining argument.
When the distribution of $\mathcal{P}$ is broad, the average in Eq.~\eqref{eq:qexprpurity} is dominated by atypical realizations of the  $m_i$ which are much more highly purifying than the typical.

As a result, the timescale for purification is in general determined not by the properties of a typical timestep $m_i$, but instead by atypical timesteps. 
Note that, because the relevant timescales are very long, the chain ${m_tm_{t-1}\cdots m_1}$ will contain \textit{many} ``atypical'' blocks, so there is no paradox here.

In order to argue for universality, let us first consider more general unitarily invariant\footnote{The protocols discussed so far [(a) and (b) in Fig.~\ref{fig:protocols}] are statistically invariant under the insertion of an arbitrary unitary at the start of any timestep [such a unitary can be absorbed into a Ginibre matrix in (a) or into a Haar-random unitary gate in (b)]. This unitary invariance allowed for the exact reduction to the permutation subspace spanned by $\{\kket{\sigma}\}_{\sigma\in S_N}$.}
models, and then spatially local models where there is no unitary invariance property. 

\subsubsection{More general invariant ensembles}
\label{sec:generalinvariantensembles}

In this section we will discuss the case of independent random matrices, but the discussion carries over to the measurement case, after replacing the ${N\to 0}$ limit with an ${N\to 1}$ limit (corresponding to replacing independent-random-matrix averages with Born-rule averages).

As in Sec.~\ref{sec:mea}, let us think of the Hilbert space as that of $\mathcal{V}$ qubits, with dimension $q_{\rm tot} = 2^\mathcal{V}$.
We take a product of independent random ${q_{\rm tot}\times q_{\rm tot}}$ matrices of the form
\begin{equation}\label{eq:mLambdaV}
m_i =  \Lambda_i V_i,
\end{equation}
where $V_i$ is a Haar-random unitary, and $\Lambda_i$ is a diagonal (positive) matrix of singular values.
Previously, in Sec.~\ref{sec:mea}, $\Lambda_i$ was a projection operator which was chosen by Born's rule.
Here we are instead considering the situation where $\Lambda_i$ is simply another independent random matrix, whose distribution we are free to choose. A trivial mapping shows that this class of models includes the Ginibre model (Sec.~\ref{sec:ginibre}) as a special case. \footnote{It makes no difference to the statistics of the entropies if instead of Eq.~\eqref{eq:mLambdaV} we take ${m_i = W_i  \Lambda_i V_i}$, with an additional independent Haar-random unitary matrix $W_i$.
Therefore this class of models includes all the unitarily invariant models for the given $q_{\rm tot}$, including the Ginibre model.}

We will use $\mathcal{P}$ to denote the purity after a single timestep, i.e. ${\mathcal{P} = \tr \Lambda^2/(\tr \Lambda)^2}$. We will assume that ${\overline{\mathcal{P}}\ll 1}$ since this is the regime of interest.

The derivation of the transfer matrix expressions for this model is similar to Secs.~\ref{sec:ginibre},~\ref{sec:mea}, giving e.g.
\begin{align}\label{eq:generalmodelarbitraryt}
\overline{e^{-k (n-1) S_n}}
= \lim_{N\to 0}\,
\lf T \, \mathcal{T}^t \ri_{\sigma_{\text{final}},\I}
\end{align}
with ${\conj{\sigma_\text{final}} = (1^{N-kn}n^k)}$.
Explicitly, the transfer matrix $\mathcal{T}$, normalized so that $\mathcal{T}_{\sigma,\sigma}=1$, is now [cf.~\eqref{eq:measurementtransfermatrix}]
\begin{equation}
\mathcal{T}_{\sigma,\tau} =\f{ \bbraket{\sigma^*| \overline{\Lambda^{\otimes 2N}} | \tau}}
{\bbraket{\I^*| \overline{\Lambda^{\otimes 2N}} | \I }},
\end{equation}
but we will not need this explicit formula, as we can relate transfer matrix elements to more physical quantities.
Taking the average of the purity $\mathcal{P}$ for a single timestep (${t=1}$), using  Eq.~\eqref{eq:generalmodelarbitraryt}, shows the purification timescale ${t_* = q}$ scales like  ${1/\overline{\mathcal{P}}}$, as stated in Eq.~\eqref{eq:qexprpurity}.  To see this, note that we have
\begin{align}\label{eq:onetimestepaveragepurity}
\overline{\mathcal{P}}
& =
\lim_{N\to 0} \,
\sum_\mu  q_{\rm tot}^{-|\mu^{-1} (12)|} \mathcal{T}_{ \I, \mu} 
\\ \label{eq:Tplusqtinv}
& = \mathcal{T}_{ \I, (12)} + \f{1}{q_{\rm tot}} + \ldots .
\end{align}
The replica limit of $\mathcal{T}_{ \I, (12)}$  is  implicit.
The terms shown explicitly are for ${\mu=(12)}$ and ${\mu=\I}$; 
we assume other terms to be subleading (this is consistent with the scalings below).
Recalling that ${\mathcal{T}_{\I, (12)}=q^{-1}}$ in our notation, the equation above is 
\begin{equation}
q= \lf  \overline{\mathcal{P}} - \f{1}{q_{\rm tot}} + \ldots \ri^{-1}.
\end{equation}
For a more general permutation 
with $\conj{\sigma}=(1^{N-\sum nk_n},2^{k_2},3^{k_3},\ldots)$, a similar assumption about higher terms gives
\begin{align}\label{eq:onetimestepaveragegeneralpermutation}
\mathcal{T}_{ \I, \sigma}
\asymp 
\overline{e^{-\sum_{n}k_n (n-1) S_n}},
\end{align}
where again the average is for a single timestep.
This equation, relating the transfer matrix elements to entropy averages, shows that it is easy to meet the universality condition. 
Recall that for universality when ${t\sim q}$ we require that 
${\mathcal{T}_{\I,\sigma}\ll \mathcal{T}_{\I,(12)}}$ whenever ${|\sigma|>1}$.
From Eq.~\eqref{eq:onetimestepaveragegeneralpermutation} one can check\footnote{Using the inequalities ${S_2\geq S_3}$ and ${(n-1)S_n\geq 2S_3}$ for ${n\geq 3}$, we see that ${\overline{e^{-\sum_n k_n (n-1) S_n}} \leq \overline{\exp\big(-\big( k_2 + 2 \sum_{n\geq 3} k_n \big)  S_3 \big)}}$.
For a non-elementary domain wall, ${\big( k_2 + 2 \sum_{n\geq 3} k_n \big) \geq 2}$, giving 
${\overline{e^{-\sum_n k_n (n-1) S_n}}\leq \overline{e^{-2S_3}}}$.
} 
that this condition holds whenever 
\begin{equation}\label{eq:universalityconditions}
\overline{e^{- 2 S_3}}  \ll \overline{e^{-S_2}}.
\end{equation}
Eq.~\eqref{eq:universalityconditions} is satisfied except in pathological models where the average of the purity is dominated by rare events with a purity close to 1.

For a concrete example of a model,
we can take $\Lambda_i$ to be be the operator that
projects a random subset of the qubits onto spin-up, representing a ``forced'' measurement.
Each qubit has a probability $(1-s)$ to be projected, and a probability $s$ not to be projected.\footnote{If $\mathcal{S}$ qubits are unmeasured, then $\Lambda_i$ is a projector of rank $2^\mathcal{S}$. Since the basis for $\Lambda_i$ does not matter we take it to be a diagonal matrix with $2^\mathcal{S}$ ones and the remaining elements zero.} 
We fix the probability $s$, with ${0<s<1}$, when we take the limit of large $\mathcal{V}$.
The entropies $S_n$ are simply equal to ${(\sum_{\alpha=1}^\mathcal{V} \chi_\alpha) \ln 2}$, where $\chi_\alpha=1$ if qubit $\alpha$ is not projected, and $\chi_\alpha=0$ if qubit $\alpha$ is projected. 
Therefore
\begin{equation}\label{eq:toymodelresult}
\overline{e^{-\sum_n k_n (n-1) S_n}}
=
\left( (1-s) + s \, 2^{-|\sigma|} \right)^{\mathcal{V}},
\end{equation}
where we have used ${\sum_n k_n (n-1)=|\sigma|}$.
Using Eq.~\eqref{eq:onetimestepaveragegeneralpermutation}, the condition for universality is seen to be satisfied
(namely ${\mathcal{T}_{\I,\sigma}\ll \mathcal{T}_{\I,(12)}}$ whenever ${|\sigma|>1}$).

\subsubsection{Short-time crossovers: a kink in a random potential}

We now make a brief detour to consider timescales that are much shorter than the purification time ${t_*=q}$, i.e. \textit{before} the onset of the ``universal regime''.
For simplicity, we focus on the short-time crossover of the distribution of $S_2(t)$. 
We make the time argument of  $S_2(t)$ explicit to distinguish it from the second
R\'enyi entropies of \textit{individual} random matrices $m_i$, which we denote by $[S_2]_i$ for $i=1,\ldots, t$.

For simplicity we consider models in which 
Eq.~\eqref{eq:onetimestepaveragegeneralpermutation}  holds as an approximate equality, i.e.
\begin{equation}\label{eq:Tapproximateequality}
{\mathcal{T}_{ \I, \sigma}
\simeq
\overline{e^{-\sum_{n}k_n (n-1) [S_n]_i}}}
\end{equation}
(for any $i$), without the nontrivial order 1 prefactor that is allowed in Eq.~\eqref{eq:onetimestepaveragegeneralpermutation}.
This holds for the toy model above, and we expect it to hold for typical local circuit models.

We conjecture that when ${t\ll q}$ we can simplify the expression for $\overline{e^{-k S_2(t)}}$ in Eq.~\eqref{eq:generalmodelarbitraryt} to a sum over a subset of low-energy kink configurations, 
obtained\footnote{In the Ginibre model, these are the configurations with the lowest energy, i.e. all other configurations incur a larger power of $1/q$ in the Boltzmann weight.
Therefore other configurations are suppressed until $t$ becomes of order $q$ (at which point the entropy gain from additional kinks can compete with the energy cost). Though we have not proved it, we guess that configurations outside this set are energetically suppressed for $t\ll q$ in a larger class of models, given mild assumptions on the distribution of the blocks.} by   partitioning $k$ commuting transpositions, labeled by
 ${(12)}$, $(23)$, $\ldots$, ${(2k-1,2k)}$, 
 among $t$ bonds.
If a bond is occupied by a single one of these transpositions, it forms an elementary kink, while
if multiple transpositions share a bond, they form a higher kink.

After this simplification,  the expression for the generating function is equivalent to 
\begin{equation}
\overline{e^{-k S_2(t)}}
= 
\overline{\lf \sum_{i=1}^t e^{-[S_2]_{i}} \ri^k},
\end{equation} 
as we see by expanding the product on the right-hand side and averaging using Eq.~\eqref{eq:Tapproximateequality}.
Therefore, in probability distribution,
\begin{equation}\label{eq:S2randomenvironment}
S_2(t)
\overset{d}{=}
- 
\ln \lf 
\sum_{i=1}^t e^{-[S_2]_{i}}
\ri.
\end{equation}
This formula can also be understood without using the replica trick, via the logic of Ref.~\cite{zhou2020entanglement}.\footnote{In this approach we work with a fixed Hilbert space, ${\mathcal{H}^{\otimes 2} \otimes \mathcal{H}^{\ast \otimes 2}}$, which is sufficient to express $e^{-S_2(t)}$ in a fixed realization of randomness.
We then introduce the permutation states by introducing resolutions of the identity in this Hilbert space at each timestep: 
${1 = \sum_{\sigma\in S_2} \kket{\sigma}\bbra{\sigma^*} + P_\perp}$. 
Here $P_\perp$ is a projector onto the subspace orthogonal to the permutation states.
In the present regime of  times $t \ll q$ we expect that terms with $P_\perp$ can be neglected \cite{zhou2020entanglement}. This leads to the RHS of Eq.~\eqref{eq:S2randomenvironment}.}
It shows that, at short times $t\ll q$, we can think of $S_2(t)$ as the free energy for a single kink in a random potential energy landscape, where the local potential is given by $[S_2]_i$.

When the number of terms in the sum in Eq.~\eqref{eq:S2randomenvironment} is sufficiently large, we can expand in fluctuations of the sum around its mean, giving
(recall that ${1/q\simeq \overline{\mathcal{P}}}$, and that $x=t/q$):
\begin{equation}\label{eq:S2lateintermediatetime}
\overline{S_2(t)} =  - \ln (x) + \f{1}{2t} \f{
\overline{\mathcal{P}^2}^c
}{
\lf \overline{\mathcal{P}} \ri^2
} + \ldots
\end{equation}
The first term agrees with the leading term in the expansion of the scaling function for  small $x$, which is $S_2(x) = \ln 1/x + O(x^2)$ 
(the scaling functions will be discussed in detail in Sec.~\ref{sec:computations}).
We see that the finite-time correction to the universal scaling result is small once the second term in Eq.~\eqref{eq:S2lateintermediatetime} is small, giving a crossover time\footnote{In general the crossover timescale may depend on the observable we look at.}
\begin{equation}\label{eq:t1scale}
t_1 \sim 
\f{
\overline{\mathcal{P}^2}
}{
\lf  \overline{\mathcal{P}} \ri^2
}.
\end{equation}
This is equivalent to the condition in App.~\ref{app:energyentropy} for the neglect of kinks of type $\conj{(12)(34)}$.
The discussion around Eq.~\eqref{eq:universalityconditions} shows that $t_1$ is much smaller than the typical purification time $q$, i.e. Eq.~\eqref{eq:t1scale} is consistent with the existence of the universal scaling regime.

As an example, consider the toy model introduced at the end of Sec.~\ref{sec:generalinvariantensembles}, with
a probability ${s=1/2}$ for a given qubit to be subjected to forced measurement in a given timestep. 
Then the purification timescale is ${t_*=q=(4/3)^\mathcal{V}}$, while the above crossover timescale is $t_1=(10/9)^\mathcal{V}$. 
So while both timescales are exponentially large in the number $\mathcal{V}$ of qubits (in this example), 
the crossover timescale is exponentially smaller than the purification time.

On much smaller timescales, Eq.~\eqref{eq:S2lateintermediatetime} is not a good approximation.
We will be interested in models where
$[S_2]_i$ has  fluctuations that are small compared to its mean, but nevertheless much larger than unity in the large $q$ limit. 
This is the case for the toy model introduced above Eq.~\eqref{eq:toymodelresult}, and it will also be the case in typical spatially local model (where the large $q$ limit is due to the large volume limit).
In this situation, the probability distribution of 
 $e^{-[S_2]_i}$ ranges over many orders of magnitude.
 
 As a result,  the sum in Eq.~\eqref{eq:S2randomenvironment} will be dominated by its largest term for  sufficiently small values of $t$,
\begin{equation}
S_2(t) \simeq \min \{ S_2^{(1)}, \ldots, S_2^{(t)} \}.
\end{equation}
This regime was discussed in Ref.~\cite{nahum2021measurement}.
As an example, consider again the toy model 
(with forced measurement of a random subset of spins) 
in the regime where ${1\ll \ln t \ll \mathcal{V}}$.
In this regime we can treat $[S_2]_i$ as a Gaussian random variable with mean $s(\ln 2)\mathcal{V}$ and variance 
$s(1-s)(\ln 2)^2 \mathcal{V}$.
Then\footnote{For i.i.d. Gaussian $X_i \in \mathcal{N}(0,\sigma)$, the minimal $Y = \min_{i=1}^t X_i$ scales as $-\sqrt{2}\sigma \sqrt{\ln t}$.} 
\begin{equation}\label{eq:shorttimesS2}
\overline{S_2(t)} =  
s(\ln 2)\mathcal{V}
- \mathcal{O} \lf
\mathcal{V}^{1/2} \sqrt{\ln   t   }
\ri.
\end{equation}

\subsubsection{Spatially local  and $k$-local models}

We expect that models with interactions between small numbers of qubits, if they are in a ``weak-monitoring'' or entangling phase, can be coarse-grained to give effective 0+1D models.
(We could also consider random tensor networks with an appropriate geometry.)
Here we give a very heuristic discussion of this coarse-graining. We use the language of replicas, but the  coarse-graining argument could probably be formulated without them.

For concreteness, consider a 1+1D quantum circuit made from local unitaries and projectors, on a qubit chain of length ${\mathcal{V}\gg 1}$.
In the case where the local unitaries are Haar-random, such circuits can be mapped to effective 1+1D replica statistical mechanics models for local permutation degrees of freedom $\sigma_{r,\tau}$, where $(r,\tau)$ is a discrete spacetime coordinate (see Ref.~\cite{potter2022entanglement} for a review).
In the entangling phase, $\sigma_{r,\tau}$ is long-range ordered \cite{jian2020measurement, bao2020theory}.
Domain walls that span the system are exponentially rare, because their free energy cost is 
proportional to the system size $\mathcal{V}$ (before including the entropy from translation in the time direction) \cite{li2021statistical,nahum2021measurement}.

Therefore we expect that we can in principle coarse-grain the system to an effective 0+1D model, by defining blocks of dimensions $(\mathcal{V}, \tau_\mathcal{V})$ which we label by $t\in \mathbb{Z}$.
We retain an effective spin $\sigma_t$ for each block.
(The timescale $\tau_\mathcal{V}$  should be larger than the typical timescale for transverse wandering of a spanning domain wall.)
After coarse-graining we arrive at an effective one-dimensional model for permutations,
 which is in principle characterized by a transfer matrix $\mathcal{T}$.
In the replica limit, the transfer matrix elements are determined by the distribution of entropies for a single coarse-grained block, as discussed in the previous section.\footnote{A given transfer matrix element such as $\mathcal{T}_{\I,(12)}$ will depend nontrivially on $N$ (for example, the local properties of the effective 1+1D statistical mechanics problem will depend strongly on $N$) but we assume it can be continued to $N\to 1$ (if we are considering measurements) or $N\to 0$ (if we are considering forced measurements, or a network of random tensors).}
Ultimately, the key input for the effective model, determining the purification timescale, is the average purity for a single block, ${\overline{\mathcal{P}} = \mathcal{T}_{\I,(12)}}$, whose inverse defines our ``effective Hilbert space dimension'' $q$.

We note again that the average ${\overline{\mathcal{P}}}$ is not dominated by typical blocks, but by rare ones. 
In a local model in $d$ spatial dimensions, $S_2$ for a given block (before averaging) can be mapped to the free energy of a domain wall in a $(d+1)$-dimensional random environment \cite{zhou2019emergent,li2023entanglement}  (see Ref.~\cite{fisher2022random} for a review).\footnote{This mapping is analogous to Eq.~\eqref{eq:S2randomenvironment} in the 0+1D
problem, which describes the free energy of a zero-dimensional domain wall in a one-dimensional random environment. In the 1+1D case, the domain wall is one-dimensional and spans the block. } 
Writing ${S_2 = s_\text{eq} \mathcal{V}+ \Delta S}$, where $s_\text{eq}$ is the average entropy density,  the random fluctuation $\Delta S$ is typically much smaller than the mean.
In 1+1D, these typical fluctuations are governed by KPZ exponents, and $\Delta S$ is of order $\mathcal{V}^{1/3}$.
In a ``2-local'' model with all-to-all coupling, we expect that $\Delta S$ is of order $\mathcal{V}^{1/2}$, similarly to the discussion of the toy model above Eq.~\eqref{eq:shorttimesS2}.
However $\overline{\mathcal{P}}=\overline{e^{-S_2}}$ is analogous to an annealed average in a disordered system, and it is dominated by atypically large negative $\Delta S$, of order $\mathcal{V}$, so that  
 $\overline{e^{-S_2}}\sim e^{- s' \mathcal{V}}$ with ${s' < s_\text{eq}}$.
The properties of typical blocks are relevant at early times, before the scaling regime ($t\sim q$) sets in.
For the 1+1D case, KPZ scaling forms \cite{ledoussal2016large} show that the early-time scaling, for $\ln t \ll \ln q$, is ${\overline{S_2(t)} = s_\text{eq} \mathcal{V} - O(\mathcal{V}^{1/3} (\ln t)^{2/3})}$,
in analogy to Eq.~\eqref{eq:shorttimesS2} for the toy model.

\section{Computation of scaling functions}
\label{sec:computations}

We now use the effective descriptions obtained in Sec.~\ref{sec:replicaanduniversality} to compute scaling functions for entanglement.
We focus on the scaling regime, which can be obtained by taking the limit of large $q$ with the scaling variable 
\begin{equation}
x = t/q    
\end{equation}
(equivalently, $x=t/t_*$) held fixed. 
In this limit, both the independent random matrix problem (``IRM'') and the Born-rule dynamics (``BR'') involve the same simplified transfer matrix 
(Sec.~\ref{sec:simplificationscalinglimit}, Sec.~\ref{sec:mea}),
and the only difference between the two problems is in the replica limit $N\to N_*$, which involves $N_* =0$ and $N_* = 1$ respectively.

Recall that the generating function encoding the statistics of the $n$th R\'enyi entropy is given by the partition function with fixed boundary conditions (Eqs.~\ref{eq:genfnTonlyN0},~\ref{eq:BRavgenfn}) 
\begin{equation}
\label{eq:part_fun_leading}
\overline{e^{-k (n-1)S_n}}|_{\text{IRM or BR}} = \lim_{N\rightarrow N_*} \langle 1^{N-nk} n^k | T^{t+1} | 1^N \rangle.
\end{equation}
The boundary conditions are the identity permutation  ($\sigma_\text{init} = \I$, here denoted by $1^N$) and a product of $k$ disjoint $n$-cycles  (denoted $1^{N-nk} n^k$): e.g. for the case ${n=3}$, ${k=2}$ we take $\sigma_{\rm final} = (123)(456)$. In the scaling limit, we may also write the right-hand side in terms of the adjacency matrix $A$ (see Eq.~\eqref{eq:TidentityplusA}) that connects permutations differing by only a single transposition
\begin{equation}
\label{eq:part_fun_scaling_limit}
\overline{e^{-k (n-1)S_n}}|_{\text{IRM or BR}} = \lim_{N\rightarrow N_*}   \langle 1^{N-nk} n^k | e^{x A} | 1^N \rangle.
\end{equation}

In the following subsections, we will introduce our counting and resummation approaches in order of increasing technical complexity. 

First, we consider the simplest case --- the calculation of $\overline{e^{-k (n-1)S_n}}$ at the leading nontrivial order in $x$ (Sec.~\ref{subsec:deterministic_S_n}). This reduces to counting the number of minimal-energy domain wall configurations and gives the deterministic early-time result for the R\'enyi entropies stated in the introduction (Eq.~\ref{eq:introsmallxasymptotics}). 

Then in Sec.~\ref{subsec:S2_order_x2} we continue the brute force counting approach to the next order  (relative order $x^2$), for the particular case of $S_2$.
At this order, we already see nontrivial fluctuations in the entropy, as well as differences between the two universality classes.

To go beyond these lowest orders, in Sec.~\ref{subsec:characterexp} we review the diagonalization of the transfer matrix in terms of group characters using known techniques \cite{STANLEY1981255,babai1979spectra,jackson1988some,murty2020ramanujan,lando2010hurwitz,dubrovin2017classical,fulton2013representation,STANLEY1981255,3026c618-b8c5-30c6-b3cf-32559324cf7e}. This streamlines the counting problem and allows expansions of $S_2$ (Sec.~\ref{sec:s2higherorder}) and $S_1$ (Sec.~\ref{sec:s1higherorder}) up to order $x^8$, for both of the replica limits of interest.

Finally in Sec.~\ref{sec:vNresummation} we analytically resum the power series for the Born-rule average of $S_1$ to \emph{all} orders and evaluate the resulting analytic expression by truncating a sum over Young tableaux. This gives us a rapidly convergent expression for $S_1$, which agrees well with numerical results.  As a byproduct we also obtain the Shannon entropy for the Born probability distribution over measurement records.

\begin{figure}
    \centering
\includegraphics[width=0.92\columnwidth]{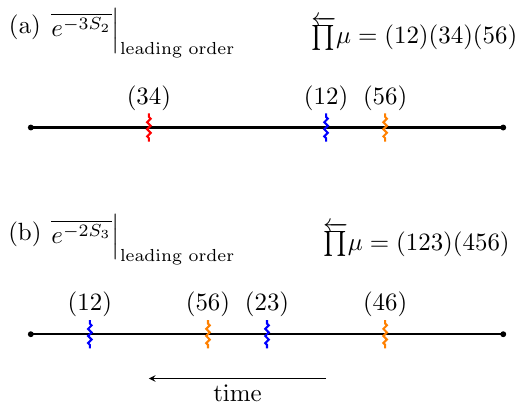}
\caption{
At leading order in the small $x$ expansion, the calculation of the generating function $\overline{e^{-k(n-1)S_n}}$ involves configurations with a minimal number of kinks consistent with the boundary conditions. (Boundary conditions fix the product ${\protect\overleftarrow{\prod}} \mu$.)
As examples, we show (a) an allowed configuration for $\overline{e^{-3 S_2}}$ and (b) one for $\overline{e^{-2S_3}}$. Lowest-order configurations have $k$ groups of $(n-1)$ kinks, with kinks from different groups commuting with each other. (Different groups have different colors in the figures.)  Since kinks from different groups commute, the partition function for $\overline{e^{-k(n-1)S_n}}$ is a product of $k$ identical factors at leading order in $x$.}
\label{fig:leadingorderillustration}
\end{figure}

\subsection{The leading-order (deterministic) part of $S_n$}
\label{subsec:deterministic_S_n}

Let us first look at the partition function in Eq.~\eqref{eq:part_fun_leading} in the limit where the scaling variable $x=t/q$ is small. 
For concreteness we will use the language of Eq.~\eqref{eq:part_fun_leading}, where we imagine kinks living at particular locations on the 1D lattice.

We start with $n = 2$.  In this case, the final state,  $\sigma_\text{final}$, is of type $1^{N - 2k } 2^k$, i.e. a product of $k$ commuting transpositions:
\begin{equation}
\sigma_\text{final} = (12)(34)(45)\cdots (2k-1,2k).
\end{equation}
Recall that a kink associated with the transfer matrix element $T_{\sigma, \tau}$ is labelled by the type $\sigma \tau^{-1}$: the boundary condition requires that the product of all kinks is equal to $\sigma_{\rm final}$. 
The lowest-energy configurations involve $k$ kinks, which are just the transpositions appearing in the expression above for $\sigma_\text{final}$.
This is illustrated in Fig.~\ref{fig:leadingorderillustration} (top).

\begin{figure*}
    \centering
    \includegraphics[width=\textwidth]{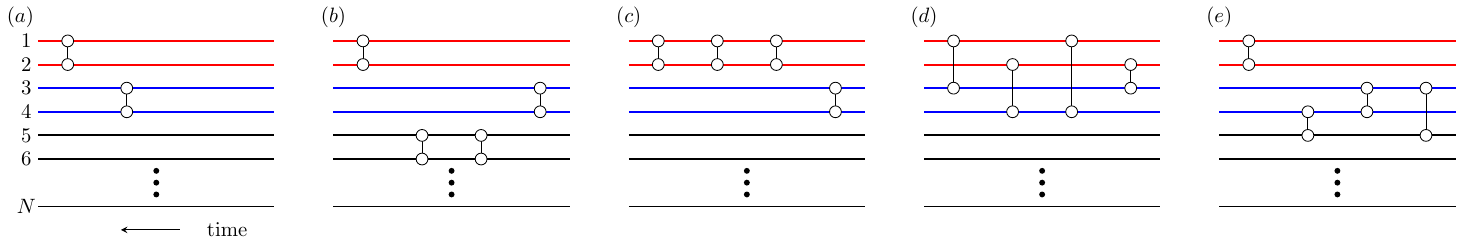}
    \caption{(a) A kink configuration contributing to $\overline{e^{-k S_2}}$ (with $k=2$)  at leading order. 
    (b-e) Examples of the four classes of configurations that contribute at the first subleading order (relative order $x^2$).
    In this figure, a kink of type $\mu=(ab)$ is represented as a bond $\fineq[-0.8ex][1.2][1]{\swap}$ connecting horizontal lines $a$ and $b$.}
    \label{fig:S2_x2}
\end{figure*}

In the scaling limit, these $k$ kinks are transpositions dispersed among  $t+1$ possible links (Fig.~\ref{fig:S2_x2}). At large $t$, there are approximately $(t+1)^k \sim t^k$ ways to distribute the kinks, and each kink costs $\frac{1}{q}$; we therefore have
\begin{equation}
\label{eq:exp_k_S_2}
    \overline{e^{-k S_2}} \approx \frac{t^k}{q^k}^{\rotatebox{30}{$\scriptstyle=x^k$}} 
    \hspace{-3pt}
    + 
    \mathcal{O}(x^{k+2}).
\end{equation}
Here the next order is $\mathcal{O}(x^{k+2})$ as the group multiplication rule requires at least two additional transpositions to ensure the overall product remains equal to $\sigma_{\rm final}$.

Eq.~\eqref{eq:exp_k_S_2} indicates at the leading order, both  $\overline{e^{-kS_2}}$ and  $\overline{e^{-S_2}}^k$ equal $x^k$.  Consequently, we have a deterministic expression at the leading order for both ${N = 0}$ and ${N = 1}$,
\begin{equation}
    S_2 = \ln \frac{1}{x} + \ldots,
\end{equation}
where remainder ``\ldots'' has cumulants that are of higher-order in $x^2$, as we will discuss below.

The kink gas picture and calculations above can be easily repeated for a general integer R\'enyi index $n \ge 2$: we split the  boundary spin $\sigma_{\rm final}$ into kinks and disperse them on the $t+1$ links.  
Now $\sigma_{\rm final}$ contains $k$ commutating $n$-cycles (a transposition is a $2$-cycle) 
and each $n$-cycle can be further decomposed into a product of $n-1$ transpositions (the minimal number).
Fig.~\ref{fig:leadingorderillustration} (bottom) shows an example for the third R\'enyi entropy. However, we must now take into account the fact that an $n$-cycle can be decomposed into $n-1$ transpositions in $n^{n-2}$ distinct ways \cite{denes1959representation}. Furthermore, there are $\sim\frac{t^{k(n-1)}}{((n-1)!)^k}$ ways to distribute these $k(n-1)$ transpositions into $t+1$ links, where the division by $(n-1)!$ reflects the fact that the ordering of the transpositions within a given $n$-cycle must be respected.
Altogether, at the leader order, we have
\begin{align}\label{eq:leadingorderanyn}
    \overline{e^{-k (n-1) S_n}} &\approx \frac{1}{q^{k(n-1)}} \frac{t^{k(n-1)}}{[(n-1)!]^k} (n^{n-2})^k( 1 + \mathcal{O}(x^{2}) ) \\
    &= x^{k(n-1)} \left( \frac{n^{n-2}}{(n-1)!} \right)^k ( 1 + \mathcal{O}(x^2 )).
\end{align}
Again we have $\overline{e^{- k (n-1) S_n}} = \left( \overline{e^{- (n-1) S_n}} \right)^k$ in this order. 
This is because the first $kn$ replicas have decoupled into $k$ independent subsets in this calculation.
We have thus derived the deterministic value of $S_n$ in Eq.~\eqref{eq:introsmallxasymptotics} of the introduction: 
\begin{equation}
    S_n = \ln \frac{1}{x}  + \frac{1}{n-1} \ln \left[\frac{n^{n-2}}{(n-1)!} \right] + 
\ldots
\end{equation}

\subsection{Direct counting for $S_2$ at order $x^2$}
\label{subsec:S2_order_x2}

Here we continue the direct counting to the next order (relative $x^2$ order) in the small-$x$ expansion for the partition function in the case $n=2$. 

In the scaling limit Eq.~\eqref{eq:part_fun_scaling_limit}, the partition function can be expanded as
\begin{align}
\label{eq:expansion}
\braket{  1^{N-2k} 2^k |  T^{t+1}  |\id}  & = 
  \sum_{\ell = 0}^{\infty} \frac{x^{k+2\ell}}{ (k+2l)!}\braket{  1^{N-2k} 2^k |  A^{k+2\ell}   |\id}
  \\\label{eq:S2_part_fun}
   & \equiv 
x^k \sum_{\ell=0}^\infty a_{N,k,\ell}\, x^{2\ell}.
\end{align}
In the kink gas language of the left-hand side, we decompose $\sigma_\text{final}$ into $k + 2\ell$ transpositions dispersed on $\sim t$ links, hence the powers of $x$ appearing on the RHS. 
(The amplitudes 
$\braket{  1^{N-2k} 2^k |  A^{k+2\ell}   |\id}$ are related to ``Hurwitz numbers'': we will discuss more systematic counting at higher orders in subsequent sections.)

The case $\ell=0$, corresponding to writing $\sigma_\text{final} =  (12)(34)\ldots (2k-1~2k)$ as a product of $k$ commuting kinks, was discussed above, and we have
 $a_{N,k,0} = 1$. 
 We now analyze the case $\ell = 1$, corresponding to writing $\sigma_\text{final}$ as a product of $k+2$ kinks.

There are several classes of configurations, which are illustrated in Fig.~\ref{fig:S2_x2}.
In this figure,  transposition $(a\, b)$ is represented by a vertical bond connecting the horizontal lines numbered $a$ and $b$, which represent elements $a,b\in \{1,2,\ldots, N\}$.
Fig.~\ref{fig:S2_x2}~(a) is an example of the kind of configuration that we had previously, at order $\ell=0$: all of the transpositions are those in  the cycle decomposition  of $\sigma_\text{final}$, 
$\sigma_\text{final} =  (12)(34)\ldots (2k-1~2k)$.
Below, when we refer to a ``2-cycle'', we mean one of the 2-cycles in this decomposition of $\sigma_\text{final}$. 
Similarly, by ``1-cycle'' we mean one of the elements with $a>2k$ that are left invariant by $\sigma_\text{final}$.

When we consider $\ell=1$ (i.e. $k+2$ kinks), the two extra transpositions can:
exchange elements between two 1-cycles [Fig.~\ref{fig:S2_x2}~(b)];
exchange two elements within the same 2-cycle  
[Fig.~\ref{fig:S2_x2}~(c)];
exchange elements between two 2-cycles   [Fig.~\ref{fig:S2_x2}~(d)];
or exchange elements  between a 2-cycle and a one-cycle [Fig.~\ref{fig:S2_x2}~(d)].
We therefore write
\begin{equation}
\label{eq:coeffaexp}
 a_{N,k,1} = 
 \binom{\hspace{-1pt} N\hspace{-1.5pt}-\hspace{-1.5pt}2k \hspace{-1pt}}{2} m_{1,1} 
  + k m_2 
   + \binom{k}{2} m_{2,2}
 + k(N-2k)m_{1,2} 
\end{equation}
where the coefficient $m_{r_1, r_2}$ refers to the case where the extra transpositions connect two cycles of length $r_1$ and $r_2$, while 
$m_2$ refers to the case where the extra transpositions belong to the same $2$-cycle. 
The combinatorial factors in Eq.~\eqref{eq:coeffaexp} account for the ways of choosing the cycles $C_1$ and $C_2$ of length $r_1$ and $r_2$ and contain all the dependence on $k$ and $N$. 
The remaining combinatorial numbers $m$ involve counting paths in a permutation group $S_M$ for a fixed value of $M$, independent of $k$ or $N$. 
(For example for $m_{2,2}$ we must consider permutations of $M=4$ elements.)
They may be written in terms of the combinatorial number~\cite{lando2010hurwitz, dubrovin2017classical}
\begin{equation}\label{eq:defofm}
m(l;[\mu]) = \braket{\mu | A^\ell | \id },
\end{equation}
which is the number of ways of writing a given permutation ${\mu\in S_M}$, of type $[\mu]$, as an ordered product of $l$ transpositions in $S_M$. 
On the right hand side, the adjacency matrix $A$ is implicitly that for $S_M$. 
($M$ is not written explicitly on the left hand side because it is determined by $[\mu]$: for example $[\mu]=1^2 2^1$ is by definition a permutation in $S_4$.)
The quantity $m(l;[\mu])$ is proportional\footnote{The {(disconnected)} simple Hurwitz number is $h^\circ_{l;[\mu]} = (d_\mu/M!) m(l;[\mu])$, where $d_\mu$ is the size of the conjugacy class of $\mu$ \cite{lando2010hurwitz}.}  to the ``simple Hurwitz number'' \cite{lando2010hurwitz}: we will refer to it as a ``Hurwitz multiplicity''.
[We will often suppress the conjugacy-class notation and write the Hurwitz multiplicity as $m(l;\mu)$. Clearly, we have ${m(\ell; \mu) = 0}$ for ${\ell < |\mu|}$. The number of minimal-length factorizations is $m(|\mu|; \mu)$.]
\begin{enumerate}[leftmargin=*]
    \item {}[Fig.~\ref{fig:S2_x2}(b)]
    If the additional pair of transpositions connect 1-cycles, they are simply of the form $(ab)(ab)$ for some $a,b>2k$.
    Once $a$, $b$, are chosen, there is no further choice of these 
    transpositions.
    We must simply sum over their possible positions on the 1D line, which gives $t^2/2$. 
    Together with the Boltzmann weight $q^{-2}$ for the added kinks, this gives a contribution $x^2/2$ to the sum in Eq.~\ref{eq:S2_part_fun}.
    Therefore we have 
     $m_{1,1}=1/2$.
     
    Equivalently, in terms of the Hurwitz multiplicity, 
    \begin{equation}
      m_{1,1} = \frac{{m(2; 1^2)}}{2!}^{\rotatebox{30}{$\scriptstyle=1$}} = \frac{1}{2}.
    \end{equation}
    \item {}[Fig.~\ref{fig:S2_x2}(c)] If the extra transpositions are assigned to the same $2$-cycle
    \begin{equation}
    \label{eq:m2dec}
        m_2 = \frac{m(3; 2^1)}{3!}^{\rotatebox{30}{$\scriptstyle=1$}} = \frac 1 6
    \end{equation}
    where $m(3;2^1) = 1$ is the number of ways to write a single transposition, e.g. $(12)$, as the ordered product of three transpositions in $S_2$.
    (The denominator accounts for the fact that when we integrate over the positions of the three kinks, their order is fixed.)

    \item {}[Fig.~\ref{fig:S2_x2}(d)] If the extra transpositions mix two distinct $2$-cycles, e.g. $(12)(34)$, within $S_4$, then
        \begin{equation}
        m_{2,2} = \frac{m(4; 2^2)^{\rotatebox{30}{$\scriptstyle=104$}}\hspace{-17pt}}{4!}
        \,\,\,\,\,\,\,
        - 2 \,
         \f{m(3; 2^1)}{3!} 
        \f{m(1; 2^1)}{1!} 
        = 4,
    \end{equation}
    where in the first term we sum over all ways of writing $(12)(34)$ as a product of 4 transpositions in $S_4$, and in the second term we subtract off factorized cases already considered in Eq.~\eqref{eq:m2dec}, in which the two cycles are not mixed.

    \item {}[Fig.~\ref{fig:S2_x2}(e)] Finally, if the extra transpositions mix a $2$-cycle with a $1$-cycle, we have
    \begin{equation}
        m_{1,2} = \frac{m(3; 1^1 2^1)^{\rotatebox{30}{$\scriptstyle=9$}} - m(3; 2^1)}{3!} = \frac 4 3
    \end{equation}
    where we subtracted the case already considered in Eq.~\eqref{eq:m2dec}.
\end{enumerate}
Summing these possibilities, the combinatorial factor is 
\begin{equation}
    { a_{N,k,1} } = \frac{k^2}{3} + \frac{1}{3}(N-4) k + \frac{1}{4}( N^2 - N ).
\end{equation}
Therefore the generating functions \eqref{eq:part_fun_leading} for the second R\'enyi entropy, for the two cases of independent random matrices (IRM) and Born rule dynamics (BR), are 
\begin{align}
\left.\overline{e^{-k S_2}}\right|_\mathrm{IRM}
& = 
x^k \lf 1 + \f{k(k-4)}{3} x^2 + \ldots \ri 
\\
\left.\overline{e^{-k S_2}}\right|_\mathrm{BR}
& = 
x^k \lf 1 + \f{k(k-3)}{3} x^2 + \ldots \ri.
\end{align}
The corresponding $O(x^2)$ results for the cumulants of $S_2$  are recorded below in Eqs.~\eqref{eq:S2_asymp},~\eqref{eq:S2_asympvar}, together with the next order.

This analysis can be generalized to higher $\ell$,
again decomposing 
 $a_{N,k,\ell} = m(k + 2\ell; 1^{N-2k} 2^k)/(k + 2\ell)!$,
which appears in the summand of Eq.~\eqref{eq:expansion},
into a sum of combinatorial factors that are polynomials in $N$ and $k$ and counting paths over fewer than $2\ell$ cycles. 
In particular, the degree of $a_{N,k,\ell}$ in $N$ and $k$ will be at most $2\ell$.
 (Similarly we may generalize to higher $n$.)

\subsection{Diagonalization of the transfer matrix}
\label{subsec:characterexp}

This subsection reviews some standard group theory results that allow the diagonalization of the transfer matrix and the determination of the Hurwitz multiplicities $m(\ell, [\mu])$  [see Eq.~\eqref{eq:defofm}],
following Stanley \cite{STANLEY1981255}.
Recall that these multiplicities appear in the expansion of the partition function in the scaling limit:
\begin{equation}
\label{eq:TexpHurw}
    \braket{\mu | T^{t+1} |\id} \simeq
    \sum_{\ell = 0}^\infty  \frac{m(\ell; \mu)}{\ell!}  x^\ell.
\end{equation}
We will denote the size of the permutation group by $N$. 
Although we we are mostly interested in calculating the {left}-hand side 
of  (\ref{eq:TexpHurw}) 
in a replica limit,
we saw in the previous section that this may require us to know the multiplicities $m(\ell, \mu)$
for values of $N$ with $N>1$.
In the present section, $N$ is an arbitrary fixed value.

Regarding $S_N$ as a set of vertices, $A$ is an adjacency matrix that connects vertices that differ by a transposition. 
We will also need a slightly generalized adjacency matrix, $A_{\delta}$,  that connects permutations through elements in the conjugacy class $\conj{\delta}$. 
The analogous Hurwitz multiplicity is
\begin{equation}
    m_{{\delta}}(\ell;\mu) \equiv \braket{\mu | (A_{\delta})^\ell | \id }.
\end{equation}
By definition, we have ${A = A_{\delta}}$ and ${m_{\delta}(\ell;\mu) = m(\ell;\mu)}$ when ${\conj{\delta} = 1^{N-2} 2^1}$, i.e. when $[\delta]$ is the class of transpositions.

The action of  $A_{\delta}$  is most easily thought about by defining the  group action
\begin{equation}\label{eq:groupaction}
    \hat{\sigma} |\mu \rangle \equiv | \sigma \mu \rangle.
\end{equation}
 We easily check that $\hat{\sigma} \hat{\nu} =  \widehat{\sigma\nu}$. 
From $\langle \mu | \hat{\sigma} |\nu \rangle = \langle \mu | \sigma \nu \rangle = \langle \sigma^{-1} \mu | \nu \rangle$, we can deduce the group action on the bra is
\begin{equation}
    \langle \mu | \hat{\sigma} = \langle \sigma^{-1} \mu |. 
\end{equation}
The adjacency matrix $A_\delta$ is then written as an element of the group algebra, in which we allow superpositions of group elements \cite{stone_mathematics_2009}:
\begin{equation}
    A_{\delta} = \sum_{\nu \in \conj{\delta}} \hat{\nu}.
\end{equation}
The adjacency matrices $A_\delta$ are in the center of the group algebra (they commute with all $\hat \sigma$) and can be simultaneously diagonalized.

The  $N!$--dimensional linear space of permutation states $|\sigma \rangle$ is reducible  under the action of Eq.~\eqref{eq:groupaction}.
It splits \cite{stone_mathematics_2009} into irreducible representations of $S_N$ that are indexed by partitions of $N$, labelled by $\lambda$.\footnote{These  are sequences of non-increasing integers $\{\lambda_1, \lambda_2, \cdots, \lambda_R\}$ that satisfy $\sum_{j=1}^R \sum_j \lambda_j = N$. The partitions can be represented as Young tableaux (see examples in App.~\ref{appsec:review_group}).}
Each irrep $\lambda$ appears with multiplicity $d^\lambda$, where $d^\lambda$ is the dimension of the irrep.

[From the physics point of view, this multiplicity is related to the fact that  
the transfer matrix has a global $S_N\times S_N$ symmetry.
(In addition to the  ``left'' action of $S_N$ in Eq.~\eqref{eq:groupaction}, 
 which is a global symmetry, there is an analogous right action: see Sec.~\ref{subsub:global_symm}.)
In an arbitrary  statistical mechanics model 
with $S_N\times S_N$ global symmetry, the state space could contain arbitrary irreps  ${(\lambda, \lambda')}$ of ${S_N\times S_N}$. 
But here the $N!$-dimensional space of permutation states decomposes into representations only of the form $(\lambda,\lambda)$ of $S_N\times S_N$,
with unit multiplicity for each $\lambda$. As a result, when we consider only the left $S_N$ action (as above), each representation $\lambda$ occurs  $d^\lambda$ times.]

The operator $P^\lambda$ that projects onto irrep $\lambda$ may be written in terms of the adjacency matrices \cite{stone_mathematics_2009},
\begin{equation}
\label{eq:P_in_A}
   P^{\lambda} = \frac{d^{\lambda}}{|G|}\sum_{\delta \vdash N}  \chi^{\lambda}_{\delta} A_{\delta},
\end{equation}
where ${|G|=N!}$ is the cardinality of the group (these formulas apply to more general groups).
Schur orthogonality relations~\cite{fulton2013representation} imply that ${P^\lambda P^{\lambda'} = \delta^{\lambda \lambda'} P^\lambda}$. Using the orthogonality of characters, ${\sum_\lambda \chi^{\lambda\ast}_\mu \chi_{\mu'}^\lambda = \delta_{\mu \mu'} |G| / d_\mu}$ ($\chi_\mu^{\lambda \ast}$ is the complex conjugate and $d_\mu$ is the size of the conjugacy class $[\mu]$) one can invert the relation:
\begin{equation}
\label{eq:A_in_P}
   A_{\delta} =  \sum_{\lambda \vdash N} \frac{  \chi^{\lambda\ast}_{\delta} d_{\delta} }{d^\lambda} P^{\lambda}. 
\end{equation}
This allows us to read off the eigenvalues $\nu^\lambda$ of $A_\delta$.
For the case $[\delta]=1^{N-1}2^1$
relevant to the transfer matrix of the kink gas we have
\begin{equation}
 \nu^\lambda = \frac{\chi_{1^{N-2} 2}^{\lambda\ast} d_{1^{N-2} 2^1}}{d^\lambda} = \frac 1 2 \left( \sum_{j=1} \lambda_j^2 - \sum_{j'}(\lambda_{j'}^{\top})^2 \right)
\end{equation}
where the second equality uses the Frobenius formula~\cite{3026c618-b8c5-30c6-b3cf-32559324cf7e} 
and where $\lambda^{\top}$ is the partition dual\footnote{$\lambda^{\top}= (\lambda_1', \lambda_2',\ldots)$, with $\lambda'_i = \# \{ \lambda_j | \lambda_j \geq i\}$ (see examples in App.~\ref{appsec:first_hook}).}
to $\lambda$
(the transposed Young tableau).
The total multiplicity of a given eigenvalue $\nu^\lambda$ is $(d^\lambda)^2$.

Next we need the matrix elements of powers of the adjacency matrices (the Hurwitz multiplicities):
\begin{align} 
   \braket{\mu | (A_{\delta})^\ell | \id } 
   &= \frac{1}{d_{\mu}}\sum_{\nu \in \conj{\mu}}  
   \braket{ \nu | (A_{\delta})^\ell | \id } \\
    &= \frac{1}{d_{\mu}}\sum_{\nu \in \conj{\mu}}  \braket{ \id | \widehat{\nu^{-1}} (A_{\delta})^\ell | \id } \\
   &=  \frac{1}{d_{\mu}} \braket{ \id | A_{\mu}^\dagger (A_{\delta})^\ell | \id } \\
    &= \frac{1}{N!} \frac{1}{d_{\mu}}\Tr( A^\dagger_{\mu} (A_{\delta})^\ell ).
\end{align}
In the last line we have used 
${A_{\delta}=\widehat{\sigma^{-1}} A_{\delta}  \hat{\sigma}}$ in order to replace the diagonal matrix element with the trace. 
Using the eigenvalues and their multiplicities, 
\begin{equation}
\label{eq:m_ell_mu}
    m_{\delta}( \ell; \mu ) = \frac{1}{N!}  \sum_{\lambda \vdash N}\chi^{\lambda}_{\mu} d^{\lambda} \left( \frac{\chi^{\lambda\ast}_{\delta} d_{\delta}}{d^\lambda} \right)^{\ell},
\end{equation}
and for the particular case of interest with ${[\delta]=1^{N-1}2^1}$,
\begin{equation}
\label{eq:mchar}
    m(\ell; \mu) = \frac{1}{N!}\sum_{\lambda \vdash N} d^\lambda \chi_\mu^\lambda (\nu^\lambda)^\ell.
\end{equation}
Finally, summing over $\ell$, we have the expression for the matrix elements of the transfer matrix in the scaling limit,
\begin{equation}
\label{eq:expsum}
    \braket{\mu | T^{t+1} |\I} \to
\frac 1 {N!}\sum_{\lambda\vdash N} d^{\lambda} \chi_{\mu}^\lambda e^{x \nu^\lambda}.
\end{equation}
In Sec.~\ref{sec:s2higherorder} we will use the Hurwitz multiplicity formula in Eq.~\eqref{eq:mchar} to compute the higher-order small $x$-expansions of $S_2$ . 
In Sec.~\ref{sec:vNresummation} we will resum the partitions in the expression for appropriate transfer matrix elements (Eq.~\eqref{eq:expsum}) in order to compute the mean of the von Neumann entropy.

\subsection{Small--$x$ expansion for $S_2$}
\label{sec:s2higherorder}

We can now use the expansions in terms of Hurwitz multiplicities  \eqref{eq:TexpHurw} and the explicit formulas (\ref{eq:mchar}, \ref{eq:expsum}) involving characters to compute the coefficients $a_{N,k,\ell}$ (Eq.~\eqref{eq:S2_part_fun}) for a few larger values of $\ell$. 
Given a sufficiently large table of values for $a_{N,k,\ell}$, we can determine its explicit form analytically via interpolation. We start by writing
\begin{equation}
    a_{N,k,\ell} = \sum_{i,j=0}^{2\ell} c_{ij} N^i k^j.
\end{equation}
In order to evaluate the $(\ell+1)^2$ coefficients $c_{ij}$, it is sufficient to evaluate $m(\ell; 1^{N-2k} 2^k)$ for $k = 0,\ldots, 2\ell$ and $N = 4\ell, \ldots, 6\ell$ (recall that $N\geq 2k$). 
For a few values of $\ell$, the sum over the integer partitions $\lambda$ of $N$ in Eq.~\eqref{eq:mchar} can be explicitly carried out. 
The characters $\chi^\lambda_{1^{N-2k}2^k}$ can be computed via the Schur polynomials~\cite{lando2010hurwitz} -- we outline the method in Appendix~\ref{app:hurw}. 

Once the coefficients $c_{ij}$ have been obtained, 
we may take the replica limit ${N\to 0}$ 
to give the  generating function $\overline{e^{-k S_2}}|_\mathrm{IRM}$ (Eq.~\eqref{eq:genfnfirst}) for a product of independent random matrices, or the limit  ${N\to 1}$ to give the  generating function 
$\overline{e^{-k S_2}}|_\mathrm{BR}$ (Eq.~\eqref{eq:genfnagain}) for the measurement process.
As usual, we can then extract the cumulants of $S_2$ by expanding the logarithm of the generating function in $k$.

For the mean values we find (for brevity we write only the lowest orders, results to order $x^8$ are in App.~\ref{app:hurw})
\begin{subequations}\label{eq:S2_asymp}
\begin{align}
& \mathrm{IRM:} \quad \overline{S_2}
= -\ln x +\frac{4}{3} x^2-\frac{637}{90}x^4+O(x^6),\\
& \mathrm{BR:} ~~\quad
\overline{S_2}
= -\ln x +x^2-\frac{949}{180}x^4+O(x^6),
\end{align}
\end{subequations}
and for the variance $C_2$, 
\begin{subequations}\label{eq:S2_asympvar}
\begin{align}
& \mathrm{IRM:} &  \overline{(S_2)^2} - (\overline{S_2})^2 & =
\frac{2}{3}x^2-\frac{32}{3}x^4+O(x^6),\\
& \mathrm{BR:} & \overline{(S_2)^2} - (\overline{S_2})^2 & =
\frac{2}{3}x^2-\frac{29}{3}x^4+O(x^6).
\end{align}
\end{subequations}
The right-hand sides of the above equations are the power-series expansions of the appropriate scaling functions.  
Higher cumulants may also be computed and show that the distribution is non-Gaussian. The $p$th cumulant is of order $x^{2p-2}$.

Numerical results will be shown for comparison in Sec.~\ref{sec:numerics}.

\subsection{Small--$x$ expansion of $S_1$}
\label{sec:s1higherorder}

The approach explained above based on interpolation can be generalized to arbitrary R\'enyi index $n$.
In  Eq.~\eqref{eq:TexpHurw} we must then set  $[\mu]=n^k 1^{N-kn}$.
As already discussed in Sec.~\ref{subsec:deterministic_S_n},
the length of a single $n$-cycle is ${|n^1| = n-1}$, so that
the expansion in Eq.~\eqref{eq:TexpHurw} starts at order $x^{k(n-1)}$, with the coefficient 
\begin{equation}
    \frac{m(k(n-1); 1^{N-nk}n^k)}{[k(n-1)]!} = \left(\frac{n^{n-2}}{(n-1)!}\right)^k
\end{equation}
(see the discussion around Eq.~\eqref{eq:leadingorderanyn}; we  used the fact that the number of minimal factorizations of each $n$-cycle is 
 ${m(n-1; n^1) = n^{n-2}}$~\cite{denes1959representation}).
Factoring out this first non-vanishing order in the small $x$ expansion, we define\footnote{Note that $a_{N,k,n,\ell}$ defined here is related to the previously defined $a_{N,k,\ell}$ in Eq.~\eqref{eq:S2_part_fun} by $a_{N,k,2,\ell} = 3^\ell a_{N,k,\ell}$.} coefficients $a_{N,k,n,\ell}$ by
\begin{equation}\label{eq:def_a_N_k_n_l}
\begin{aligned}
\langle \mathbb{I} |T^{t+1}|1^{N-nk}n^k\rangle
&=
\left(\frac{n^{n-2}}{(n-1)!}\right)^k x^{k(n-1)}\\
&\qquad \times \sum_{\ell=0}^\infty
a_{N,k,n,\ell}\frac{x^{2\ell}}{{ (n+1)^\ell}} \;.
\end{aligned}
\end{equation}
From an argument analogous to that in Sec.~\ref{subsec:S2_order_x2} it follows that for fixed $N$ and $\ell$, $a_{N,k,n,\ell}$ is a polynomial in $k$ and $n$. 
In App.~\ref{app:hurw} we determine these polynomials for the lowest orders up to $\ell=4$.\footnote{Here we check the consistency of the polynomiality conjecture by evaluating $a_{N,k,n,\ell}$ for ${\ell=1,2,3,4}$ on an overdetermined set of values. While we do not have a proof for the polynomiality of $a_{N,k,n,\ell}$ in $k$ and $n$ for general $\ell$, we expect it can be checked via the (proved) polynomiality of the \emph{connected} simple Hurwitz numbers~\cite{ekedahl2001hurwitz,lando2010hurwitz}.}.

Using these results, we compute the entanglement entropy by analytic continuation as explained in the previous section. Taking the derivative of Eq.~\eqref{eq:def_a_N_k_n_l} with respect to $k$ at ${k=0}$ and then the derivative with respect to $n$ at ${n = 1}$, we obtain, for ${N= 0}$ and ${N=1}$, respectively,
\begin{subequations}\label{eq:S1_asymp}
\begin{align}
& \mathrm{IRM:} \quad \overline{S_1}
= -\ln x + 1-\gamma + \frac{11}{24} x^2-\frac{1739}{2880}x^4+O(x^6),
\label{eq:S1IRM}
\\
& \mathrm{BR:} ~~\quad
\overline{S_1}
= -\ln x + 1-\gamma + \frac{5}{24} x^2-\frac{239}{2880}x^4+O(x^6)
\label{eq:S1BR}
\end{align}
\end{subequations}
where $\gamma\approx 0.577216$ is the Euler-Mascheroni constant. The scaling results up to $x^8$ are given in App.~\ref{app:hurw}.

\subsection{Comment on large $x$}\label{sec:largex}

So far we have discussed the small $x$ expansion. In the opposite limit of large ${x=t/q}$, we enter a regime where the entropies are much smaller than one, and are distributed over many orders of magnitude. It is then  more useful to consider the typical value of $S_2$ 
than its average.

In principle, one should be able to extract asymptotics in this regime using the dominant eigenvalues of the transfer matrix, but the analytical continuation to the replica limit appears to be subtle.

However, in this regime, it is sufficient to consider the two leading singular values of $\check\rho$, since others are strongly subleading. This is discussed for the IRM case (${N=0}$) in App.~E of \cite{nahum2021measurement} and 
 gives ${\overline{\ln S_2}|_{\rm IRM}\sim -t/q}$ in the present notation. 
(This calculation could be extended to the Born rule case by reweighting using the Born rule factor.)
The exponential decay constant at large $x$ appears consistent with  numerical data for large $x=t/q$ in the IRM case.\\

\subsection{Resummation of Von Neumann entropy for Born-rule dynamics}
\label{sec:vNresummation}

The small $x$ expansion obtained in the previous section becomes rather cumbersome quickly because one has to determine a polynomial of increasingly high degree in three variables $N,k,n$. However, we can use a different analytical approach for the Von Neumann entropy in the Born rule case. For a convenient normalization, let $\tilde{\rho} = {\check{\rho}}/{\overline{\tr\check\rho}}$ and expand near ${N=1}$:
\begin{subequations}
\begin{align}
\label{eq:MOmdef}
M(N,x) := \overline{\tr(\tilde\rho^N)}
& = \overline{\tr\tilde\rho}+(N-1)\,\overline{\tr(\tilde\rho \ln \tilde\rho)}+\cdots\\
\Omega(N,x) := \overline{(\tr\tilde\rho)^N}
& = \overline{\tr\tilde\rho}+(N-1)\,\overline{(\tr\tilde\rho) \ln (\tr\tilde\rho)}+\cdots \label{eq:Omdef}
\end{align}
\end{subequations}
Note that the overlines above denote independent-random-matrix averages. In terms of the transfer matrix,
\ba\label{eqMOmega}
M(N,x) & = \braket{N^1| T^{t+1} |\id}, & 
  \Omega(N,x) & = \braket{\id| T^{t+1} |\id}.
\end{align}
We can extract the Born-rule-averaged entropy by differentiating with respect to $N$, 
\begin{align}\label{vNBR}
\left.\overline{S_1}\right|_{\rm BR} 
& = 
- \partial_N (M(N, 1) - \Omega(N, 1))\Big|_{N=1}
\\
& =-
\lf {\overline{\tr\check \rho}} \ri^{-1} \,
{\overline{(\tr \check \rho)\tr( \rho \ln \rho)}}.
\end{align}
For concreteness we consider the measurement process in Sec.~\ref{sec:mea}.
The factor of ${\lf {\overline{\tr\check\rho}}\ri^{-1}  = (q_{\rm tot}/q_s)^t}$ is then the total number of possible distinct measurement records, i.e. distinct sequences of measurement outcomes  (Eq.~\ref{eq:summedtrajectories}).
We can thus reduce the calculation of the Von Neumann entropy to the two terms \eqref{eqMOmega} around $N = 1$. 
Below in Secs.~\ref{sec:singvalmoments}  and \ref{sec:tracemoments} we consider the calculation of $M$ and $\Omega$ in turn. In Sec.~\ref{sec:shannon} discuss the results of Shannon entropy of the measurement records using the continuation of $\Omega$. 

\subsubsection{Moments of the singular values}\label{sec:singvalmoments}

Using the transfer matrix formalism, we write
\begin{equation}
    M(N,x) = \braket{N^1| T^{t+1} |\id}
\end{equation}
and in the scaling limit use Eq.~\eqref{eq:expsum}. This requires the characters for an irreducible representation $\lambda$ over the $N$-cycle conjugacy class. The Frobenius formula~\cite{macdonald1998symmetric} implies that these characters have a particularly simple structure (see also~\cite{STANLEY1981255}). Specifically, they vanish unless $\lambda = \lambda_r = (1^{r-1},N - r + 1)$ for $r=1,2,...,N$, i.e. unless $\lambda$ corresponds to $L$-shaped Young diagram. In this case, one simply has $\chi^{\lambda_r}_{N^1} = (-1)^r/N$ and
\begin{equation}
d^{\lambda_r} \equiv \binom{N-1}{r-1} \;,  \quad \nu(\lambda_r) = \frac{1}{2} N(N+1 -2r).
\end{equation}
Therefore we have
\begin{equation}
\label{eq:M_N_sinh}
\begin{aligned}
    M(N,x) &= \frac{1}{N!} \sum_{r = 1}^N \binom{N-1}{r-1} (-1)^{r-1} e^{x N(N+1 - 2r)/2}
    \\ &= \frac{2^{N-1}}{N!} \left(\sinh\frac{N}{2}x\right)^{N-1}.
\end{aligned}
\end{equation}
We can now treat $N$ as a continuous variable and obtain
\begin{equation}
\label{eq:partial_N_M}
\begin{aligned}
    \partial_N M(N,x)\Big|_{N=1} &= \ln \left(2\sinh(x/2)\right)+\gamma -1 \\
    &=\ln (x)+\gamma -1+\frac{x^2}{24}-\frac{x^4}{2880}+O(x^6) \;.
\end{aligned}
\end{equation}
Comparing with \eqref{eq:S1BR}, we see that this term fully captures the singular behavior as $x \to 0$. 

\subsubsection{Moments of the trace}
\label{sec:tracemoments}

To calculate the moments of the trace of $\tilde\rho$, we once again write it using the transfer matrix
\begin{equation}
    \Omega(N,x) = \braket{\id| T^{t+1} |\id}
\end{equation}
and use Eq.~\eqref{eq:expsum} for the scaling limit. In this case, the characters are computed on the conjugacy class of the identity and thus one has $\chi^\lambda_{\id} = d^\lambda$. So, we arrive at
\begin{equation}
\label{eq:OmegaPlancherel}
    \Omega(N,x) =
\frac 1 {N!}\sum_{\lambda\vdash N} (d^{\lambda} )^2 e^{x \nu^\lambda} \;.
\end{equation}
In this expression, we recognize the Plancherel measure for the symmetric group $S_N$ associating to each irreducible representation a weight $(d^\lambda)^2/N!$~\cite{borodin2000asymptotics}. Since all possible integer partitions of $N$ are involved, the analytic continuation in $N$ is not straightforward. 

In the following,  we thus sketch the steps and defer the details to App.~\ref{app:resum}. 

\begin{enumerate}[leftmargin=*]
    \item [1] We use the identity 
\begin{equation}
    y\ln y = \int_0^{\infty} \frac{e^{-uy} - ye^{-u} + y - 1}{u^2} du
\end{equation}
to evaluate $\overline{\tr(\tilde{\rho}) \ln \tr(\tilde{\rho})}$. Setting $y = \tr( \tilde{\rho} )$ and averaging, we have
\begin{equation}
\begin{aligned}
    &\overline{\tr(\tilde{\rho}) \ln \tr(\tilde{\rho})} = \partial_N \Omega(N, x)\Big|_{N=1} \\
= &\int_0^{\infty} \frac{\overline{e^{-u\tr( \tilde{\rho} )}} - \overline{\tr( \tilde{\rho} )}e^{-u} + \overline{\tr( \tilde{\rho} )} - 1}{u^2} du.
\end{aligned}
\end{equation}
We introduce the generating function for the moments of the trace
\begin{equation}
\label{eq:d_Omega_N}
    G(u,x) = \overline{e^{-u \tr(\tilde\rho)}} = \sum_{N=0}^\infty \frac{(-u)^N}{N!} \Omega(N).
\end{equation}
Then, using $-\partial_u G(u,x)\Big|_{u=0} = \overline{\tr( \tilde{\rho} )} = \Omega(1,x) = M(1,x) = 1$, we have the integral representation for $\partial_N \Omega( N, x )\Big|_{N=1}$
\begin{align}
\label{eq:u_integral}
\partial_N \Omega(N, x)\Big|_{N=1} & = \overline{\tr(\tilde{\rho}) \ln \tr(\tilde{\rho})} \\
& = \int_0^\infty du \; \frac{G(u,x)-e^{-u}}{u^2} \;.
\end{align}
\item [2] [App.~\ref{appsec:first_hook}] To evaluate the partition sum $\sum_{\lambda \vdash N}$ in Eq.~\eqref{eq:OmegaPlancherel}, we set the maximal number of rows in the Young tableaux (see examples of Young tableaux for $S_4$ in App.~\ref{appsec:review_group}) of the partition $\lambda = (\lambda_1,\ldots,\lambda_r)$ to be $R$ (i.e. $r \le R$) and introduce a set of integer variables $h_j = R + \lambda_j - j$. The $\{h_j\}$ is the hook length of the first column. They are strictly decreasing, non-negative $h_1>h_2>\ldots>h_R\geq 0$, and  satisfy $\sum_i h_i = N + R(R-1)/2$. 

We can express all quantities in terms of $\{h_j\}$. Using standard properties of Young diagrams~\cite{macdonald1998symmetric}, we have
\begin{align}
    &d^\lambda= \frac{N!\Delta(\vec h)}{\prod_{i=1}^R h_i!}\\
    &\nu(\lambda) =  -\frac{(2R-1)N}{2} + \frac1 2\sum_{j=1}^R [h_j^2 - (j-1)^2 ]
\end{align}
where we introduce the Vandermonde determinant $\Delta(\vec h) = \prod_{1\leq i<j\leq R}(h_i - h_j) = \det(h_i^{k-1})_{i,k=1}^R$. The sum over partitions $\sum_{\lambda \vdash N}$ can thus be converted to a sum over $\{h_j\}$
\begin{equation}
\label{eq:vdash_to_h}
   \sum_{N=0}^{\infty} \sum_{\lambda \vdash N }  =\lim_{R\rightarrow \infty} \sum_{N=0}^{\infty}  \sum_{\stackrel{h_1 > h_2 > \cdots > h_R \ge 0 }{ \sum_{j=1}^R h_j = N + \sum_j (j-1)}}.
\end{equation}
\item [3] [App.~\ref{appsec:remove_N}] We fix $R$ to be a constant. In other words, we only consider partitions with at most $R$ rows. The corresponding generation function is $G_R( u, x)$, and 
\begin{equation}
    \lim_{R\to\infty} G_R(u) = G(u).
\end{equation}
When $R$ is finite, the summation over $N$ in the generation function effectively removes the sum constraint of $\sum_{j=1}^R h_j = N + \sum_j (j-1)$ in Eq.~\eqref{eq:vdash_to_h}. Additionally, since the Vandermonde determinant vanishes when any pair of $h$ coincide, the descending order of $\{h_j\}$ can be removed by including a factor $1/R!$. Setting $\tilde u = u e^{-x(R-1/2)}$, we have
\begin{equation}\label{eq:finitedeterminant}
\begin{aligned}
    G_R(u,x) =& e^{-\frac{x}{2} \sum_{j=1}^R (j-1)^2 } (-\tilde u)^{-\sum_{j=1}^R (j-1)} /R!\\
    &\sum_{h_1,\ldots,h_R=0}^\infty \Delta(\vec h)^2 \prod_{i=1}^R \frac{e^{x h_i^2/2} (-\tilde u)^{h_i}}{h_i!^2}
\end{aligned}
\end{equation}
\item[4] [App.~\ref{appsec:det}] The resulting structure is typical of determinantal processes~\cite{borodin2009determinantal} and the joint distribution of eigenvalues in random matrix ensembles~\cite{mehta2004random}, with the important difference that the $h$ variables are integers. Following the techniques of~\cite{baik2000distribution}, we can cast the resulting expression into a single determinant using Andréief's equality~\cite{forrester_meet_2019}, and absorb all constants inside it having finally
\begin{equation}
\label{eq:G_R_as_det}
G_R(u,x) = \det(I_{j,k})_{j,k=0}^{R-1},
\end{equation}
with
\begin{equation}
\label{eq:I_j_k_diverge}
I_{j,k} = e^{-x(j^2 + k^2)/4}
\sum_{h=0}^\infty \frac{e^{x h^2 /2} (-\tilde u)^{h-(j+k)/2}}{(h-j)!(h-k)!} .
\end{equation}
\item [5] [App.~\ref{appsec:regulate}] The expression \eqref{eq:I_j_k_diverge} for $I_{j,k}$ is formally correct only for negative $x<0$ because of the divergent sum over $h$. However, introducing a Gaussian integral, it can be resummed in terms of Bessel function
\begin{equation}\label{Eq:I_j_k}
\begin{aligned}
    I_{j,k} & = i^{| j-k| }e^{-\frac{1}{8} x (j-k)^2} \\ 
&\int_{-\infty}^\infty \frac{dw}{\sqrt{2 \pi x}}
    e^{-\frac{w^2}{2 x}} J_{| j-k| }\left(2 e^{(w+(j+k)x/2)/2} \sqrt{\tilde u}\right).
\end{aligned}
\end{equation}
\end{enumerate}
Collecting everything together (Eqs.~\ref{eq:G_R_as_det},\ref{Eq:I_j_k}), the scaling function for $\overline{S_1}|_\text{BR}$ is given by:
\begin{equation}
\label{eq:S1BRG}
\overline{S_1}|_\text{BR}  = 
- \ln \left[ 2 \sinh \f{x}{2} \right]
+ 1 - \gamma + \int_0^\infty \dd u 
\f{G(u,x)-e^{-u}}{u^2},
\end{equation}
where $G(u)$ is the $R\to\infty$ limit of the determinant in Eq.~\eqref{eq:finitedeterminant}.
This determinant can be evaluated numerically for fixed $R$.
\begin{figure}[t]
    \centering
    \includegraphics[width=\columnwidth]{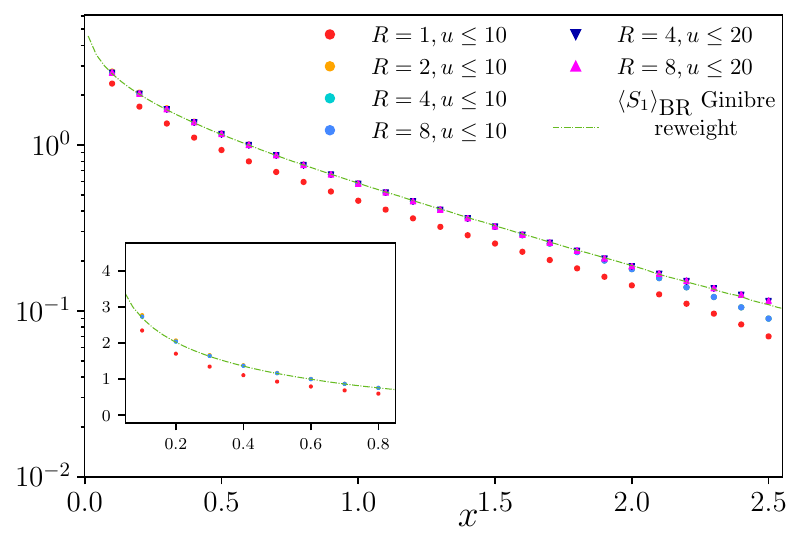}
    \caption{Scaling functions of $\overline{S_1}$ obtained by the resummation approach.  
    We truncate up to size $R=1,2,4,8$ determinants and cut off the $u$ integral in Eq.~\eqref{eq:u_integral} at $10$ or $20$. The green dashed line shows numerical results for the reweighted Ginibre ensemble discussed in Sec.~\ref{sec:numerics}. Inset shows behaviors close to ${x = 0}$.}
    \label{fig:resum_S1}
\end{figure}
Fig.~\ref{fig:resum_S1} shows the numerical evaluation of the scaling function $\overline{S_1}|_{\rm BR}$ through the determinant and the integral in Eq.~\eqref{eq:u_integral}. We truncate the determinant at size $R=1,2,4,8$ and the integral in Eq.~\eqref{eq:u_integral} at $u = 10$ or $20$ and discretize with $\Delta u = 0.1$. Fig.~\ref{fig:resum_S1} summarizes the results and the comparison with a direct numerical evaluation of $\overline{S_1}|_{BR}$ using the Ginibre ensemble (to be discussed in Sec.~\ref{sec:numerics}). Convergence is fast in $R$: The relative difference is less than $2 \times 10^{-2}$ between $R = 2$ and $R = 4$ and less than $2 \times 10^{-4}$ between $R = 4$ and $R = 8$. However, at large $x$, the integrand in Eq.~\eqref{eq:u_integral} has a steep decrease at small $u$, so the step $\Delta u$ has to be decreased to improve accuracy. 

The Von Neumann entropy must converge to zero for large $x$, due to the purification process.
Therefore the last term of Eq.~\eqref{eq:S1BRG} must diverge linearly as $x/2$ at large $x$, in order to cancel the divergence of the first term. This asymptotic form provides a check on the numerics at large $x$ (and confirms the result of the more heuristic argument at the end of Sec.~\ref{sec:shannon}).
It also provides the deficit in the Shannon entropy growth rate that was quoted above (after Eq.~\eqref{eq:shannonwithG}).

\subsubsection{Aside: Shannon entropy of measurement record}\label{sec:shannon}

We note in passing that the quantities $\Omega(N,x)$ defined in Eq.~\ref{eq:Omdef} are the moments of the Born probability.
These can be used to obtain the Shannon entropy of the distribution over measurement outcomes \cite{zabalo2022operator}
(which we compute for a fixed sequence of unitaries, before averaging over the unitaries).

Using the fact that ${P(\text{outcomes})=\tr \check\rho}$ (see Eq.~\ref{eq:Poutcomes}), we may write 
\begin{eqnarray}
\notag
\overline{S_{\rm Shannon}}  &= - (q_{\rm tot}/q_s)^{t} \lim_{N\to 1} \partial_N \overline{(\tr \check \rho)^N} \\
&= t\ln q_{\rm tot}/q_s - \partial_N \Omega( N, x)\Big|_{N=1} .\label{eq:SShannonOmega}
\end{eqnarray}
Alternatively, we may write
the Shannon entropy  as the ${N\to 1}$ limit of $S_\text{Shannon}^{(N)}$, defined by
${e^{ -(N-1) S_\text{Shannon}^{(N)}} = \sum_\text{outcomes} P(\text{outcomes})^N}$.
Then
\be
\label{eq:shannonlimit}
\overline{
e^{ -(N-1) S_\text{Shannon}^{(N)}}
}=
\lf\f{q_{\rm tot}}{q_s}\ri^{-(N-1)t}
\braket{\I | T^{t+1} |\I}.
\ee
The small--$x$ expansion of $\overline{S_\text{Shannon}}$ may therefore be computed from the amplitude $\braket{\I | T^{t+1} |\I}$ (compare Sec.~\ref{subsec:S2_order_x2})
but may also be read off from Sec.~\ref{sec:s1higherorder} and  Eq.~\ref{eq:partial_N_M} below:
\begin{equation}
\overline{S_{\rm Shannon}} = t \ln (q_{\rm tot}/q_s) - \frac{1}{4}x^2  + \frac{1}{12} x^4 + \ldots 
\end{equation}
A nonperturbative result is given by the generating function defined in Eq.~\ref{eq:d_Omega_N}:
\begin{equation} \label{eq:shannonwithG}
\overline{S_{\rm Shannon}} = t \ln (q_{\rm tot}/q_s) - \int_0^{\infty} du \frac{G(u,x) - e^{-u} }{u^2}.
\end{equation} 
From Eq.~\eqref{eq:SShannonOmega}, we see how Shannon entropy is expressed as a difference between the entropy of a uniform distribution over all possible measurement outcomes and an entropy deficit that signals a nontrivial distribution instead. The former term depends on the specific protocol while the latter becomes universal in our scaling regime. 

In particular, at large times ($x\gg1$), we enter a stationary regime in which the Shannon entropy growth rate becomes a constant and we have the particularly simple result [through a two-row expansion ($R = 2$) by the approach of Sec.~\ref{sec:tracemoments} for example]
\be\label{eq:shannonlatetime}
{\overline{S_\text{Shannon}}\sim t \ln(q_{\rm tot}/q_s) - \f{x}{2}},
\ee
Heuristically, this large-$x$ form may be understood in terms of the free energy density of the  partition function
$\braket{\I | T^{t+1} |\I}$.
If we fix $N$, then (as ${t\to\infty}$) the extensive-in-$t$ part of the free energy is independent of the boundary conditions.
Therefore we can relax the constraint on the product of the kink types, and the partition function becomes a product of independent bonds. This gives
\ba
\overline{e^{-(N-1) S^{(N)}_\text{Shannon}}}
\asymp 
\lf \f{q_{\rm tot}}{q_s} \ri^{-(N-1)t}
e^{\f{N(N-1)t}{2q}} ,
\end{align}
where $N(N-1)/(2q)$ is the total Boltzmann weight for placing a kink on a given bond, taking account of the $N(N-1)/2$ types of transposition. 
If we assume that we can now commute the large $t$ limit and the ${N\to 1}$ limit, we again obtain Eq.~\eqref{eq:shannonlatetime}.

\section{Numerics}\label{sec:numerics}

This section is about numerical verification of the results for the two universality classes discussed in Sec.~\ref{sec:ginibre} and Sec.~\ref{sec:mea}, namely  purification by independent random matrices or by a measurement process. In the replica calculation, these two universality classes correspond to taking respectively the ${N\to 0}$ or ${N\to 1}$ limit for the total degree of the permutation group. 

For the case of independent random matrices, we study two kinds of models: 
(A) the independent random Ginibre ensemble  and 
(B) Haar random unitary matrices alternated with projection operators. 
(A) is the model in Sec.~\ref{sec:ginibre}.
(B) is the version of the model in Sec.~\ref{sec:mea} in which the measurements are forced, instead of chosen with Born's rule.
These two models should give us the same scaling function in the scaling regime, so this allows a numerical test of universality.

For the measurement universality class, we again consider two different models in order to check universality.
One of these is the quantum measurement process 
described in Sec.~\ref{sec:mea} (taking account of Born's rule).
This is the  ``$N\to 1$'' analog of  (B) above. For a second model, we consider the ``$N\to 1$'' analog of (A) above. 
This is not a standard quantum measurement process, 
but on analytical grounds we expect it to be in the same universality class. 
It is defined by reweighting the averages of the independent Ginibre ensemble by an additional factor of ${\tr \check\rho}$. 
Below we will confirm numerically that, for each value of $N$, the  scaling forms agree between the (A)--type and the (B)--type model.

For simulations, we perform the following  protocols. 
First, protocols in which we start by drawing  products of independent random matrices $M$ from either model (A) or (B) as in Eq.~\eqref{eq:M_m}. 
The difference lies in how we average: 
\begin{itemize}[itemindent=5.4em, left=5.7em]
    \item[\bf Protocol (0)] $(N=0)$ We take the sample mean of the $n$th R\'enyi entropy $\langle S_n \rangle$, where $S_n$ is computed  using the normalized density matrix $\rho$ for a given sample, Eq.~\eqref{eq:renyi_entropy}.
    \item[\bf Protocol (1)] {\bf ($N=1$ by reweighting)}: We define the average of $S_n$ using the reweighted expression $\frac{\langle \tr( \check{\rho} S_n )\rangle}{\langle \tr( \check{\rho} ) \rangle}$, where the numerator and denominator are sample means.
\end{itemize}
For the type-(B) model, we have the alternative of simply 
simulating the measurement process in the standard way:
\begin{itemize}[itemindent=5.6em, left=6.1em]
\item[\bf Protocol (1')] 
     {\bf (${N=1}$ by Born rule dynamics)}: we simulate the dynamics defined in Sec.~\ref{sec:mea}, 
     for the case where a single qubit is measured in each timestep,
     generating a random unitary and a random measurement outcome in each timestep.
    For the measurement outcome, we accept $\check{\rho} = PU \rho U^\dagger P$ or $\check{\rho} = (\I - P) U \rho U^\dagger (\I - P)$ with probability $\tr(\check{\rho})$.
    Here $P$ is a projector of rank $q_{\rm tot}/2$, where $q_{\rm tot}$ is the Hilbert space dimension. 
    We sample the mean of $S_n$ in the simulation.
\end{itemize}
This should give identical results to protocol (1) for the type (B) model, as the resulting averages are equal (even for finite time).  
Note that although for convenience we label the two universality classes above by ``$N=0$'' and ``$N=1$'', the numerical results do not rely on the use of the replica trick in any way.

\begin{figure}[h]
\centering
\includegraphics[width=0.95\columnwidth]{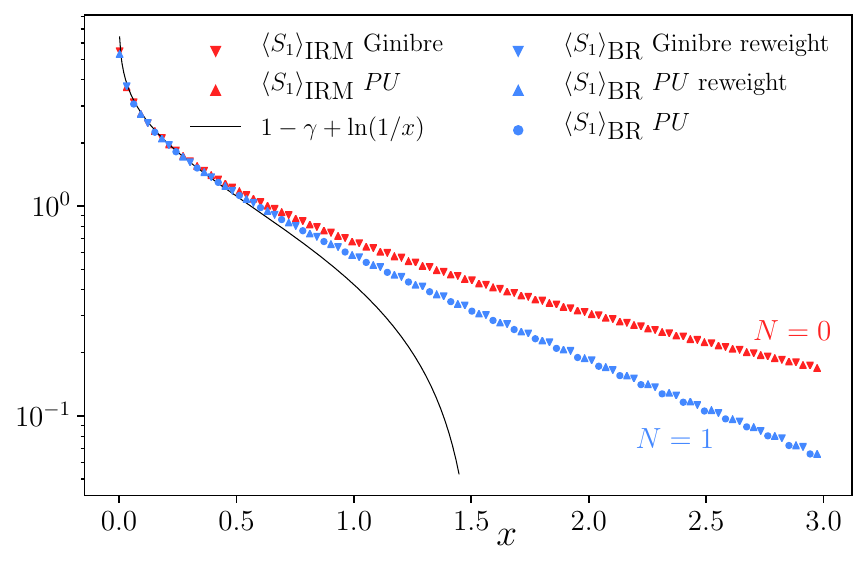}
\includegraphics[width=0.95\columnwidth]{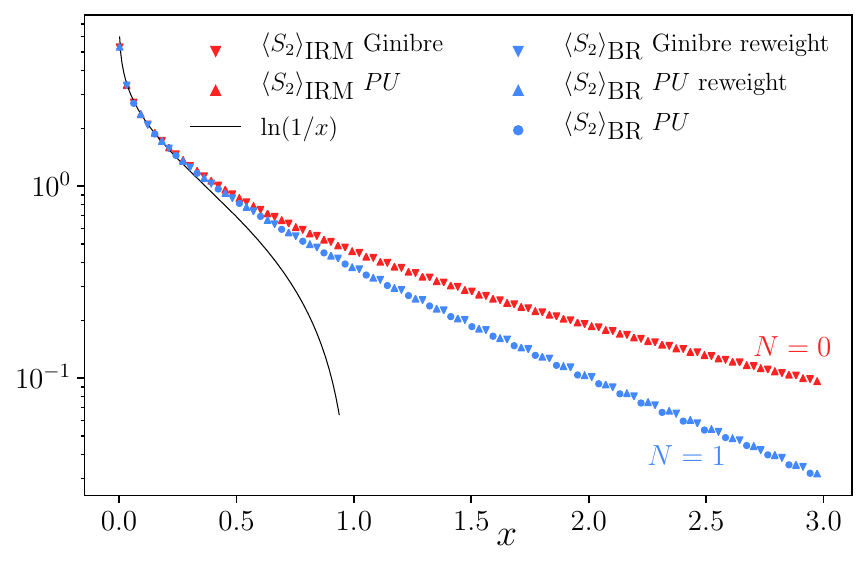}
\caption{The scaling functions of $\overline{S_1}$ (top) and $\overline{S_2}$ (bottom) for the two universality classes $N=0$ (red, independent random matrices) and $N=1$ (blue, measurement with Born rule). The agreement between the two $N=0$ protocols  (red, upward and downward triangle markers), 
and similarly the agreement between the two $N=1$ protocols (blue, upward/downward triangle and circular markers), 
is evidence for universality. ``Ginibre'' refers to the (A)--type models discussed in the text, and ``$PU$'' refers to the (B)--type models which alternate unitaries and projectors.}
\label{fig:cmp_class}
\end{figure}

First, we extract the scaling functions numerically for $\overline{S_1}$ and $\overline{S_2}$ in order to test for universality. Results are shown in Fig.~\ref{fig:cmp_class}.
Protocol (0) is marked by red upward and downward triangles. 
Protocols (1) and (1') are marked by upward/downward triangles and circles. 

For all of the simulations, the total Hilbert space dimension is taken to be $400$. 
This implies that ${q=t_*=400}$ for both the (A)-type and the (B)-type models. 
[Recall that for the (B)-type model we have $1/q=1/q_{\rm s}-1/q_{\rm tot}$, by \eqref{eq:T12_eff_q}: here ${q_{\rm tot} =400}$ and ${q_s = 200}$, giving ${q=400}$.]

The fact that (for a given $N$) results from two different models collapse into the same scaling function  (for $x$-values of order 1) is strong evidence for the existence of two well-defined universality classes. The black lines show the small-$x$ asymptotics of the scaling functions.

\begin{figure}[h]
\centering
\includegraphics[width=0.8\columnwidth]{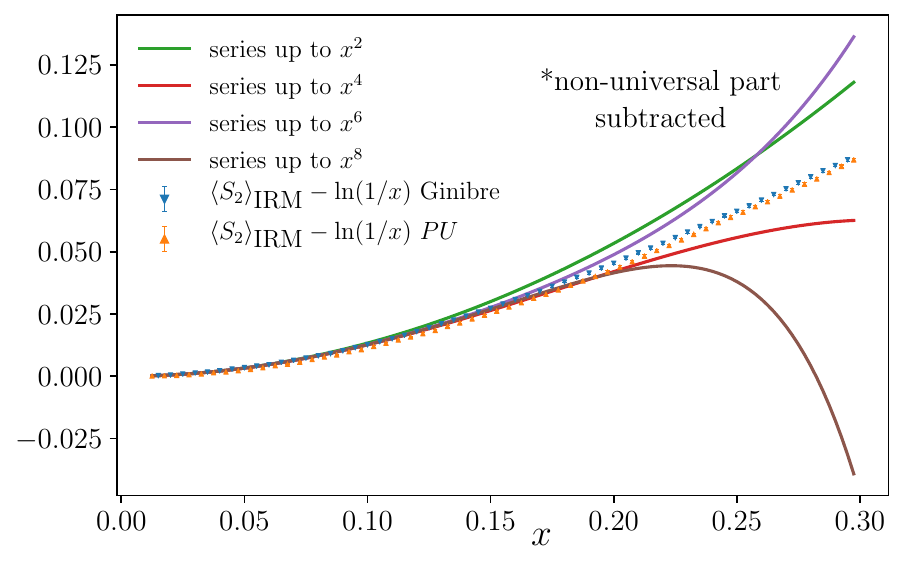}
\includegraphics[width=0.8\columnwidth]{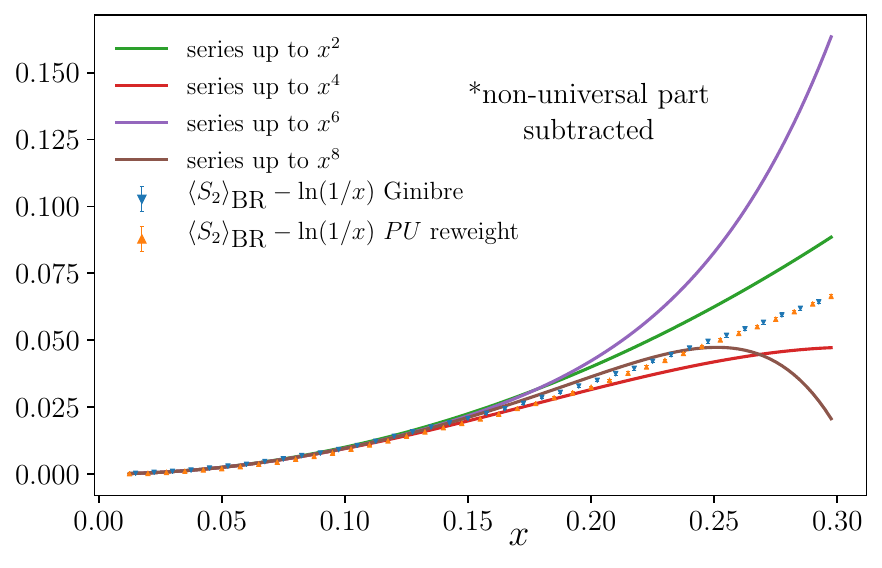}
\caption{Comparison of the numerical data of $\overline{S_2}$ with the small $x$ expansion. Non-universal parts are subtracted.}
\label{fig:N0_small_x}
\end{figure}

\begin{figure}[h]
\centering
\includegraphics[width=0.8\columnwidth]{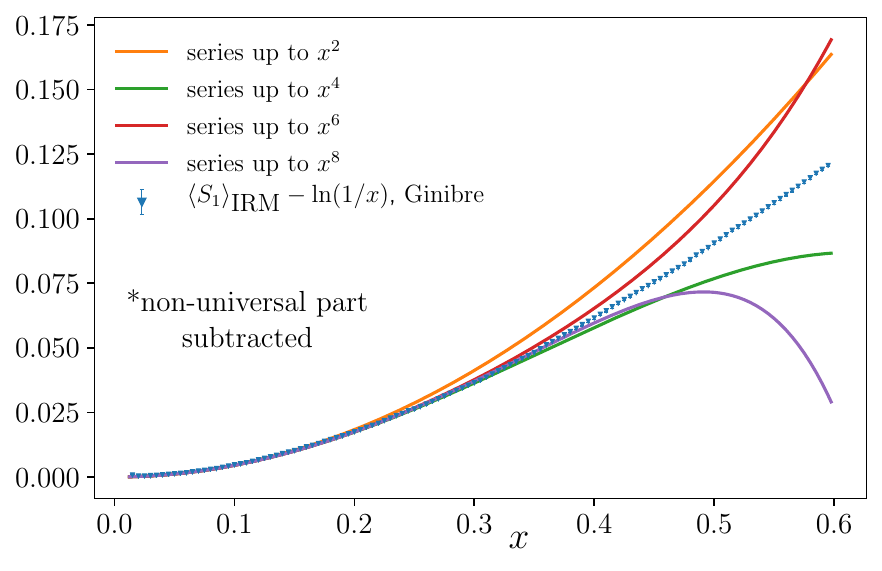}
\includegraphics[width=0.8\columnwidth]{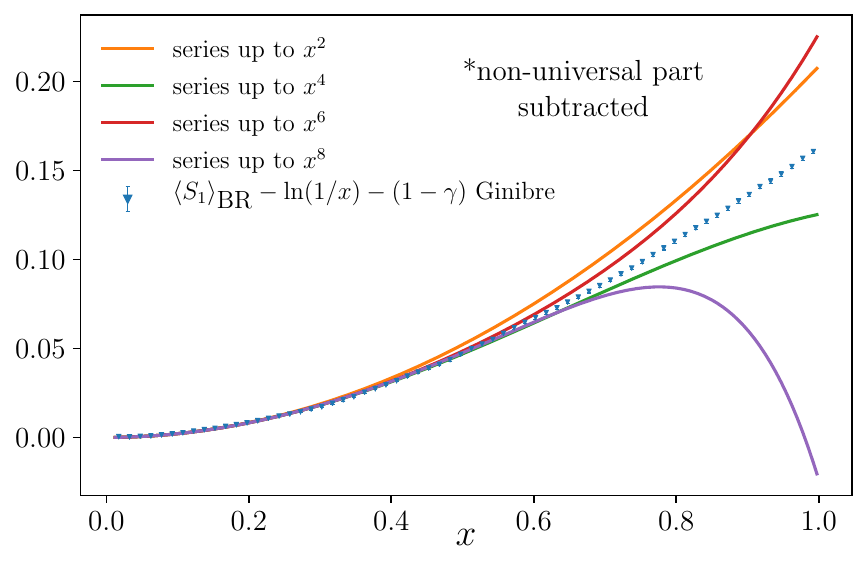}
\caption{Comparison of the numerical data of $\langle S_1 \rangle$ with the small $x$ expansion. Non-universal parts are subtracted.}
\label{fig:N1_small_x}
\end{figure}

Next, in Fig.~\ref{fig:N0_small_x} and Fig.~\ref{fig:N1_small_x} we compare the {small-$x$} expansions of the scaling functions with numerical results. Since the numerics are for a finite value of $q_{\rm tot}$, the raw data contains non-universal terms that are dominant at small $x$, and which reduce the range of $x$ where the data can be compared with the expected scaling forms. To reduce these finite size effects,  we analytically compute and subtract the non-universal terms. We then compare with the analytic results of the small $x$ expansion in Eq.~\eqref{eq:S2_IRM_BR} and \ref{eq:S1_asymp} in Fig.~\ref{fig:N0_small_x} and Fig.~\ref{fig:N1_small_x}. The agreement of the numerical curve is up to the point where the $x^6$ and $x^8$ corrections start to deviate. 
The present system size gives reasonable evidence for the correctness of the nontrivial $x^2$ terms in the power series expansions.
Note that higher powers of $x$ do not give appreciably better agreement with the data for intermediate $x$, probably indicating that the series are only asymptotic series, rather than having a nonzero radius of convergence.

\section{Conclusions and Future Extensions}

We have argued for universal regimes for purification both in monitored quantum systems and in products of random matrices.
The analytical and numerical results confirm that the R\'enyi entropies have nontrivial universal probability distributions which depend on the scaling variable $x=t/t_*$ (with $t_*=q$ in our notation) and which differ between the two classes of problem. 
In the context of the measurement phase transition, the universality discussed here holds within the weakly monitored (entangled) phase.

The corresponding universality class for purification by measurements is expected to include, for example,  generic quantum circuit or continuous-time models (with either projective or weak measurements) that lie within the weakly monitored phase. Interestingly, in standard spatially local or $k$-local circuit models, the purification timescale $t_*$ is however strongly influenced by rare temporal regions of the circuit.

To conclude, we discuss extensions of this work: first, clarifying more technical issues related to the replica formalism, and second, extensions to other universality classes.

One additional motivation for studying the present effective 1D replica models is to shed light on higher-dimensional effective models for random quantum circuits, monitored circuits, and random tensor networks. Such models have been intensely studied in the recent literature, particularly in 1+1 dimensions.  However, it is usually challenging to take the replica limit explicitly.  The present setting allows exact results in the replica limit,  clarifying, for example, the difference between  exact averages and ``annealed'' approximations to them, as well as differences in the scaling functions for the von Neumann and higher R\'enyi entropies.

Even in 1D, there are various aspects of the replica treatment that it will be interesting to examine further. At very late times one can understand $S_2(t)$ [and $S_n(t)$ for other $n$] in terms of just two singular values. At first, one would think that late times are also simple in the transfer matrix formalism and reduce to studying a small number of dominant eigenvalues of the replica transfer matrix. However, this appears to be subtle in the replica limit, and the nature of the analytical continuation requires further study.

The universality we have discussed in this paper is not associated with a critical point, but instead with low effective ``temperature'' in the replica statistical mechanics model.
However, it is possible to obtain a phase transition even in the 1D problem when the effective interactions between the kinks are made long-range: this will be discussed elsewhere~\cite{ANunpublished}. 

The present scaling results may also be extended to 
consider frame potentials in many-body quantum chaos: this will be discussed in Ref.~\cite{christopoulos2024universal}.

Finally, we discuss extensions of our results to other dynamical universality classes, in which further restrictions are imposed on the form of the matrices/circuits.

{\bf Real matrices:} Both for Born rule dynamics and for a product of independent random matrices, we expect the universality classes to change when the matrices are constrained to be real-valued. 
Within random matrix theory, a natural example is a product of real Ginibre matrices. 
Similarly, we may consider quantum circuits composed of real unitary (i.e. orthogonal) gates together with measurements. 

For such models, the basis of invariant states is enlarged from the $N!$ permutation states $\{\kket{\sigma}\}$ to a set of ${(2N-1)!!}$ states corresponding to all possible pairings of $2N$ elements (these generalize Eq.~\eqref{eq:permutationstatesdefn} by allowing pairings between two barred indices, or between two unbarred indices). Correspondingly, there is an enlarged universal transfer matrix generator $A$ (cf. Eq.~\eqref{eq:transfmatlimit}). We expect that a universal regime again exists for real matrices but with different scaling functions from the complex case. 

{\bf Other random matrix ensembles:} We have described criteria for the universality of our results both in the context of invariant random matrix ensembles and in the context of random circuits (which have a natural tensor product structure for the Hilbert space). It would be interesting to investigate random matrix models outside these classes: for example, various models of sparse random matrices.
We recall that the key timescale $t_*$ (which in the discrete-time setting we can also think of as an effective dimensionality,  $q=t_*$) is not necessarily proportional to the size $d$ of the matrix: for example in the random circuit case $t_*$ scales as $d^\alpha$, where $\alpha$ is smaller than one.

The measurement problem also yields a large class of natural ensembles for products of random matrices,  differing from products of independent random matrices by correlations of the ``Born rule'' form.

{\bf Clifford circuits:} We may consider purification by monitored Clifford circuits~\cite{gullans2020dynamical} or quasi-1D Clifford tensor networks.  In this case, there is again an enlarged space of invariant states \cite{gross2021schur,li2021statisticalclifford,leone2023clifford}, and we expect an enlarged transfer matrix generator $A$ and a distinct universality class. The fact that the scaling functions for Clifford must differ from the generic case is also clear from the fact that in the Clifford system, all R\'enyi entropies are equal, and are given by an integer times a basic unit. For example, the latter point implies that statistical fluctuations in the entropy are of order 1 in the small $x$ regime, whereas in the generic case fluctuations are small at small $x$ [see Eq.~\eqref{eq:introsmallxasymptotics}].

{\bf Charge conservation:}  Circuits that respect a symmetry, such as U(1) charge conservation, may also be considered. Generically, we expect the scaling functions in this paper to apply in this case also, with the caveat that $t_*$ will in general depend on the charge sector.  If the charge is initially undetermined, we must wait until it has been revealed by measurements (``charge sharpening'' \cite{agrawal_entanglement_2022}) before knowing the appropriate value of $t_*$ to use in the scaling form. However, in the weak-monitoring phase, this sharpening timescale is typically much shorter than the purification time.

{\bf Verification with small $\mathcal{V}$:} Finally, it would be interesting to simulate generic quantum circuits or continuous-time models with a small number $\mathcal{V}$ of qubits. Since the characteristic timescale $t_*$ grows very fast (exponentially) with $\mathcal{V}$, one might anticipate that the universal scaling functions can match well to data even for fairly small values of~$\mathcal{V}$.\\
 
 \textit{Note added:}
While we were completing this manuscript, the related work  Ref.~\cite{bulchandani_random-matrix_2023} appeared on the arXiv, which studies similar random matrix models to the present paper and argues for universality. The methods in Ref.~\cite{bulchandani_random-matrix_2023} are complementary to those used here.

\addtocontents{toc}{\protect\setcounter{tocdepth}{-1}}

\acknowledgements
We thank J. Chalker for discussions.
ADL is extremely grateful for discussions and suggestions to G. Barraquand and Pierre Le Doussal and especially A. Borodin for suggesting the references~\cite{stanley2009combinatorial,olshanski2009plancherel} about the polynomial behavior of the coefficients.
ADL acknowledges support by the ANR JCJC grant ANR-21-CE47-0003 (TamEnt). CL acknowledges the fellowship support from the Gordon and Betty Moore Foundation through the Emergent Phenomena in Quantum Systems (EPiQS) program.
TZ was supported by NTT Research Award AGMT DTD 9.24.20. at the start of the project. TZ acknowledges the Engaging Cluster from MIT for providing the computational resources for this project.

\appendix

\addtocontents{toc}{\protect\setcounter{tocdepth}{+1}}

\section{Suppression of composite domain walls}
\label{app:energyentropy}

We present here a simple physical argument regarding the suppression of composite domain walls which was referred to in Sec.~\ref{sec:universality}.

Let us start with a gas of elementary kinks on a chain of length $t$ (with some fixed boundary conditions).
In other words, we start with a transfer matrix in which higher kinks have fugacity zero.
Let us take $t$ to be at most of order $q$, so that the number of kinks remains of order $q^0$ as ${q\to \infty}$.

Now compare this partition function with one in which higher kinks are allowed. For concreteness, consider a higher kink of type ${[(12)(34)]}$, which contributes a fugacity ${T_{\I,(12)(34)}}$ to the Boltzmann weight. 
This kink is free to dissociate into \textit{two} elementary kinks, with combined fugacity $q^{-2}$.
However the additional kink increases the entropy by a factor of $\ln t$.\footnote{With the present scaling $t\lesssim q$, where the number of kinks is of order $1$, each extra kink contributes an entropy $\ln t + O(1)$.}
Therefore (comparing free energies) the composite kink is entropically suppressed whenever 
\begin{equation}\label{eq:higherkinksuppressioncondition2}
T_{\I,(12)(34)} \ll q^{-2} t,
\end{equation}
where the final factor is the exponential of the added entropy.
In order for the scaling limit to make sense, the composite kink must be absent on scales $t$ of order $q$, i.e we require $T_{\I,(12)(34)} \ll q^{-1}$. This is easily generalized to give Eq.~\eqref{eq:higherkinksuppressioncondition} for an arbitrary composite kink of type $\conj{\mu}$.

Even if this condition holds, higher kinks may still be important at timescales much shorter than $q$.
For simplicity we consider only kinks of type $\conj{(12)(34)}$.  Eq.~\eqref{eq:higherkinksuppressioncondition2} gives us a crossover time $t_1$, above which we can neglect these kinks:
 \begin{equation}
 \label{eq:appendixt1}
 t_1 = q^2 T_{\I,(12)(34)}.
 \end{equation}
 Given $T_{\I,(12)(34)}\ll 1/q$, then ${t_1\ll q}$,  so this large crossover time is consistent with the existence of a scaling regime.  Eq.~\eqref{eq:appendixt1} is equivalent to Eq.~\eqref{eq:t1scale} in the main text.

\section{Character expansion and the generating function for entropy \label{app:hurw}}

In this appendix we provide more details on extracting the scaling functions for the R\'enyi entropy $S_n|_{\rm IRM,BR}$, and the analytical continuation to $n=1$ to get the von Neumann entropy $S_1\Big|_{\rm IRM, BR}$. We start by providing a practical way to evaluate the character expansion using \emph{Schur functions}. The Schur function $s_\lambda(p)$ associated with a partition $\lambda=(\lambda_1,\lambda_2,...,\lambda_\ell) \vdash N$ is a multivariable function of (the ``symmetric polynomial'') indeterminates $p=(p_1,p_2,...)$. 

First define the one-part partition Schur polynomial $s_n(p)$ via the series 
\begin{equation} \sum_{n=0}^\infty s_n z^n = e^{\sum_{n=1}^\infty \frac{p_n}{n} z^n},
\end{equation}
then define the Schur function $s_\lambda(p)$ via the Jacobi-Trudi identity \begin{equation}
s_\lambda = \mathrm{Det}||s_{\lambda_j-j+i}||,
\end{equation}
i.e. the Schur function $s_\lambda(p)$ is defined via the determinant of an $\ell \times \ell$ matrix whose $(i,j)$ entry is the one-part partition Schul polynomial $s_{\lambda_j-j+i}(p)$. We list a few examples here:  $s_0=1$, $s_{1^1}=p_1$, $s_{2^1} = \frac{1}{2}(p_1^2+p_2)$, $s_{3^1} = \frac{1}{6}(p_1^3+3p_1p_2+2p_3)$ etc.

The Schur function is a convenient tool to calculate various quantities associated with irreps of the symmetric group. For example, for a permutation element  $\sigma \in \conj{1^{r_1}2^{r_2}...\ell^{r_\ell} }\subset S_N$, the character of an irrep $\lambda$ evaluated on $\sigma$ can be calculated via
\begin{equation}
\chi^\lambda_\sigma = 1^{r_1}2^{r_2}\cdots \ell^{r_\ell}\left(\frac{\partial^{r_1}}{\partial p_1^{r_1}}\frac{\partial^{r_2}}{\partial p_2^{r_2}}\cdots\frac{\partial^{r_\ell}}{\partial p_\ell^{r_\ell}}s_\lambda(p)\right)\Bigg|_{p=0},
\end{equation}
and in particular, the dimension of the irrep $\lambda$, $d^\lambda$, can be extracted
\begin{equation}
d^\lambda = 
\chi^\lambda_{\mathbb{I}}
=\frac{\partial^N}{\partial p_1^N} s_\lambda(p)\Big|_{p=0}
=N!s_\lambda(1,0,0,...).
\end{equation}

Using Schur function, we can rewrite Eq.~\eqref{eq:expsum}:
we have 
\begin{widetext}
\begin{equation}\label{eq:Schur_expansion}
\begin{aligned}
\overline{(\tr(\check \rho^1))^{r_1}
(\tr(\check \rho^2))^{r_2}\cdots
(\tr(\check \rho^\ell))^{r_\ell}} &= \langle 1^N|T^{t+1}|1^{r_1}2^{r_2}\cdots \ell^{r_\ell}\rangle \\
&\rightarrow H^\circ_{1^{r_1}2^{r_2}\cdots \ell^{r_\ell}}(x) \equiv  1^{r_1}2^{r_2}\cdots \ell^{r_{\ell}} \cdot \sum_{\lambda \vdash N} \frac{d^\lambda}{N!}
\frac{\partial^{r_1}}{\partial p_1^{r_1}}
\frac{\partial^{r_2}}{\partial p_2^{r_2}}
\cdots
\frac{\partial^{r_\ell}}{\partial p_{\ell}^{r_{\ell}}}
s_\lambda(p)\Bigg|_{p=0}
e^{\frac{x}{2} \sum_i (\lambda^2_i-\lambda^{\prime 2}_i)}.
\end{aligned}
\end{equation}
\end{widetext}
where the last arrow takes the scaling limit $t\rightarrow \infty$, $q\rightarrow \infty$, $x=t/q$ finite.

As an example, we consider two simple cases: when $r_2=r_3=\cdots=0$, the derivative is simply $\frac{\partial^N}{\partial p_1^N} s_\lambda(p)|_{p=0} = d^\lambda$ and we obtain a formula for computing $\overline{(\tr(\check  \rho))^N}$ \cite{hurwitz1891riemann,dubrovin2017classical}. When $k=1$ and $n=N$ the derivative is $\frac{\partial}{\partial p_N} s_\lambda(p)$ and simple calculation gives $\overline{\tr(\check  \rho^N)}= \frac{2^{N-1}}{N!}\left(\sinh \frac{N}{2}x\right)^{N-1}$, in agreement with Eq.~\eqref{eq:M_N_sinh}.

In Eq.~\eqref{eq:def_a_N_k_n_l} in the main text we defined coefficients $a_{N,k,n,\ell}$ through $\langle \mathbb{I}|T^{t+1}|1^{N-nk}n^k\rangle = \left(\frac{n^{n-2}}{(n-1)!}\right)^k x^{(n-1)k} \sum_{\ell=0}^\infty a_{N,k,n,\ell}\frac{x^{2\ell}}{(n+1)^\ell}$. We used Eq.~\eqref{eq:expsum} to compute these coefficients. First, we have $a_{N,k,n,0}=1$. For an overly determined set of values of $(N,k,n)$: $0\leq N\leq 41$, and $0\leq kn \leq N$, we have verified that the values of $a_{N,k,n,\ell}$, $\ell=1,2,3,4$, satisfy the following polynomial expression: define \begin{equation}
Y\equiv \frac{n(n-1)k}{2},
\end{equation}
we have
\begin{widetext}
\begingroup
\allowdisplaybreaks
\begin{align}
a_{N,k,n,1}=&\frac{(1+n) (-1+N) N}{4}-\frac{(1+n) (6+5 n)}{12} Y+N Y+Y^2,\\                        
a_{N,k,n,2}=&\frac{(1+n)^2 (-1+N) N (-12+N+3 N^2)}{96}+\frac{(1 +n)^2 (360+360 n+360 n^2+359 n^3+300 n^4) }{1440}Y\notag\\
&-\frac{26+79 n+64 n^2+35 n^3}{48} N Y-\frac{6-7 n+16 n^2+5 n^3}{48} N^2 Y+\frac{1+n}{4}  N^3 Y\notag\\
&-\frac{156+636 n+779 n^2+586 n^3+143 n^4}{288} Y^2-\frac{ 3-10 n+5 n^2}{12} N Y^2+\frac{3+n}{4} N^2 Y^2\notag\\
&-\frac{2-9 n+n^2}{12} Y^3+N Y^3+\frac{Y^4}{2},\\                        
a_{N,k,n,3}=&\frac{(1+ n)^3 (-1+N) N (1344-700 N-105 N^2+30 N^3+15 N^4)}{5760}\notag\\
&-\frac{(1+n)^3 (42336+42336 n+42336 n^2+42336 n^3+42336 n^4+42335 n^5+41958 n^6+33516 n^7) }{90720}Y\notag\\
&+\frac{8176+ 32764 n+52012 n^2+46869 n^3+37331 n^4+25599 n^5+16053 n^6+4500 n^7}{5760}N Y\notag\\
&+\frac{-3570- 15785 n-23655 n^2-12316 n^3-5258 n^4+2337 n^5+1259 n^6+300 n^7}{5760}N^2 Y\notag\\
&-\frac{108+277 n+201 n^2+339 n^3+115 n^4}{576} N^3 Y-\frac{ (-2+n) (1+n) (5+n) (1+5 n)}{384} N^4 Y +\frac{(1+n)^2}{32}  N^5 Y\notag\\
&+\frac{24528+123000 n+263376 n^2+348366 n^3+376127 n^4+333573 n^5+238041 n^6+106001 n^7+20388 n^8}{17280}Y^2\notag\\
&+\frac{-71 40-41000 n-82765 n^2-69339 n^3-45803 n^4-6889 n^5+1200 n^6 }{5760}N Y^2\notag\\
&-\frac{648+ 1536 n-13 n^2+2565 n^3+729 n^4+143 n^5 }{1152}N^2 Y^2-\frac{-5-41 n-7 n^2+5 n^3}{48} N^3 Y^2\notag\\
&+\frac{(1+n) (5+n)}{32}  N^4 Y^2-\frac{42840+29826 0 n+753750 n^2+866911 n^3+710157 n^4+278481 n^5+40585 n^6 }{51840}Y^3\notag\\
&-\frac{216+522 n-409 n^2+676 n^3+143 n^4}{288} N Y^3-\frac{-10-105 n+2 n^2+n^3}{48} N^2 Y^3+\frac{5+3 n}{12} N^3 Y^3\notag\\
&-\frac{108+292 n-325 n^2+202 n^3+87 n^4}{288} Y^4-\frac{-5-63 n+2 n^2}{24} N Y^4+\frac{5+n}{8} N^2 Y^4\notag\\
&+\frac{2+29 n+3 n^2}{24} Y^5+\frac{N Y^5}{2}+\frac{Y^6}{6},\\                        
a_{N,k,n,4}=&\frac {(1+n)^4 (-1+N) N (-519168+434224 N-61208 N^2-16065 N^3+35 N^4+525 N^5+105 N^6)}{645120} \notag\\
&+\frac{1}{4838400} (1+n)^4(7787520+7787520 n+7787520 n^2+7787520 n^3+7787520 n^4+7787520 n^5\notag\\
&\qquad~\qquad~\qquad\qquad +7787520 n^6+7787519 n^7+7785480 n^8+7646640 n^9+5798160 n^{10}) Y\notag\\
&-\frac{1}{725760}(429 0264+21465432 n+45301896 n^2+55572327 n^3+50802624 n^4+43651748 n^5+36354968 n^6\notag\\
&\quad~\quad~\qquad +30288963 n^7+22899212 n^8+15503978 n^9+7286580 n^{10}+1541736 n^{11}) N Y\notag\\
&-\frac{1}{2903040}(-133 76664-68323920 n-141058680 n^2-155562393 n^3-114310224 n^4-77086094 n^5\notag\\
&\qquad \qquad\quad~ -41619428 n^6-21317001 n^7-1790996 n^8+3290104 n^9+1408176 n^{10}+268128 n^{11}) N^2 Y\notag\\
&+\frac{-38 694-243121 n-542128 n^2-505901 n^3-136818 n^4+9173 n^5+132148 n^6+64777 n^7+14100 n^8}{69120}N^3 Y\notag\\
&+\frac{-1150 0-48070 n-76360 n^2-61421 n^3-48284 n^4-12796 n^5+3596 n^6+1559 n^7+300 n^8 }{46080}N^4 Y\notag\\
&-\frac{(1+n) (42-49 n-219 n^2+381 n^3+125 n^4) }{4608}N^5 Y-\frac{(1+n)^2 (-42-103 n+16 n^2+5 n^3) }{4608}N^6 Y\notag\\
&+\frac{(1+n)^3}{384}  N^7 Y\notag\\
&-\frac{1}{87091200}(514831680+309 2376960 n+8301683520 n^2+13916566800 n^3+17871889680 n^4\notag\\
&~\quad~\qquad\qquad +20111002320 n^5+20598887659 n^6+19630971796 n^7+16896121794 n^8+12523347916 n^9\notag\\
&~\quad~\qquad\qquad+7058306539 n^{10}+2495528280 n^{11}+397436400 n^{12}) Y^2\notag\\
&-\frac{1}{1451520}( -13376664-82802496 n-216083742 n^2-322706370 n^3-339652635 n^4-305610748 n^5\notag\\
&\quad\,\qquad\qquad -222928152 n^6-137445342 n^7-52652287 n^8-7567308 n^9+536256 n^{10}) N Y^2\notag\\
&+\frac{1}{138240}(- 232164-1939740 n-5649137 n^2-7399688 n^3-4360520 n^4-1968112 n^5+859003 n^6\notag\\
&\qquad~\quad~\quad +817684 n^7+252778 n^8+40776 n^9) N^2 Y^2\notag\\
&+\frac{-6 9000-307980 n-482165 n^2-364352 n^3-525066 n^4-166796 n^5-18497 n^6+3600 n^7 }{69120}N^3 Y^2\notag\\
&-\frac{420 -1604 n-8041 n^2+2132 n^3+4654 n^4+872 n^5+143 n^6 }{9216}N^4 Y^2\notag\\
&-\frac{(1+n) (-21-83 n-9 n^2+5 n^3)}{384}  N^5 Y^2+\frac{(1+n)^2 (7+n)}{384}  N^6 Y^2\notag\\
&+\frac{1}{2903040}(17835552+129256512 n+405758808 n^2+754035324 n^3+998224836 n^4+1083271249 n^5\notag\\
&\qquad\,\quad\qquad +956480096 n^6+687801882 n^7+354824152 n^8+109472321 n^9+14747124 n^{10}) Y^3\notag\\
&+\frac{1}{207360}(- 464328-4743396 n-16867686 n^2-28442171 n^3-25182992 n^4-17181798 n^5-4423418 n^6\notag\\
&\qquad~\quad~\quad +338557 n^7+244656 n^8) N Y^3\notag\\
&-\frac{41 4000+2008080 n+3098790 n^2+1641793 n^3+3855112 n^4+1382910 n^5+297466 n^6+40585 n^7 }{207360}N^2 Y^3\notag\\
&-\frac{14 0-974 n-4615 n^2+1019 n^3+823 n^4+143 n^5 }{1152}N^3 Y^3-\frac{-70-385 n-207 n^2+13 n^3+n^4}{384} N^4 Y^3\notag\\
&+\frac{(1+n) (7+3 n)}{96}  N^5 Y^3\notag\\
&-\frac{1}{2488320}(2785968+330 55200 n+137503656 n^2+280806312 n^3+320603419 n^4+268123420 n^5\notag\\
&\qquad\,\qquad\quad +130071978 n^6+31882612 n^7+2892283 n^8) Y^4\notag\\
&-\frac{1 03500+552780 n+884025 n^2+252346 n^3+923142 n^4+368526 n^5+40585 n^6 }{51840}N Y^4\notag\\
&-\frac{210 -2012 n-10109 n^2+1481 n^3+575 n^4+87 n^5 }{1152}N^2 Y^4\notag\\
&-\frac{-105-640 n-169 n^2+6 n^3}{288} N^3 Y^4+\frac{35+30 n+3 n^2}{192} N^4 Y^4\notag\\
&-\frac{41400+24570 0 n+436170 n^2+102553 n^3+352641 n^4+206763 n^5+35845 n^6 }{51840}Y^5\notag\\
&-\frac{42-491 n-2713 n^2-5 n^3+87 n^4}{288} N Y^5+\frac{42+283 n+28 n^2+3 n^3}{96} N^2 Y^5+\frac{7+3 n}{24} N^3 Y^5\notag\\
&+\frac{-28+372 n+2293 n^2+502 n^3+n^4}{576} Y^6+\frac{7+52 n+3 n^2}{24} N Y^6+\frac{7+n}{24} N^2 Y^6\notag\\
&+\frac{6+49 n+7 n^2}{72} Y^7+\frac{N Y^7}{6}+\frac{Y^8}{24}.
\end{align}
\endgroup

From these expressions, we conjecture the general form of $a_{N,k,n,\ell}$ to be
\begin{equation}
a_{N,k,n,\ell} = \sum_{0\leq i+j\leq 2\ell}
 a_{i,j}(n) N^i Y^j,\quad
 \mathrm{deg}_n(a_{i,j}) \leq [i/2]+2(2\ell-i-j),
\end{equation}
where $[i/2]$ denotes the largest integer equal to or less than $i/2$. From this, we see that for each $\ell$, the number of coefficients in order to determine the polynomial $a_{N,k,n,\ell}$ is $\frac{1}{3} (1 + 2 l) (3 + 8 l + 5 l^2)$. For $\ell=1,2,3,4$, this number is $16, 65, 168, 345$ respectively. This means that the number of constraints that we used to determine $a_{N,k,n,\ell}$, namely, the number of constraints $0\leq N\leq 41$, $0\leq kn\leq N$, exceeds the number of coefficients, providing convincing evidence for the polynomiality of the coefficients $a_{N,k,n,\ell}$ in $N$, $k$, and $n$ for given $\ell$.

The polynomials $a_{N,k,n,\ell}$ are analytic in $k$, hence allowing us to take derivative with respect to $k$. Using 
\begin{equation}
\begin{aligned}
\overline{S_2}(N)
&=-\overline{(\tr(\check\rho))^N \ln(\tr(\rho^2))}\\
&=
-\lim\limits_{k\rightarrow 0}
\frac{d}{dk}\overline{(\tr(\check \rho^n))^k(\tr(\check \rho))^{N-nk}}\Big|_{n=2}\\
&=-\lim\limits_{k\rightarrow 0}
\frac{d}{dk}\langle
\mathbb{I}|T^{t+1}|1^{N-nk} n^k\rangle\Big|_{n=2},
\end{aligned}
\end{equation}
we get the 2nd R\'enyi entropies in the scaling limit for the two classes:
\begin{subequations}
\label{eq:S2_IRM_BR}
\begin{align}
& \mathrm{IRM:} \quad \overline{S_2}(N=0)
= -\ln x +\frac{4}{3} x^2-\frac{637}{90}x^4
+\frac{301328 }{2835}x^6-\frac{108056999}{37800}x^8+O(x^{10}),\\
& \mathrm{BR:} ~~\quad
\overline{S_2}(N=1)
= -\ln x +x^2-\frac{949}{180}x^4+\frac{1900303 }{22680}x^6-\frac{26053301}{11200}x^8 + O(x^{10}).
\end{align}
\end{subequations}

A similar procedure can be employed to obtain the von Neumann entropy. We have
\begin{equation}
\begin{aligned}
\overline{S_1}(N)
&=\lim\limits_{n\rightarrow 1}\overline{S_n}(N)\\
&=\lim\limits_{n\rightarrow 1}\frac{1}{1-n}\overline{(\tr(\check \rho))^N \ln (\tr (\rho^n) )}\\
&=-\lim\limits_{n\rightarrow 1} \frac{d}{dn}\lim\limits_{k\rightarrow 0}
\frac{d}{dk}\overline{(\tr(\check{\rho}^n))^k(\tr(\check{\rho}))^{N-nk}}.
\end{aligned}
\end{equation}
The von Neumann entropies in the scaling limit  for the two classes:
\begin{subequations}
\label{eq:S1_IRM_BR}
\begin{align}
& \mathrm{IRM:} \quad \overline{S_1}(N=0)
= -\ln x + 1-\gamma + \frac{11}{24} x^2-\frac{1739}{2880}x^4+\frac{329489}{181440}x^6 - \frac{83530439}{9676800}x^8+O(x^{10}),
\\
& \mathrm{BR:} ~~\quad
\overline{S_1}(N=1)
= -\ln x + 1-\gamma + \frac{5}{24} x^2-\frac{239}{2880}x^4+\frac{3679}{36288}x^6 - \frac{2423279}{9676800}x^8+O(x^{10}),
\end{align}
\end{subequations}
 where $\gamma\approx 0.577216$ is the Euler-Mascheroni constant.

\section{Exact results for product of Ginibre matrices at finite $t$}

\subsubsection{Finite $t$ results for $\overline{\tr(\check \rho^n)}$}

In this appendix, we review the exact random matrix results for the statistics of product of Ginibre matrices from Refs~ \cite{PhysRevE.83.061118} and \cite{Akemann_2013}. The exact results contain non-universal terms away from the scaling limits. 

We use the notations defined in Sec.~\ref{sec:ginibre}. The leading in $q$ behavior of the normalized eigenvalues of $\check \rho = M M^\dag$ for finite $t$ has the form
\begin{equation}\label{ptx}
P_t(x)=\sum_{k=1}^t
\Lambda_{k,t} x^{\frac{k}{t+1}-1}{}_tF_{t-1}\left\{\left[\left(1-\frac{1+j}{t}+\frac{k}{t+1}\right)_{j=1}^t\right],
\left[\left(1+\frac{k-j}{t+1}\right)^{k-1}_{j=1},\left(1+\frac{k-j}{t+1}\right)^t_{j=k+1}\right];\frac{t^t}{(t+1)^{t+1}}x\right\},
\end{equation}
where ${}_pF_q(a_1,...,a_p,b_1,...,b_q,z)$ is the generalized hypergeometric function, and the prefactor is $\Lambda_{k,t} = t^{-\frac{3}{2}}\sqrt{\frac{t+1}{2\pi}}
\left(\frac{t^{\frac{t}{t+1}}}{t+1}\right)^k\frac{\prod_{j=1}^{k-1}\Gamma(\frac{j-k}{t+1})\Gamma_{j=k+1}^t\Gamma\left(\frac{j-k}{t+1}\right)}{\prod_{j=1}^t\Gamma\left(\frac{j+1}{t}-\frac{k}{t+1}\right)}$. One has
\begin{equation}\label{cat1}
\mathbb{E}[\varepsilon^n] = \int^{\frac{(t+1)^{t+1}}{t^t}}_0 \varepsilon^n P_t(\varepsilon) d\varepsilon = \frac{1}{tn+1}\binom{nt+n}{n},
\end{equation}
where the right-hand side defines the Fuss-Catalan number. Note that here $\varepsilon$ is the normalized limiting eigenvalue; it is related to the original (un-normalized) limiting eigenvalue $\lambda$ by $\varepsilon = \lambda/q^t$. The actual eigenvalue distribution of $\check \rho$, which we denote by $R_t(\lambda)$, is related to the normalized distribution by $R_t(\lambda) = q^{1-t}P_t(\frac{\lambda}{q^t})$, and we have
\begin{equation}
\mathbb{E}[\lambda^n] = \frac{1}{q}\int^{ \frac{(t+1)^{t+1}}{t^t} q^t}_0\lambda^n R_t(\lambda)d\lambda = q^{tn}\cdot \frac{1}{tn+1}\binom{nt+n}{n}.
\end{equation}
Note that the prefactor $1/q$ is necessary for consistency.

The finite $q$, finite $t$ case was studied by Akemann et al. \cite{Akemann_2013}. The main objective is to calculate the moments of the eigenvalues of $\check \rho$: 
\begin{equation}\label{rmtr}
\overline{\tr(\check \rho^n)}= q \mathbb{E}[\lambda^n]
=\sum_{l=0}^{q-1}\frac{(-1)^{q-l-1}}{n!}\left(\frac{(n+l)!}{l!}\right)^{t+1}\binom{n-1}{q-l-1}.
\end{equation}

We list the first few terms in orders of $q^{-1}$:

\begin{equation}\label{numeratr}
\begin{aligned}
\overline{\tr(\check \rho^n)}
=&q^{nt+1}\left(\frac{1}{nt+1}\binom{nt+n}{n}+
\frac{(n-1)(t+1)(nt-2)}{24}\binom{nt+n-2}{n-1}q^{-2}\right.\\
&\left.+\frac{(n-1) (t+1) (n t-4) (5 n^3 t^2+5 n^3 t-7 n^2 t^2-31 n^2 t-6 n^2+24 n t+30 n-36)}{5760}\binom{nt + n -4}{n-1} q^{-4}+ O(q^{-6})\right)
\end{aligned}
\end{equation}

\subsubsection{Finite $t$ expansion for $\overline{(\tr(\check \rho))^n}$}

We use the combinatorial interpretation to expand $\overline{\left(\tr(\check \rho)\right)^n}$ as
\begin{equation}
\overline{\left(\tr(\rho)\right)^n}
=
\sum_{\sigma_1,...,\sigma_t \in S_n} q^{n(t+1)-h_n(\sigma_1,...,\sigma_t)},
\end{equation}
where we define
\begin{equation}
h_n(\sigma_1,...,\sigma_t):=|\sigma_1|+|\sigma_2\sigma_1^{-1}|+\cdots + |\sigma_t\sigma_{t-1}^{-1}| + |\sigma_t| .
\end{equation}
Note, here $\sigma_1,...,\sigma_t$ are not necessarily transpositions. Calculating the first few orders gives
\begin{equation}\label{denonontr}
\overline{(\tr(\check \rho))^n}
=
q^{n(t+1)} \left(1 + 
\frac{t(t+1)n(n-1)}{4}q^{-2}
+
\frac{t(t+1)n(n-1)(4 + 3 (4 - 5 n + n^2) t + (-12 + n + 3 n^2) t^2)}{96} q^{-4}+O(q^{-6})\right).
\end{equation}

\subsubsection{Non-universal terms for the von Neumann entropy}
\label{subsec:S1_nonuniversal}

Plugging in Eqs.~\eqref{numeratr} and \eqref{denonontr} to Eq.~\eqref{vNBR}, we obtain the von Neumann entropy for the BR case ($N=1$):
\begin{equation}
\overline{S_1}|_{\text{BR}}
=
-\ln x+1-\gamma+\frac{5}{24} x^2
-\frac{3}{2t} + \frac{13}{12 t^2}
+ \frac{6 x^2}{24 t}+ \frac{x^2}{12 t^2}+\cdots
\end{equation}

\end{widetext}

\section{Non-universal contributions at small $x$}
\label{app:small_x}

There are non-universal contributions to numerical data computed at finite $q$ and finite $t$. For a fixed $q$, these contributions are significant at small enough $x$, when compared with e.g. the universal $O(x^2)$ term in the small $x$ expansion.

In order to make the comparison with small $x$ asymptotic series, we have computed and subtracted these non-universal terms, or more precisely their $q\to\infty$ limit. Here we describe our approach to analytically compute these terms. 

The non-universal terms for $S_2$ can be analytically computed through a $q\rightarrow\infty$ expansion. In this limit, the universal part, aside from the leading constant + $\ln 1/x$, all vanish. It thus isolates out the leading piece of the non-universal part at small $x$. 

In the domain wall picture, $q\rightarrow\infty$ makes the creation cost of new domain walls prohibitively large. Therefore, for the Ginibre ensemble, 
\begin{equation}
    \langle 1^N | T^{t+1} | 1^{N-2k} 2^k \rangle \sim  q^{-k} ( t + 1)^k 
\end{equation}
where $q^{-k}$ is the weight of $k$ domain walls and $(t+ 1)^k$ counts different ways to distribute them into $t+1$ links. Analytically continuing to $k = 0$ gives
\begin{equation}
    -\partial_k  q^{-k} ( t + 1)^k \Big|_{k=0} = \ln \frac{q}{t+1} 
     = \ln \frac{1}{x} + \ln ( 1 + \frac{1}{t} ).
\end{equation}
The non-universal term is thus
\begin{equation}
\label{eq:S2_non_universal_Ginibre}
    \ln ( 1 + \frac{1}{t} ).
\end{equation}
The non-universal term is the same for Haar unitary followed by the projective measurement for our parameter choice of $q_{\rm tot} = 2q_s = q$. In terms of the transfer matrix, we compute
\begin{equation}
    \langle \mu | \mathbb{P} | \sigma_t^* \rangle \langle \sigma_t | \mathbb{P} | \sigma_t^*\cdots \langle \sigma_{2}| \mathbb{P} | \sigma_t^* \rangle \langle \sigma_t | \rangle .
\end{equation}
In the $q_{\rm tot} \rightarrow \infty$ limit while keeping $q_{\rm tot} = 2q_s$, the domain wall cost for the $t$ links connected by $\mathcal{T}$ is $\frac{1}{q}$ and the $\frac{1}{q_{\rm tot}}$ for $T$. The cost happens to be the same and the non-universal term is again given by the entropy of the domain wall, i.e. Eq.~\eqref{eq:S2_non_universal_Ginibre}. 

We believe the non-universal terms for the measurement process are also the domain wall configurational entropy. However accessing the non-universal terms in $S_1$ requires $n \rightarrow 1$ limit, at which the counting of distributing the domain walls is more involved, even when the number of domain walls is kept minimized at $(n-1)k$ when $q\rightarrow \infty$. We have a conjectured closed-form expression for the non-universal term of the measurement process
\begin{equation}
    1 - \gamma + \texttt{PolyGamma}[0, t+2] - \ln t 
\end{equation}
which is consistent with the large $t$ series expansion obtained in Sec.~\ref{subsec:S1_nonuniversal}
\begin{equation}
\label{eq:S1_non_universal}
    1 - \gamma - \frac{3}{2t} + \frac{13}{12t^2} - \frac{1}{t^3} + \frac{119}{120t^4} + \cdots. 
\end{equation}

Eq.~\eqref{eq:S2_non_universal_Ginibre}, and Eq.~\eqref{eq:S1_non_universal} are the terms subtracted in Fig.~\ref{fig:N0_small_x} and Fig.~\ref{fig:N1_small_x} for the comparison with the small $x$ expansion. 

\section{Resummation over partitions}
\label{app:resum}
In Sec.~\ref{sec:vNresummation} of the main text, we sketch the steps to analytically continue (the $N$ derivative of) the moments of the trace
\begin{equation}
\label{eq:tr_rho_n_char}
\overline{\tr(\rho )^n} = \Omega( N, x ) = \frac{1}{N!} \sum_{\lambda \vdash N} (d^{\lambda})^2 \exp( x \nu( \lambda ) ) 
\end{equation}
to $N = 1$ through its generating function
\begin{equation}
G( u, x ) = \sum_{N = 0}^{\infty} \frac{(-u)^N}{N!} \Omega( N, x ).  
\end{equation}
To obtain results that are at least numerically tractable, we approximate $G(u,x)$ via $G_R(u, x)$ with a cutoff parameter $R$, which limits the maximal number of rows in the sum over partitions. Eventually, $\partial_N \Omega(N, x)\Big|_{N-1}$ is expressed as a determinant of an $R\times R$ matrix, which is exact at $R \rightarrow \infty$ yet converges fast enough that a $R \sim 8$ approximation gives excellent agreement with the numerical results in Sec.~\ref{sec:numerics}.  

In this Appendix, we substantiate the steps of the resummation. Necessary group theory concepts will be reviewed along with simple examples. We follow the steps below: 
\begin{enumerate}
\item In Sec.~\ref{appsec:review_group}, we review basic facts about irreducible representations of the symmetric group. 
\item In Sec.~\ref{appsec:first_hook}, we organize the summation over partitions in Eq.~\eqref{eq:tr_rho_n_char} in terms of partitions of $R$ rows, where each can be written as a summation over the hook-length variable of the first column: $h_1, \cdots, h_R$. 
\item In Sec.~\ref{appsec:remove_N}, we truncate the summation to partitions over at-most $R$ rows and convert the sum of $N$ into an unconstrained sum over $h_1, \cdots, h_R$. 
\item In Sec.~\ref{appsec:det}, we convert the unconstrained over $h_1, \cdots, h_R$ into a single determinant of a  $R \times R$ matrix with divergent elements.
\item Finally in Sec.~\ref{appsec:regulate}, we regulate the divergent $R\times R$ divergent elements through an integral representation. 
\end{enumerate}

\subsection{Dimensions of the Irreps of $S_N$}
\label{appsec:review_group}

It is well known that the irreducible representations (irreps) of the symmetric group $S_N$ can be denoted as semi-standard Young tableaux. The number of boxes $\lambda_j$ of $j$th row is a non-negative integer that is non-increasing for $j$. The total number of boxes is $N$. Hence $\lambda := \{ \lambda_1 \ge \lambda_2 \ge \cdots \ge \lambda_R \}$ forms a partition of $N$ , which is denoted as $\lambda \vdash N$. See examples of irreps in terms of Young Tableaux for $S_4$ in Tab.~\ref{tab:irrep_S4}.

\ytableausetup{centertableaux,boxsize = 1em}
\begin{table}[h]
\centering
\begin{tabular}{ |c|c|c|c|c|c| } 
 \hline
$\lambda$& \ytableaushort{4 3 2 1} & \ytableaushort{4 2 1, 1} & \ytableaushort{3 2, 2 1}
& \ytableaushort{4 1, 2 1} &  \ytableaushort{4, 3, 2, 1} \\ \hline
$d^{\lambda}$ & 1 & 3 & 2 & 3 & 1 \\ \hline
\end{tabular}
\caption{Examples of Young tableaux: Irreducible representations of $S_4$. The numbers in the boxes are the hook length.}
\label{tab:irrep_S4}
\end{table}

A hook $h_{s,t}$ at position $s, t$ of a Young tableau includes all the boxes to its right, below, and itself. The hook length $|h_{s,t}|$ is the number of boxes in a hook. It equals the number of boxes to its right and below, plus one (itself). In the first row of Tab.~\ref{tab:irrep_S4}, we fill the hook length in each box of the Young tableau.

The dimension of an irrep can be computed from the hook length formula
\begin{equation}
d^{\lambda} = \frac{N!}{ \prod_{s,t} |h_{s,t}|}, 
\end{equation}
namely the number of group elements $N!$ divided by all the hook lengths in a Young tableau. For example, the dimension of the irrep $\ytableaushort{4 1, 2 1}$ is $\frac{4!}{4 \times 1 \times 2 \times 1} = 3$. We display the dimensions of $S_4$ irreps in the second row of Tab.~\ref{tab:irrep_S4}. It is easy to check that the total dimension identity $\sum_{\lambda} \left( d^{\lambda} \right)^2 = N!$ is satisfied.

\subsection{Summation in terms of the hook length of the first column}
\label{appsec:first_hook}

One challenge to the summation over $N$ and all partitions is the constraints over valid partitions: 
\begin{equation}
\sum_{ \lambda \vdash N} 
 = \sum_{R=0}^{\infty} \sum_{\stackrel{\lambda_1 \ge \lambda_2 \ge \cdots \ge \lambda_R}{ \sum_{j=1}^R \lambda_j = N}}. 
\end{equation}
In addition, the factor $d^{\lambda}$ in the sum Eq.~\eqref{eq:tr_rho_n_char} is not directly expressed in the partition variable $\lambda_j$. 

We proceed by changing the summation variables from $\lambda_j$ to the hook lengths of the first column, calling them $h_j$:
\begin{equation}
h_j \equiv h_{j,1}, \quad j = 1, \cdots, R. 
\end{equation}
The variables $h_j$ are related to $\lambda_j$ through the definition of the hook length,
\begin{equation}
\label{eq:h_j_lambda_j}
h_j = h_{j,1} = \underbrace{\lambda_j}_{\text{self + boxes to the right}} + \underbrace{R - j}_{\text{boxes below}}. 
\end{equation}
They satisfy the following constraints: 
\begin{equation}
\begin{aligned}
  h_1 >  &h_2  > \cdots > h_R \ge 0 ,  \\
  \sum_{j =1}^{R} h_j &= N + \sum_{j=1}^R ( R - j) = N + \frac{R(R-1)}{2} \\
&= N + \sum_{j=1}^{R} (j-1).
\end{aligned}
\end{equation}
Hence we can rewrite the sum over partitions as a constrained sum over $\{h_j\}$:
\begin{equation}
\sum_{ \lambda \vdash N} 
 = \sum_{R=0}^{\infty} \sum_{\stackrel{h_1 > h_2 > \cdots > h_R \ge 0 }{ \sum_{j=1}^R h_j = N + \sum_j (j-1)}}. 
\end{equation}
Next, we show that all the factors in the sum (Eq.~\eqref{eq:tr_rho_n_char}) can be expressed in  $\{h_j\}$. 

{\bf Dimension of the irrep} $d^{\lambda}$: 

It is a standard result (see e.g. Chap.~7 of Ref.~\cite{stanley_enumerative_1999}) that a (non-negative) integer smaller than $h_{s,t}$ can either be a hook length on its right ($h_{s,u}, u>t$), or $h_{v,t} - h_{s, t}$ for $v < s$. Therefore
\begin{equation}
\prod_{t<u} h_{s,t} \prod_{v<s} (h_{v,1} - h_{s,1} )  = h_{s,1}! . 
\end{equation}
We can then rewrite the hook length formula in $\{h_j\}$:
\begin{equation}
d^{\lambda} = \frac{N!}{ \prod_{j=1}^R h_j! } \prod_{j<k} ( h_j - h_k).
\end{equation}

{\bf The exponent}: $\nu ( \lambda )$

The exponent is the difference of sum between a Young tableau and its transpose
\begin{equation}
\begin{aligned}
\nu( \lambda) &= \frac{1}{2} \sum_{\lambda \vdash N} \lambda^2 - \lambda^{\top 2} \\
 &\equiv \frac{1}{2} \sum_{\lambda \vdash N} \sum_{j=1}^R \lambda_j^2  -  \frac{1}{2} \sum_{j =1}^{R'} (\lambda_j')^2,
\end{aligned}
\end{equation}
where $\lambda_{j'}$ is the number of boxes in the transpose of $\lambda$. For instance, \ydiagram{4, 3, 2} is the transpose of \ydiagram{3,3,2,1}. Since the total number of boxes is invariant after transposition, we also have $\sum_{j'=1}^{R'} \lambda_j' = N$. We then subtract $\frac{1}{2} \sum_j \lambda_j$ and add $\frac{1}{2} \sum_{j'} \lambda_{j'}$, which gives
\begin{equation}
\nu(\lambda ) \equiv \sum_{\lambda \vdash N} \sum_{j=1}^R {\lambda_j \choose 2}  -  \sum_{j =1}^{R'} {\lambda_j' \choose 2}. 
\end{equation}
The binomial number ${n \choose 2}$ can be interpreted as $\sum_{j=0}^{n-1} j$. In the transposed Young tableau, we insert numbers $0$ to $\lambda_{j'} - 1$ from the left to the right in each row, for instance \ytableaushort{ 0 1 2 3, 0 1 2, 0 1 }. The sum $\sum_{j =1}^{R'} {\lambda_j' \choose 2}$ can be identified as the sum of all the numbers in the box. We then transpose back to the original Young tableau; for the example above, it is \ytableaushort{ 0 0 0, 1 1 1, 2 2, 3}. The sum of numbers in the box can also be 
\begin{equation}
\sum_{j'=1}^{R'}  {\lambda_{j'} \choose 2 } = \sum_{j=1}^{R}  ( j - 1 ) \lambda_j. 
\end{equation}
The exponent can be written entirely in $\{\lambda_j\}$
\begin{equation}
\begin{aligned}
\nu( \lambda ) &= \sum_{j=1}^R ( \frac{1}{2} (\lambda_j^2 - \lambda_j) - ( j - 1) \lambda_j )\\
&= \sum_j \frac{1}{2} \lambda_j^2 - ( j - \frac{1}{2} ) \lambda_j \\
&= \frac{N}{2} + \sum_j \frac{1}{2} \lambda_j^2 - j \lambda_j.
\end{aligned}
\end{equation}
We then substitute the $\lambda_j$ by the hook length of the first column $h_j$ (Eq.~\eqref{eq:h_j_lambda_j}) and get
\begin{equation}
\begin{aligned}
  \nu( \lambda ) &= \frac{N}{2} + \sum_j \frac{1}{2} ( \lambda_j - j + R )^2 - \frac{1}{2} (j-R)^2 - R \lambda_j \\
  &= \frac{N}{2}  + \frac{1}{2}\sum_j h_j^2 - RN - \frac{1}{2} \sum_{j=1}^R ( j - R)^2 \\
  &= \frac{1}{2} \sum_j h_j^2 + \frac{N}{2} - RN - \frac{(R - 1) R ( 2R - 1)}{12} \\
  &= (\frac{1}{2} - R) N  +  \frac{1}{2} \sum_{j=1}^R h_j^2 - \frac{1}{2} \sum_{j = 1}^R  (j - 1)^2. 
\end{aligned}
\end{equation}

With the expression of $d^{\lambda}$ and $\nu(\lambda)$, we finally have 
\begin{equation}
\begin{aligned}
&\Omega( u, x ) 
 = \frac{1}{N!} \underbrace{\lim_{R\rightarrow\infty} \sum_{\stackrel{h_1 > h_2 > \cdots > h_R \ge 0}{ \sum_{j=1}^R h_j = N + \sum_j (j-1)}}}_{\sum_{\lambda \vdash N}}\\
&\qquad \qquad \underbrace{\frac{N!^2}{ \prod_{j=1}^R (h_j!)^2 } \prod_{j<k} ( h_j - h_k)^2}_{(d^{\lambda})^2}\\
&\times\underbrace{\exp\Big\{ x (\frac{1}{2} - R) N  +  \frac{x}{2} \sum_{j=1}^R [( h_j^2 - (j - 1)^2 ] \Big\}}_{\exp(\nu(\lambda) x)}. 
\end{aligned}
\end{equation}
and
\begin{equation}
\label{eq:G_u_x_full}
\begin{aligned}
&G( u, x ) 
 = \sum_{N=0}^{\infty} \underbrace{\lim_{R\rightarrow\infty} \sum_{\stackrel{h_1 > h_2 > \cdots > h_R \ge 0 }{ \sum_{j=1}^R h_j = N + \sum_j (j-1)}}}_{\sum_{\lambda \vdash N}}\\
&\qquad \qquad \underbrace{\frac{1}{ \prod_{j=1}^R (h_j!)^2 } \prod_{j<k} ( h_j - h_k)^2}_{(d^{\lambda})^2 / (N!)^2}\\
&\times\underbrace{\exp\Big\{ x (\frac{1}{2} - R) N  +  \frac{x}{2} \sum_{j=1}^R [( h_j^2 - (j - 1)^2 ] \Big\}}_{\exp(\nu(\lambda) x)} (-u)^N. 
\end{aligned}
\end{equation}

\subsection{Fix $R$ and resum over $N$}
\label{appsec:remove_N}

Eq.~\eqref{eq:G_u_x_full} is an exact expression of $G(u,x)$ in terms of summation over the hook length of the first column. With $R \rightarrow \infty$ inside the sum, it can hardly be further simplified. To proceed, we fix $R$ to a finite integer and define the truncated generating function
\begin{equation}
\label{eq:G_R_u_x_N}
\begin{aligned}
&G_R( u, x ) 
 = \sum_{N=0}^{\infty} \sum_{\stackrel{h_1 > h_2 > \cdots > h_R \ge 0 }{ \sum_{j=1}^R h_j = N + \sum_j (j-1)}} \\
&\qquad \qquad \frac{1}{ \prod_{j=1}^R (h_j!)^2 } \prod_{j<k} ( h_j - h_k)^2\\
&\times\exp\Big\{ \underline{x (\frac{1}{2} - R) N}  +  \frac{x}{2} \sum_{j=1}^R [( h_j^2 - (j - 1)^2 ] \Big\} \underline{(-u)^N}. 
\end{aligned}
\end{equation}
Obviously
\begin{equation}
G(u, x) = \lim_{R\rightarrow \infty} G_R(u, x). 
\end{equation}
The underlined parts in Eq.~\eqref{eq:G_R_u_x_N} are the only explicit $N$ dependence in the summand. Let $\tilde{u} = u\exp( (\frac{1}{2} - R )x )$, the $N$ dependent term is $(-\tilde{u})^{N}$. We remove $N$ through the constraint, $\sum_{j=1}^R h_j = N + \sum_{j=1}^{R} (j-1)$, which gives
\begin{equation}
\label{eq:G_R_u_x_imp_N}
\begin{aligned}
&G_R( u, x ) 
 = \sum_{N=0}^{\infty} \sum_{\stackrel{h_1 > h_2 > \cdots > h_R \ge 0 }{ \sum_{j=1}^R h_j = N + \sum_j (j-1)}} \\
&\qquad \qquad \frac{1}{ \prod_{j=1}^R (h_j!)^2 } \prod_{j<k} ( h_j - h_k)^2\\
&\times\exp\Big\{ \frac{x}{2} \sum_{j=1}^R [( h_j^2 - (j - 1)^2 ] \Big\} (-\tilde{u})^{\sum_{j=1}^R [h_j - (j-1)]}. 
\end{aligned}
\end{equation}
The only $N$ dependence is in the constrained sum of $\{h_j\}$, but the sum over $N$ effectively removes the constraint $\sum_{j=1}^R h_j = N + \sum_j (j-1)$. Hence we have
\begin{equation}
\begin{aligned}
&G_R( u, x ) 
 =  \sum_{h_1 > h_2 \cdots > h_R \ge 0 } \frac{1}{ \prod_{j=1}^R (h_j!)^2 } \prod_{j<k} ( h_j - h_k)^2\\
&\times\exp\Big\{ \frac{x}{2} \sum_{j=1}^R [h_j^2 - (j - 1)^2 ] \Big\} (-\tilde{u})^{\sum_{j=1}^R [h_j - (j-1)]}. 
\end{aligned}
\end{equation}
The variables $\{h_j\}$ are ordered, and due to the factor $\prod_{j<k}( h_j - h_k)$ it is zero whenever any pair of them are equal. Switching the order of  $\{h_j\}$  to $h_2 > h_1 > h_3 > \cdots > h_R \ge 0 $ does not change the sum. Thus the over-counting by removing the strict descending order can be accounted by a $\frac{1}{R!}$ factor. We thus obtain an unconstrained sum for $G_R( u, x)$:
\begin{equation}
\label{eq:G_R_u_x_free}
\begin{aligned}
&G_R( u, x ) 
 =  \sum_{h_1 , h_2, \cdots , h_R= 0}^{\infty}  \frac{1}{R!}\frac{1}{ \prod_{j=1}^R (h_j!)^2 } \prod_{j<k} ( h_j - h_k)^2\\
&\times\exp\Big\{ \frac{x}{2} \sum_{j=1}^R [h_j^2 - (j - 1)^2 ] \Big\} (-\tilde{u})^{\sum_{j=1}^R [h_j - (j-1)]}
\end{aligned}
\end{equation}
where the variables $\{h_j\}$ are now independent.

\subsection{A single determinant}
\label{appsec:det}

The unconstrained sum in Eq.~\eqref{eq:G_R_u_x_free} contains a factor of
\begin{equation}
\prod_{j< k} (h_j - h_k )^2
\end{equation}
which can be recognized as the square of the Vandermonde determinant
\begin{equation}
\Delta( \vec{h} ) 
 \equiv \det( h_j^{k} )_{j,k+1=1}^R \equiv \prod_{j< k}(  h_j - h_k ).
\end{equation}
Summations in the form of
\begin{equation}
\label{eq:sum_F_delta_2}
\sum_{h_1, h_2, \cdots, h_R} \left( \prod_{k=1}^{R} F( h_k )  \right) \left( \prod_{j=1}^R G_{j} \right)  \Delta( \vec{h} )^2 
\end{equation}
can be converted into a single determinant using {Andr{\'e}ief} identity
\cite{forrester_meet_2019}, whose discrete version is also known as the Cauchy-Binet formula: 
\begin{equation}
\label{eq:andreief}
\begin{aligned}
&\int_I dx_1 \cdots \int_I d x_{R} \det [ f_{j-1}( x_k)]_{j,k=1}^{R} \det[ g_{j-1}( x_k)]_{j,k=1}^{R} \\
&= R! \det\left[ \int_I f_j(x) g_k ( x) dx \right]_{j,k=0}^{N-1}. 
\end{aligned}
\end{equation}

Summation with $F = G_j= 1$ in Eq.~\eqref{eq:sum_F_delta_2} corresponds to taking
\begin{equation}
f_{j-1}(h_k) = g_{j-1}(h_k)  = h_k^{j-1}.
\end{equation}
An intuitive understanding is to imagine a product of a $R \times \infty$ Vandermonde matrix times a $\infty \times R$ Vandermonde matrix:
\begin{equation}
\det
\underbrace{\begin{bmatrix}
1 & 1 & 2 &\cdots \infty \\
0 & 1^2 & 2^2 &\cdots \infty^2 \\
\vdots & \vdots & \vdots & \cdots \\ 
0 & 1^R & 2^R & \cdots \infty^R
\end{bmatrix}}_{A}
\underbrace{
\begin{bmatrix}
1 & 0 & \cdots  & 0 \\
1 & 1^2 &  \cdots &1^R \\
2 & 2^2 &  \cdots & 2^R \\
\vdots & \vdots & \vdots  & \cdots \\ 
\infty & \infty^2 & \cdots & \infty^R\\
\end{bmatrix}}_{B}
\end{equation}
Eq.~\eqref{eq:andreief} enables us to evaluate the determinant through the product of determinants by all possible samplings of $R$ columns of $A$ and the same $R$ rows of $B$:
\begin{equation}
\det( AB) = \sum_{h_1, \cdots h_R} \det(A_{h_1, \cdots, h_R} ) \det( B_{h_1, \cdots, h_R } ), 
\end{equation}
where the sub-indices $\{h_j\}$ here denote to take out the corresponding columns in $A$ and rows in $B$. 

The additional factor $F$ and $G$ can be distributed symmetrically to a factor multiplying the row ($G_j$) and column ($F(h_k)$)
\begin{equation}
f_{j-1}(h_k) = h_k^{j-1} \sqrt{F_k( h_k) G_j },
\end{equation}
which suggests
\begin{equation}
f_{j-1} (h_k) = ( -\tilde{u})^{\frac{1}{2}  h_k - \frac{1}{2} ( j - 1 )}  e^{ -\frac{x}{4} (j-1)^2} \frac{h_k^{j-1}}{h_k!} e^{ \frac{1}{4} x h_k^2 }. 
\end{equation}
The function $f$ has a factor $\frac{h_k^{j-1}}{h_k!}$. We can further do a series of row transformations (whose determinant is $1$ and does not change the overall determinant) to change it to $\frac{h_k( h_k - 1) \cdots (h_k -(j-2))}{h_k!} = \frac{1}{(h_k - (j-1))!}$. We end up with
\begin{equation}
  f_j(h_k) = ( - \tilde{u} )^{ \frac{1}{2} h_k - \frac{1}{2} j } e^{ - \frac{x}{4}j^2 }  e^{ \frac{1}{4}x h_k^2 } \frac{1}{(h_k - j)!}.
\end{equation}
Due to the symmetric choice to distribute the $F$ and $G$ factors, the function $g$ is the same. The truncated function $G_R$ thus becomes a single determinant: 
\begin{equation}
\label{eq:G_R_det}
  G_R( u, x ) = \det\left(  \sum_{h=0}^{\infty}  \frac{( -\tilde{u} )^{h - \frac{1}{2}( j+ k ) } e^{ - \frac{x}{4} (j^2 + k^2 ) } e^{ \frac{1}{2}x h^2 }}{(h-j)! (h-k)!} \right)_{j,k=0}^{R-1}.
\end{equation}
However, in the physical region, $x$ is positive and the $e^{\frac{x}{2} h^2}$ factor grows faster than a factorial. The matrix elements in Eq.~\eqref{eq:G_R_det} are divergent. 

\subsection{Regulate the sum through a Gaussian integral}
\label{appsec:regulate}

We introduce a Gaussian integral to regulate the Gaussian factor
\begin{equation}
\begin{aligned}
e^{\frac{1}{2} x h^2 } &= \frac{1}{\sqrt{2\pi}}\int_{-\infty}^{\infty} dw e^{- \frac{w^2}{2}} e^{ \sqrt{x} wh } \\
&= \frac{1}{\sqrt{2\pi x}} \int_{-\infty}^{\infty} dw e^{- \frac{w^2}{2x} + wh} .
\end{aligned}
\end{equation}
so that the variable $h$ becomes linear in the exponent. The sum over $h$ inside the integral is convergent: the resummed matrix element is related to a Bessel function. We will use the form below:
\begin{equation}
\sum_{h=0}^{\infty} \frac{z^{h- \frac{j+k}{2}}}{(h-j)!(h-k)!}
 = i^{|j-k|} J_{|j-k|} ( -2i \sqrt{z} ). 
\end{equation}
(One can check by matching the standard power series of the Bessel function $J_m( x ) = \sum_{l = 0}^{\infty} \frac{(-1)^l}{l! (m+l)!} \left( \frac{x}{2} \right)^{m + 2l}$.)

Consequently $G_R$ is the determinant of an $R\times R$ matrix
\begin{equation}
  G_R(u,x) = \det(I_{jk})_{j,k=0}^{R-1}
\end{equation}
whose elements are
\begin{equation}
\begin{aligned}
  I_{jk} = \frac{1}{\sqrt{2\pi x} }\int dw e^{- \frac{w^2}{2x}} \sum_{h=0}^{\infty} e^{ - \frac{x}{4} ( j^2 + k^2 ) } \\
   \frac{( - \tilde{u} )^{h  - \frac{j + k}{2}}  e^{ w ( h - \frac{j + k}{2} ) }  e^{w \frac{j + k}{2}}}{( h-j)! (h-k)!}.
\end{aligned}
\end{equation}
We can further shift the Gaussian integral to absorb $e^{w \frac{j+k}{2}}$: 
\begin{equation}
e^{- \frac{x}{4} (j^2 + k^2 ) } e^{ - \frac{w^2}{2x} +  w \frac{j+k}{2} } 
 = e^{ - \frac{1}{2x} ( w - \frac{(j+k)x}{2} )} e^{- \frac{x}{4} ( j - k )^2}
\end{equation}
and obtain
\begin{equation}
\begin{aligned}
&I_{jk}=
\int \frac{dw}{\sqrt{2\pi x} } e^{- \frac{w^2}{2x}} e^{ - \frac{x}{8} ( j - k )^2  } \\
&\qquad \qquad \times \sum_{h=0}^{\infty} \frac{( - \tilde{u} )^{h  - \frac{j + k}{2}}  e^{ (w + \frac{(j + k)x}{2}) ( h - \frac{j + k}{2} ) } }{( h-j)! (h-k)!}\\
&= \int \frac{dw}{\sqrt{2\pi x} } e^{- \frac{w^2}{2x}} e^{ - \frac{x}{8} ( j - k )^2  } i^{|j-k|} J_{|j-k|} ( 2 \sqrt{\tilde{u} e^{w + \frac{(j + k)x}{2}}} ) 
\end{aligned}
\end{equation}
which is the expression \eqref{Eq:I_j_k} in the main text. 

\bibliographystyle{apsrev4-1}
\bibliography{biblio.bib}

\end{document}